\pdfoutput=0
\documentclass[11pt]{article}
\usepackage{jheppub}
\usepackage{axodraw}
\usepackage{epsfig}
\usepackage{amsfonts}
\usepackage{bm,bbm}
\usepackage{xcolor}
\usepackage{cancel}
\newcommand{\la}[1]{\label{#1}}
\newcommand{\be}{\begin{equation}}
\newcommand{\ee}{\end{equation}}
\newcommand{\ba}{\begin{eqnarray}}
\newcommand{\ea}{\end{eqnarray}}
\newcommand{\fig}{fig.~}

\newcommand{\eq}{eq.~}
\newcommand{\eqs}{eqs.~}
\newcommand{\se}{sec.~}
\newcommand{\ses}{secs.~}
\newcommand{\app}{appendix~}
\newcommand{\nr}[1]{(\ref{#1})}
\newcommand{\tr}{{\rm tr\,}} 
\newcommand{\nn}{\nonumber \\}
\newcommand{\fr}[2]{{\frac{#1}{#2}\,}}

\renewcommand{\vec}[1]{{\bf #1}}
 
\newcommand{\tfr}[2]{{\textstyle \frac{#1}{#2}\,}}

\newcommand{\gammaE}{\gamma_\rmii{E}}

\newcommand{\rmO}{{\mathcal{O}}}

\def\lsi{\raise0.3ex\hbox{$<$\kern-0.75em\raise-1.1ex\hbox{$\sim$}}}
\def\gsi{\raise0.3ex\hbox{$>$\kern-0.75em\raise-1.1ex\hbox{$\sim$}}}

\newcommand{\rmi}[1]{{\mbox{\scriptsize #1}}}
\newcommand{\rmii}[1]{{\mbox{\tiny\rm{#1}}}}

\newcommand{\re}{\mathop{\mbox{Re}}}
\newcommand{\im}{\mathop{\mbox{Im}}}
\newcommand{\Tint}[1]{{\hbox{$\sum$}\!\!\!\!\!\!\!\int\,}_{\!\!\!\!\raise-0.9ex\hbox{$\scriptstyle{#1}$}}}
\newcommand{\Tinti}[1]{{{\Sigma}\!\!\!\!\raise0.3ex\hbox{$\int$}_\rmii{${#1}$}}}
\newcommand{\bi}{\begin{itemize}}
\newcommand{\ei}{\end{itemize}}
\newcommand{\hide}[1]{ }

\newcommand{\deltabar}{\raise-0.02em\hbox{$\bar{}$}\hspace*{-0.8mm}{\delta}}
\newcommand{\ddeltabar}{\raise-0.18em\hbox{$\bar{}$}\hspace*{-0.8mm}{\delta}}
\newcommand{\A}{\mathcal{A}}

\newcommand{\F}{\mathcal{F}}
\renewcommand{\H}{\mathcal{H}}
\newcommand{\I}{\mathcal{I}}
\newcommand{\iI}{\rmii{$I$}}
\newcommand{\J}{\mathcal{J}}

\renewcommand{\L}{\mathcal{L}}

\renewcommand{\P}{\mathcal{P}}

\newcommand{\X}{\mathcal{X}}
\newcommand{\Y}{\mathcal{Y}}

\newcommand{\mpl}{m_\rmii{pl}} 

\newcommand{\varh}{\mathbbm{h}} 

\newcommand{\tensor}{\rmi{t}}

\newcommand{\vac}{\rmi{vac}} 
\newcommand{\cl}{\rmi{\hspace*{0.3mm}cl}}

\newcommand{\zero}{{\scriptscriptstyle (0)}}
\newcommand{\one}{{\scriptscriptstyle (1)}}
\newcommand{\meas}[1]{\frac{2 k^d_{ } {#1} }{(4\pi)^{\frac{d}{2}}_{ }
 \Gamma(\frac{d}{2})}}
\newcommand{\bit}{\hspace*{0.4mm}}

\newcommand{\lift}{\vphantom{ |^b_q }}

\def\Lwidth{1}

\def\Agl(#1,#2)(#3,#4,#5){\PhotonArc(#1,#2)(#3,#4,#5){\Lwidth}
{6.283 #3 mul 360 div #4 #5 sub #4 #5 sub mul sqrt mul Ldensity mul}}
\def\Lgl(#1,#2)(#3,#4){\Photon(#1,#2)(#3,#4){\Lwidth}
{#1 #3 sub #1 #3 sub mul #2 #4 sub #2 #4 sub mul add sqrt Ldensity mul}}
\def\Aqu(#1,#2)(#3,#4,#5){\ArrowArc(#1,#2)(#3,#4,#5)}
\def\Aaqu(#1,#2)(#3,#4,#5){\ArrowArcn(#1,#2)(#3,#5,#4)}
\def\Lqu(#1,#2)(#3,#4){\ArrowLine(#1,#2)(#3,#4)}
\def\Laqu(#1,#2)(#3,#4){\ArrowLine(#3,#4)(#1,#2)}
\def\Aqq(#1,#2)(#3,#4,#5){\CArc(#1,#2)(#3,#4,#5)}
\def\Lqq(#1,#2)(#3,#4){\Line(#1,#2)(#3,#4)}
\def\Agh(#1,#2)(#3,#4,#5){\DashArrowArc(#1,#2)(#3,#4,#5){1}}
\def\Aagh(#1,#2)(#3,#4,#5){\DashArrowArcn(#1,#2)(#3,#5,#4){1}}
\def\Lgh(#1,#2)(#3,#4){\DashArrowLine(#1,#2)(#3,#4){1}}
\def\Lagh(#1,#2)(#3,#4){\DashArrowLine(#3,#4)(#1,#2){1}}
\def\TAsc(#1,#2)(#3,#4,#5)%
{\SetWidth{2.0}\ArrowArc(#1,#2)(#3,#4,#5)\SetWidth{1.0}}
\def\Adsc(#1,#2)(#3,#4,#5){\DashCArc(#1,#2)(#3,#4,#5){3}}
\def\Ldsc(#1,#2)(#3,#4){\DashLine(#1,#2)(#3,#4){3}}
\def\Asc(#1,#2)(#3,#4,#5){\CArc(#1,#2)(#3,#4,#5)}
\def\Lsc(#1,#2)(#3,#4){\Line(#1,#2)(#3,#4)}
\def\Ahh(#1,#2)(#3,#4,#5){\GluonArc(#1,#2)(#3,#4,#5){\Lwidth}
{6.283 #3 mul 360 div #4 #5 sub #4 #5 sub mul sqrt mul Ldensity mul}}
\def\Lhh(#1,#2)(#3,#4){\Gluon(#1,#2)(#3,#4){\Lwidth}
{#1 #3 sub #1 #3 sub mul #2 #4 sub #2 #4 sub mul add sqrt Ldensity mul}}
\def\Ahh(#1,#2)(#3,#4,#5){\PhotonArc(#1,#2)(#3,#4,#5){\Lwidth}
{6.283 #3 mul 360 div #4 #5 sub #4 #5 sub mul sqrt mul Ldensity mul}}
\def\Lhh(#1,#2)(#3,#4){\Photon(#1,#2)(#3,#4){0.6} 
{#1 #3 sub #1 #3 sub mul #2 #4 sub #2 #4 sub mul add sqrt Ldensity mul}}

\newcommand{\opicx}[1]{\;\parbox[c]{45pt}{\begin{picture}(70,30)(-10,5)
\SetWidth{1.0}\SetScale{0.8} #1 \end{picture}}\; }
\def\AmplAgauge{\opicx{%
 \Line(0,30)(70,30)%
 \PhotonArc(35,30)(25,180,270){1.5}{5}%
 \PhotonArc(35,30)(25,270,360){-1.5}{5}%
 \PhotonArc(35,30)(22,180,270){1.5}{5}%
 \PhotonArc(35,30)(22,270,360){-1.5}{5}%
 \def\Lwidth{0.5}
 \Line(35,0)(35,14)
 \Line(35,0)(33,-2)
 \Line(37,16)(35,14)
}}
\def\AmplBgauge{\opicx{%
 \Line(0,30)(70,30)%
 \PhotonArc(35,30)(25,180,270){1.5}{5}%
 \PhotonArc(35,30)(25,270,360){-1.5}{5}%
 \PhotonArc(35,30)(22,180,270){1.5}{5}%
 \PhotonArc(35,30)(22,270,360){-1.5}{5}%
 \CCirc(16.5,14.5){7}{White}{White}
 \GCirc(21,10){2.5}{0}
 \GCirc(12,20){2.5}{0}
 \Adsc(16.5,14.5)(7,225,45)
 \Adsc(16.5,14.5)(7,45,225)
 \def\Lwidth{0.5}
 \Line(35,0)(35,14)
 \Line(35,0)(33,-2)
 \Line(37,16)(35,14)
}}
\def\AmplCgauge{\opicx{%
 \Line(0,30)(70,30)%
 \PhotonArc(35,30)(25,180,270){1.5}{5}%
 \PhotonArc(35,30)(25,270,360){-1.5}{5}%
 \PhotonArc(35,30)(22,180,270){1.5}{5}%
 \PhotonArc(35,30)(22,270,360){-1.5}{5}%
 \CCirc(53.5,14.5){7}{White}{White}
 \GCirc(49,10){2.5}{0}
 \GCirc(58,20){2.5}{0}
 \Adsc(53.5,14.5)(7,225,45)
 \Adsc(53.5,14.5)(7,45,225)
 \def\Lwidth{0.5}
 \Line(35,0)(35,14)
 \Line(35,0)(33,-2)
 \Line(37,16)(35,14)
}}
\def\AmplDgauge{\opicx{%
 \Line(0,30)(70,30)%
 \PhotonArc(35,30)(25,180,270){1.5}{5}%
 \PhotonArc(35,30)(25,270,360){-1.5}{5}%
 \PhotonArc(35,30)(22,180,270){1.5}{5}%
 \PhotonArc(35,30)(22,270,360){-1.5}{5}%
 \CCirc(35,5){7}{White}{White}
 \GCirc(27,8){2.5}{0}
 \GCirc(43,8){2.5}{0}
 \Adsc(35,7)(7,0,180)
 \Adsc(35,7)(7,180,360)
 \def\Lwidth{0.5}
 \Line(35,-2)(35,16)
 \Line(35,-2)(33,-4)
 \Line(37,18)(35,16)
}}
\def\AmplEgauge{\opicx{%
 \Line(0,30)(70,30)%
 \PhotonArc(35,30)(25,180,270){1.5}{5}%
 \PhotonArc(35,30)(25,270,360){-1.5}{5}%
 \PhotonArc(35,30)(22,180,270){1.5}{5}%
 \PhotonArc(35,30)(22,270,360){-1.5}{5}%
 \GCirc(16.5,14.5){2.5}{0}
 \Adsc(11,8)(7,45,405)
 \def\Lwidth{0.5}
 \Line(35,0)(35,14)
 \Line(35,0)(33,-2)
 \Line(37,16)(35,14)
}}
\def\AmplFgauge{\opicx{%
 \Line(0,30)(70,30)%
 \PhotonArc(35,30)(25,180,270){1.5}{5}%
 \PhotonArc(35,30)(25,270,360){-1.5}{5}%
 \PhotonArc(35,30)(22,180,270){1.5}{5}%
 \PhotonArc(35,30)(22,270,360){-1.5}{5}%
 \GCirc(53.5,14.5){2.5}{0}
 \Adsc(59,8)(7,135,495)
 \def\Lwidth{0.5}
 \Line(35,0)(35,14)
 \Line(35,0)(33,-2)
 \Line(37,16)(35,14)
}}

\makeatletter \@addtoreset{equation}{section} \makeatother
\renewcommand{\theequation}{\arabic{section}.\arabic{equation}}
\makeatletter
\renewcommand\section{\@startsection {section}{1}{\z@}%
                                   {-5.5ex \@plus -1ex \@minus -.2ex}
                                   {2.3ex \@plus.2ex}%
                                   {\normalfont\large\bfseries}}
\renewcommand\subsection{\@startsection{subsection}{2}{\z@}%
                                     {-3.25ex\@plus -1ex \@minus -.2ex}%
                                     {1.5ex \@plus .2ex}%
                                     {\normalfont\normalsize\bfseries}}
\renewcommand\thesection {\@arabic\c@section}
\renewcommand\thesubsection   {\thesection.\@arabic\c@subsection}
\renewcommand{\@seccntformat}[1]{%
\csname the#1\endcsname.\hspace{1.0em}}
\makeatother

\begin{document}



\begin{flushright}
November 2026
\end{flushright}

\arxivnumber{2605.27910}

\vspace*{-0.95cm}


\title{\boldmath\Large
 Matching second-order classical and 1-loop quantum \\[1mm]
 tensor power spectra in de Sitter spacetime
}


\author[a]
{A.~Hauser,} 
\author[a]
{M.~Laine}


\affiliation[a]{
AEC, 
Institute for Theoretical Physics, 
University of Bern, \\ 
Sidlerstrasse 5, CH-3012 Bern, Switzerland}

\emailAdd{armando.hauser@students.unibe.ch}
\emailAdd{laine@itp.unibe.ch}


 

 
%
\abstract{
Large corrections to the inflationary tensor power spectrum have been
speculated to emerge either as second-order scalar-induced classical
effects, or as 1-loop quantum corrections. These two sources are not
independent of each other. Choosing the example of a massless
minimally coupled scalar field, we show how the full 1-loop result 
can be divided into its classical and vacuum parts. Working first 
in dimensional regularization, we show that the classical part is IR
divergent, with IR referring to small comoving momenta that have an
influence for a very long time. In the full 1-loop quantum result,
these divergences cancel. Introducing then a momentum cutoff that
permits for a numerical evaluation of the classical contribution, 
we show that the IR sensitivity manifests itself as a cubic
divergence. We suggest a procedure of ``non-perturbative
renormalization'' for extracting physical information not affected by
the divergence. If this can be implemented in realistic systems, 
it could consolidate numerical studies of inflationary 
scalar-induced gravitational waves.
}







\maketitle
\flushbottom



\section{Introduction}
\la{se:intro}

While inflationary cosmology has traditionally focussed on the 
physics of very large wavelengths, visible in the 
cosmic microwave background radiation (CMB) and large-scale
structure formation (LSS), recent years have enlarged the domain
of interest into smaller wavelengths. This concerns both scalar 
(i.e.\ density) and tensor (i.e.\ gravitational-wave)
perturbations. While the post-inflationary evolution 
of the scalar perturbations is quite complicated, gravitational waves 
propagate almost unperturbed till present day. Therefore, 
with the advent of gravitational-wave astronomy, tensor perturbations 
may offer for a direct probe 
of the inflationary and reheating epochs. 

A topic of particularly intense activity has been 
that of scalar-induced gravitational waves (SIGW).
This is a broad notion, which could be used either for
generating the frozen-out tensor power spectrum during
inflation, or the post-inflationary gravitational-wave
background as scalar perturbations re-enter inside
the Hubble horizon; 
cf., e.g., refs.~\cite{kt,mollerach,matarrese,%
nakamura,wands,sigw} for foundations, 
refs.~\cite{md1,racco,%
terada,inomata,gong,sasaki,ng,wni,white,%
minnesota,chicago,cas,sigw_theory,%
munich,garcia,xx,angelo,zurich}
for selected applications, 
and refs.~\cite{domenech,polter} for reviews.
In either context, 
the basic idea is that enhanced scalar fluctuations
source classical gravitational waves 
at second order in perturbations. 
To study such dynamics reliably, will ultimately 
necessitate large-scale numerical simulations 
(cf., e.g., refs.~\cite{xx,angelo,zurich} and references therein
for steps in that direction).

Issues somewhat similar to SIGW generation 
have also been discussed in another language, 
namely as 1-loop corrections~\cite{sw1,sw2,classic,baumann,explicit,mb0,mb} 
to the inflationary tensor power spectrum. On the theoretical side, 
two separate lines of research are worth underlining. One is 
an effective-theory inspired approach where, rather than computing
the tensor power spectrum directly, the goal is to construct an 
effective action for soft tensor perturbations. It can then be 
shown that 1-loop corrections cancel in the soft graviton
self-energy, as required by a Ward identity, 
and this in turn forbids the presence of 
overly large loop corrections 
(cf., e.g., refs.~\cite{ema,beijing}).
The second line of research is the determination of 
partial (cf., e.g., refs.~\cite{log1,log2,usr}) or
full 1-loop corrections to the tensor power spectrum
(cf., e.g., refs.~\cite{conformal,seoul,madrid}). 
Full 1-loop computations
are more demanding than partial 
arguments, 
and have only been completed 
for simple systems, notably massless 
conformally or minimally coupled spectator scalar fields. 
The full 1-loop corrections are non-vanishing, however a late-time
logarithm in them has been found to cancel, and this was again  
connected to a general symmetry argument~\cite{madrid}. 

The goal of the present paper is to demonstrate,
in the inflationary context,
that SIGW-like contributions represent a subpart of the full
quantum corrections. We refer to SIGW-like contributions
as {\em classical} ones. However, it
is appropriate to stress that  
we adopt a notion  
in which classicality refers to a general-relativistic 
gravitational-wave solution, 
while the scalar source is treated as consisting of  
quantum-mechanical mode functions, which 
requires a complexification of the scalar field space. 
The physical thinking is that 
scalar fields maintain their quantum properties, unless decohered 
by additional phenomena, such as thermal damping and noise. 
A benefit of this approach is that 
the ``classical'' term can be naturally identified 
as corresponding to specific graviton propagators
within the Keldysh $r/a$ basis 
of the full quantum-mechanical in-in computation.
 Our notion of classicality agrees with the general SIGW idea in which 
 gravitational waves originate from vacuum-generated
 scalar spectra~\cite{sigw},
 and specifically with one of the works with which we compare~\cite{ema}.
That said, classicality has also been employed in 
a broader sense in the literature, 
incorporating the source dynamics
(cf., e.g., refs.~\cite{clas1,clas14,clas15,clas2}), 
 and in \app\ref{app:def_cl} we compare with one such
 enlarged notion. 

Once the classical contributions have been identified, 
we analyze their divergence structure. If we consider
a gravitational perturbation with a comoving
momentum $\vec{k}$, it can be produced from scalar modes with comoving
momenta $\vec{p}$ and $\vec{q} \equiv \vec{p-k}$. One possible source
of divergences is that there is no upper bound on how large
$p \equiv |\vec{p}|$ can be. We refer to this as
the UV domain. But there is also another source of divergences. 
The smallest momenta producing the mode $\vec{k}$ are parallel
ones, $\vec{p} \parallel \vec{q} \parallel \vec{k}$, with $p + q = k$. 
We refer to this as the IR domain. 
These smallest momenta (largest wavelengths) decrease below any given
physical momentum 
farthest in the past, so that we have to integrate for a long time, 
to fully account for their effect on late-time observables. 
This turns out to yield another divergence, which for the 
classical contribution is in fact the dominant one. 
In the full 1-loop quantum result, 
this IR divergence gets cancelled by the vacuum contribution. However,
if we investigate the classical part on its own, for instance in 
a numerical study of an inflationary SIGW signal, a method is needed
for cancelling the unphysical divergence. 

Our presentation is organized as follows. After defining the
tensor power spectrum (cf. \se\ref{se:definition}) and the interactions
of spectator scalars with gravitons (cf.\ \se\ref{se:vertices}), 
we turn to our first main ingredient, a demonstration of how 
the full 1-loop computation can be split into its 
classical and vacuum parts (cf.\ \se\ref{se:ingredients}).  
We then take an opportunity to compare with some
developments in
recent literature~\cite{madrid,ema} (cf.\ \se\ref{se:checks}).
Subsequently, we turn to the
divergences affecting the classical and vacuum contributions, 
demonstrating how the IR divergences cancel between 
these two sets, and suggesting how the classical part could 
be ``renormalized'' on its own (cf.\ \se\ref{se:match}). We conclude with 
an outlook on many possible future extensions (cf.\ \se\ref{se:concl}). 
The main text is complemented by five appendices, 
elaborating on various
technical details. 

%
\section{Quantum-field-theoretic definition of a tensor power spectrum}
\la{se:definition}

Via the Einstein equations, the expansion history
of the universe is determined by the expectation value of 
the energy-momentum tensor, properly renormalized.
The energy-momentum tensor
consists of quadratic and higher powers 
of elementary quantum fields.
Therefore, the simplest observables relevant for cosmology 
are given by {\em 2-point expectation values}. 
Before thermalization, the minimal working assumption
is that the state with
respect to which the expectation value is taken, 
is unitarily related to an earlier vacuum state, 
which is often referred to as the Bunch-Davies vacuum. 
 
To be concrete, let us denote by $h^\tensor_{ij}$ 
the transverse-traceless (or tensor) part of 
a metric perturbation (the definition
is given in \eq\nr{g_mn}, with $i,j$
denoting spatial indices). Working in the interaction picture, 
where both states and operators are time-dependent, and assuming
translational invariance in spatial directions, 
the equal-time 2-point vacuum expectation value
of $h^\tensor_{ij}$ has the form
\be
 \frac{ 
 \langle\, 0(t) \,| \,
                 h^\tensor_{ij}(t,\vec{x})
               \,
                 h^\tensor_{ij}(t,\vec{y}) 
          \,  |\, 0(t) \,\rangle 
 }
 {
 \langle\, 0(t) \,| \, 0(t) \,\rangle 
 }
 \; \equiv \; 
 \int \! \frac{{\rm d}^3_{ }\vec{k}}{(2\pi)^3_{ }}
 \, e^{i \vec{k}\cdot(\vec{x-y}) }_{ }
 \, G^{ }_{\tensor}(t,k)
 \;, \la{G_t}
\ee
where a sum over repeated indices is understood; 
$t$ denotes physical time; 
$\vec{k}$ comoving momentum; 
and $k \equiv |\vec{k}|$. 
Assuming rotational invariance, 
we have written $ G^{ }_{\tensor} = G^{ }_{\tensor}(t,k)$.
The power spectrum is 
defined from the Fourier transform as 
\be
 \P^{ }_{\tensor}(t,k) 
 \; \equiv \; 
 \frac{k^3_{ }}{2\pi^2_{ }} \, 
 G^{ }_{\tensor}(t,k)
 \; 
    \xrightarrow[]
                {3\;\to\;d}
 \; 
 \, 
 \meas{} \, 
 G^{ }_{\tensor}(t,k)
 \;, \la{P_t}
\ee
where, for future reference, we have also indicated
the form applying in dimensional regularization in 
$d$ spatial dimensions. 
The fact that the underlying definition originates
from a point-split correlator (cf.\ \eq\nr{G_t}), 
suggests that the observable
should be UV finite, once local counterterms are 
included in the action.
In practice, the power spectrum is normally extracted
by setting $\vec{x} = \vec{y}$ in \eq\nr{G_t}, because
the same integrand can be identified in this limit, 
even though the integral as a whole is divergent. 

In the interaction picture, the vacuum state evolves
with the time-evolution operator defined by the interaction
Hamiltonian, 
\be
 |\, 0(t) \,\rangle
 \; = \; 
 \hat U^{ }_\iI(t;0) 
 \, | 0 \rangle
 \;, \quad 
 \hat U^{ }_\iI(t;0) 
 \; = \; 
 \hat T \, \exp 
 \biggl[\,
 -i \int_{0}^t \! {\rm d} t' \, 
    \hat{H}^{ }_\iI (t') 
 \,\biggr]
 \;. \la{0t}
\ee
Instead, operators like $h^\tensor_{ij}$ 
evolve with the equations of motion 
determined by the free Hamiltonian, $\hat{H}^{ }_{\scriptscriptstyle 0}$.
The free equations of motion are most simply determined
from the Lagrangian, as we recall in \app\ref{app:A}, and likewise the 
interaction Hamiltonian is related to the interaction Lagrangian, 
as specified in \se\ref{se:vertices}. 
Therefore, we can base our considerations on an action, 
denoted by $S$ (cf.\ \eq\nr{S}).
Furthermore, we mostly employ conformal time, which takes
values $\tau \in (-\infty,0)$ during the inflationary epoch.

%
\section{Interactions of gravitons with spectator scalars}
\la{se:vertices}

We now consider a real scalar field, $\varphi$, 
coupled to gravity according to the action
\be
 S \; \supset \; 
 \int_\X \sqrt{-g} 
 \, 
 \biggl\{\,
   \frac{R}{16\pi G} 
 - 
   \biggl[\,   
  \Lambda 
  + 
  \frac{1}{2} {\varphi}^{ }_{,\alpha} \varphi^{,\alpha}_{ }
  + 
  \frac{1}{2} \bigl( m^2_{ } + \xi R  \bigr)\, \varphi^2_{ }
 \,\biggr]
  \; + \; 
  \ldots
 \,\biggr\}
 \;, \la{S}
\ee
where $\X \equiv (\tau,\vec{x})$, 
$g \equiv \det g^{ }_{\mu\nu}$, $R$ is the Ricci scalar, 
$G$ is the Newton constant,
$\Lambda$ is the cosmological constant, 
$m^2_{ }$ is the mass-squared of the scalar field,  
and $\xi$ is a non-minimal coupling.
Here and in the following,  
we employ 
the metric signature ($-$$+$$+$$+$). 
The dots in \eq\nr{S} stand for higher-dimensional 
gravitational operators~\cite{eft1,eft2},  
or interactions of $\varphi$ with itself
or other fields.
We refer to the coefficients of
the higher-dimensional operators as {\em counterterms}, 
even though they also include finite parts.
In the present paper, for simplicity, 
the scalar field is assumed to have no background value, and  
its energy density, $\sim m^4_{ }$, is $\ll \Lambda$, 
so that it acts as a ``spectator'' during inflation, 
having little effect on the overall expansion. 
In this philosophy, $\varphi$ 
is a representative of Standard Model fields, 
which due to their large multiplicity could actually have a dominant
effect on loop corrections~\cite{sw1}.

We expand the metric as~\cite{jm} 
(cf.\ \app\ref{app:A} for a reminder of different conventions)
\be
 g_{\mu\nu}^{ }
 \; = \; 
 a^2_{ } \, 
 \biggl(\, 
   \begin{array}{cc}
   -1  \hspace*{3mm} &  0 \\ 
   0  &  \exp( h^\tensor_{ij})
   \end{array}
 \,\biggr)
 \; = \; 
 a^2_{ } \, 
 \biggl(\, 
   \begin{array}{cc}
   -1 \hspace*{3mm} &  0 \\ 
   0  &  \delta^{ }_{ij} + h^\tensor_{ij}
   + \frac{1}{2} h^\tensor_{ik} h^\tensor_{kj}
   \end{array}
 \,\biggr)
 + \rmO(h^{\tensor}_{ij})^3_{ }
 \;, \la{g_mn_nl}
\ee
where $h^\tensor_{ij}$ is symmetric, traceless, and transverse. 
The inverse metric is  
\be
 g^{\mu\nu}_{ }
 \; = \; 
 \frac{1}{a^2_{ }} \, 
 \biggl(\, 
   \begin{array}{cc}
   -1  \hspace*{3mm} &  0 \\ 
   0  &  \exp( - h^\tensor_{ij})
   \end{array}
 \,\biggr)
 \; = \; 
 \frac{1}{a^2_{ }} \, 
 \biggl(\, 
   \begin{array}{cc}
   -1  \hspace*{3mm} &  0 \\ 
   0  &  \delta^{ }_{ij} - h^\tensor_{ij}
   + \frac{1}{2} h^\tensor_{ik} h^\tensor_{kj}
   \end{array}
 \,\biggr)
 + \rmO(h^{\tensor}_{ij})^3_{ }
 \;. \la{g^mn_nl}
\ee
The corresponding Ricci scalar is given in \eq\nr{R_nl}, and 
contains no term linear in $h^\tensor_{ij}$.

Considering the quadratic part of the Ricci scalar, 
and going over to $d$ spatial dimensions for future reference, 
we find the kinetic terms (cf.\ \eq\nr{del_L_nl})
\be
 S 
 \; 
 \overset{ }{\supset} 
 \; 
 \frac{1}{32\pi G}
 \int_\X a^{d + 1}_{ }
 \,\biggl(\,
  - \frac{1}{2} \, \partial^{ }_\alpha h^\tensor_{ij} 
                \, \partial^\alpha_{ } h^\tensor_{ij} 
 \,\biggr)
 \;. 
\ee
The bare $G$ has the mass dimension $1-d$, while 
$h^\tensor_{ij}$ is dimensionless.
A bare canonically normalized graviton field, $\varh^\tensor_{ij}$
(with mass dimension $(d-1)/2$), 
is obtained by
writing 
\be
 h^\tensor_{ij} 
 \; \equiv \;
 \sqrt{32\pi G} \,\varh^\tensor_{ij}
 \;. \la{rescale}
\ee
A solution (mode expansion) for $\varh^\tensor_{ij}$, 
and likewise for an appropriately normalized scalar
field, is worked out in \app\ref{app:mode}.

After the rescaling in \eq\nr{rescale}, 
it is conventional to introduce a renormalization scale, $\mu$, 
by inserting $ 1 = \mu^{d-3}_{ }\mu^{3-d}_{ }$ next to appearances
of $G$, and by defining
the combination $G \mu^{d-3}_{ } \equiv 1/\mpl^2$ to be a renormalized 
Newton constant, which has mass dimension $-2$. Then 
the fields $\varh^\tensor_{ij}$ appear together
with a factor $\mu^{(3-d)/2}_{ }$.
In the paragraph below \eq\nr{tilde_phik_1}, we recall how this
guarantees that, apart from powers of $H^2_{ }/\mpl^2$, 
physical results only depend on 
the dimensionless 
combinations $\mu/H$ and $k\tau$. 
However, to streamline the notation, 
we work with bare parameters in the body of the text. 

We next turn to the interactions between gravitons and scalar fields. 
The leading term is linear in $h^\tensor_{ij}$. It originates by
inserting \eq\nr{g^mn_nl} in the kinetic terms in \eq\nr{S}, 
and rescaling the graviton field according to \eq\nr{rescale}, 
leading to 
\be
 S^{ }_{\iI}
 \;
 \underset{\rmii{\nr{g^mn_nl},\nr{rescale}}}{
 \overset{\rmii{\nr{S}}}{\supset}}
 \; 
 \int_\X \sqrt{8\pi G} \, a^{d-1}_{ }  
 \,\varh^\tensor_{ij}
 \,\varphi^{ }_{,i}
 \,\varphi^{ }_{,j}
 \;. \la{S_I_1}
\ee
We also need quartic vertices. 
One originates from 
the quadratic part
of $g^{\mu\nu}_{ }$ in \eq\nr{g^mn_nl}, producing
\be
 S^{ }_{\iI}
 \;
 \underset{\rmii{\nr{g^mn_nl},\nr{rescale}}}{
 \overset{\rmii{\nr{S}}}{\supset}}
 \; 
 - \int_\X 8\pi G\, a^{d-1}_{ } 
 \,\varh^\tensor_{ik}
 \,\varh^\tensor_{kj}
 \,\varphi^{ }_{,i}
 \,\varphi^{ }_{,j}
 \;. \la{S_I_2b}
\ee
A second quartic vertex originates from the non-minimal coupling, 
$\xi$, in \eq\nr{S}, 
\be
 S^{ }_{\iI}
 \;
 \underset{\rmii{\nr{R_nl},\nr{rescale}}}{
 \overset{\rmii{\nr{S}}}{\supset}}
 \; 
 - \int_\X \xi \, 4 \pi G\, a^{d-1}_{ }
 \bigl(\, 
 \,\varh^{\tensor\hspace*{0.3mm}\prime}_{ij}
 \,\varh^{\tensor\hspace*{0.3mm}\prime}_{ij}
 \; - \; 
 \,\varh^\tensor_{ij,k}
 \,\varh^\tensor_{ij,k}
 \,\bigr)
 \varphi^2_{ }
 \;, \la{S_I_2a}
\ee
where 
$
 \varh^{\tensor\hspace*{0.3mm}\prime}_{ij} 
 \equiv 
 \partial^{ }_\tau 
 \varh^{\tensor}_{ij}
$.
Even though we focus on minimal coupling in this paper
($\xi = 0$), we show the vertex in \eq\nr{S_I_2a},
the reason being that it can lead to loop corrections proportional
to $1/\xi$, which combine with the prefactor to a finite effect
(cf.\ the paragraph below \eq\nr{res_L_dr}).

While we have specified the interactions 
with the Lagrangian formalism, 
cf.\ \eqs\nr{S_I_1}--\nr{S_I_2a}, 
our physical observable has been defined 
in the canonical language, 
cf.\ \eqs\nr{G_t} and \nr{0t}. If the interactions 
contain no time derivatives, 
an interaction Hamiltonian amounts 
to minus the interaction Lagrangian. More generally, following
ref.~\cite{sw1}, we need to carry out a Legendre transform, 
$
 \mathcal{H} = 
 \pi^\tensor_{ij} \varh^{\tensor\hspace*{0.3mm}\prime}_{ij}
 - \mathcal{L}
$,
with 
$
 \pi^\tensor_{ij} \equiv \partial \mathcal{L} / 
 \partial \varh^{\tensor\hspace*{0.3mm}\prime}_{ij}
$,
which can lead to non-polynomial terms. Given that in the interaction 
picture, the time evolution of the fields is determined by the free 
Hamiltonian, $\mathcal{H}^{ }_\rmii{$0$}$, 
we can subsequently re-express canonical momenta 
in terms of time derivatives via the inverse transformation, 
$
 \varh^{\tensor\hspace*{0.3mm}\prime}_{ij}
 = 
 \partial \mathcal{H}^{ }_\rmii{$0$} / 
 \partial \pi^\tensor_{ij}
$.
Applied to \eqs\nr{S_I_1}--\nr{S_I_2a}, and going over to  
dimensionless (``conformal'') 
operators normalized as
\be
 \widehat\varphi 
 \; \equiv \; 
 a^{\frac{d-1}{2}}_{ } \hspace*{0.3mm} \varphi
 \;, \quad
 {\widehat\varh}^\tensor_{ij}
 \; \equiv \; 
 a^{\frac{d-1}{2}}_{ } \hspace*{0.3mm} \varh^\tensor_{ij}
 \;, \la{hat_phi}
\ee
this leads to 
\ba
 \int^t_0 \! {\rm d}t' \, \hat H^{ }_\iI 
 &
  \underset{\rmii{\nr{hat_phi}}}{
  \overset{\rmii{\nr{S_I_1}--\nr{S_I_2a}}\vphantom{\big |}}
  {\longleftrightarrow}}
 & 
 \int_\X 
 \; \biggl\{ \, 
 \Bigl(\, 
   - \sqrt{8 \pi G a^{1-d}_{ }}\; \widehat \varh^\tensor_{ij}
   + 8 \pi G a^{1-d}_{ }\; \widehat \varh^\tensor_{ik} 
                           \widehat \varh^\tensor_{kj}
 \,\Bigr)\, 
 \widehat\varphi^{ }_{,i} 
 \widehat\varphi^{ }_{,j} 
 \nn[3mm]
 & & 
 \quad + \, 
 \xi\; 4 \pi G a^{1-d}_{ } 
 \biggl[\, 
  \frac{ 
  (
   \widehat\varh^{\tensor\hspace*{0.3mm}\prime}_{ij}
   + \frac{1-d}{2}\H\, \widehat \varh^{\tensor}_{ij} 
  )^2_{ }
  }{
  1 - \xi \, 8 \pi G a^{1-d}_{ }\, \widehat\varphi^{\hspace*{0.3mm}2}_{ }
  }
  \; - \; 
    \widehat \varh^{\tensor}_{ij,k}
    \widehat \varh^{\tensor}_{ij,k}
 \,\biggr]
 \, \widehat\varphi^{\hspace*{0.3mm}2}_{ }
 \; + \; ... 
 \,\biggr\}
 \;. \hspace*{6mm} \la{S_to_H}
\ea
Restricting ourselves to vertices 
relevant for 1-loop contributions, 
the first term on the second line can be expanded up to 
leading order in $G a^{1-d}_{ }$. 
For \eq\nr{0t}, we are faced with 
\be
 \hat U^{ }_\iI (t;0) \; = \; 
 \mathbbm{1} 
 \; - \; 
 i  
 \int_{0}^t \! {\rm d}t' \, \hat H^{ }_\iI (t') 
 \; - \;  
 \int_{0}^t \! {\rm d}t' 
 \int_{0}^{t'} \! {\rm d}t'' 
 \, \hat H^{ }_\iI (t') 
 \, \hat H^{ }_\iI (t'') 
 \; + \;
 \rmO(\hat H^3_\iI)
 \;, \la{H_I}
\ee
where we always replace physical times, 
$
 \int_{0}^t \! {\rm d}t'
$,
with conformal ones, cf.\ \eq\nr{S_to_H}. 

%
\section{Ingredients for a 1-loop computation of the tensor power spectrum}
\la{se:ingredients}

%
\begin{figure}[t]
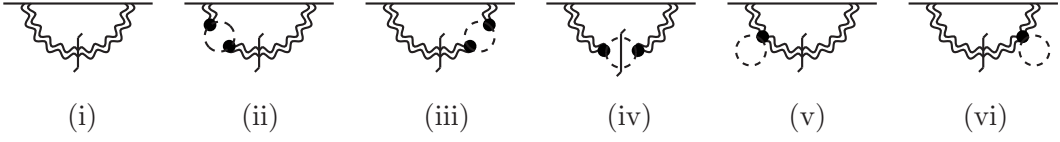


\def\Lwidth{2}

\begin{eqnarray*}
  && 
  \hspace*{-10mm}
  \AmplAgauge  
  \hspace*{6.0mm}
  \AmplBgauge 
  \hspace*{6.0mm}
  \AmplCgauge 
  \hspace*{6.0mm}
  \AmplDgauge 
  \hspace*{6.0mm}
  \AmplEgauge 
  \hspace*{6.0mm}
  \AmplFgauge 
  \hspace*{6.0mm}
 \\[5mm]
 &&
 \hspace*{3mm}
 \mbox{(i)}
 \hspace*{19mm}
 \mbox{(ii)}
 \hspace*{19mm}
 \mbox{(iii)}
 \hspace*{18mm}
 \mbox{(iv)}
 \hspace*{18mm}
 \mbox{(v)}
 \hspace*{18mm}
 \mbox{(vi)}
\end{eqnarray*}

\caption[a]{\small
  The tree-level diagram~(i) and the 1-loop diagrams (ii)--(vi)
  contributing to the tensor power spectrum, 
  with the loop order referring to the scalar fields. 
  The horizontal line corresponds to the end-of-inflation hypersurface, 
  at conformal time $\tau\to 0^-_{ }$, 
  at which the operator from \eq\nr{G_t} is evaluated. 
  Wiggly lines are graviton propagators, 
  and dashed lines are scalar propagators.
  The vertical ``cut'' symbolizes the vacuum initial state, 
  from which the evolution 
  starts at $\tau = -\infty$.
  }

\la{fig:diags}
\end{figure}

Making use of \eqs\nr{S_to_H} and \nr{H_I}, 
we can compute corrections from the scalar field 
to the tensor power spectrum, 
defined by \eqs\nr{G_t} and \nr{P_t}. 
For reference, let us recall that the tree-level
result, obtained by setting the time-evolution operator to unity, 
$
 \hat U^{ }_\iI (t;0) \to \mathbbm{1} 
$,  
and evaluated outside of the Hubble horizon, 
i.e.\ for $| k\tau | \ll 1$, 
reads 
\be
 \P^{ }_\tensor(\tau\to 0^-_{ },k) 
 \; \overset{d\;=\;3}{=} \; \frac{16 H^2_{ } }{\pi \mpl^2 }
 \; + \; 
 \rmO\biggl( \frac{H^4_{ }}{\pi^2_{ } \mpl^4} \biggr)
 \;, \la{P_t_0}
\ee
where we have re-expressed the Newton constant through the Planck
mass, as specified in the paragraph below \eq\nr{rescale}. 

The tree-level contribution as well as the 
1-loop corrections to $\P^{ }_\tensor$
are illustrated in \fig\ref{fig:diags}. Only ``connected'' 
contributions are kept, as the disconnected ones cancel, 
and tadpole diagrams have been omitted, 
assuming $\vec{k}\neq\vec{0}$.


%
\subsection{Contractions and their splitup into classical and vacuum parts}

We start by describing the results for the individual contractions
shown in \fig\ref{fig:diags}. Let us  
first consider the contributions denoted by (ii), (iii) and (iv). 
These originate from the vertex linear in 
the metric perturbation, given in \eq\nr{S_to_H}. Diagrams (ii) and
(iii) originate from the quadratic appearances of $\hat H^{ }_\iI$
in \eq\nr{H_I}, whereas diagram (iv) has two appearances of the linear
term in \eq\nr{H_I}, on both sides of the observable. 

Once we insert the mode expansions from \eqs\nr{mode_exp_phi} 
and \nr{mode_exp_hij} into the vertex in \eq\nr{S_I_1}, with
$\Phi_p$ denoting a scalar mode function and 
$\widetilde \phi_k \equiv a^{\frac{1-d}{2}} \Phi_k$ 
a rescaled graviton mode function; integrate
over the spatial directions, which implements momentum conservation; 
and make use of the commutation relations in \eq\nr{w_x}, we are left
over with integrals over two three-momenta, $\vec{k}$ and $\vec{p}$, 
as well as sums over vector indices and helicities
(or polarization states). 
The integral over $\vec{k}$
is not carried out, but we rather identify the integrand 
as the power spectrum
defined in \eq\nr{G_t}. The helicity sums appear as 
\be
 \sum_{i,j}
 \sum_{\lambda,\lambda'}
 \epsilon^{\lambda}_{ij,\vec{k}}
 \epsilon^{\lambda'}_{ij,\vec{k}}
 \epsilon^{\lambda*}_{mn,\vec{k}}
 \epsilon^{\lambda'*}_{uv,\vec{k}}
 \quad
 \mbox{or}
 \quad
 \sum_{i,j}
 \sum_{\lambda,\lambda'}
 \epsilon^{\lambda}_{ij,\vec{k}}
 \epsilon^{\lambda'}_{ij,-\vec{k}}
 \epsilon^{\lambda*}_{mn,\vec{k}}
 \epsilon^{\lambda'*}_{uv,-\vec{k}}
 \;. 
\ee
Making use of \eq\nr{eps_props}, both yield
$
 \sum_{ij} \mathbbm{T}^{ }_{mn;ij}
           \mathbbm{T}^{ }_{ij;uv}
 = 
 \mathbbm{T}^{ }_{mn;uv}
$,
where $\mathbbm{T}$ is the projector to the tensor channel
(cf.\ \eq\nr{def_T}).
Recalling that $\mathbbm{T}$ is orthogonal to $\vec{k}$, 
the remaining sum then gives
\be
 \sum_{m,n,u,v}
 \mathbbm{T}^{ }_{mn;uv}
 \, p^{ }_m\, p^{ }_n \, p^{ }_u\, p^{ }_v 
 \; 
 \overset{\rmii{\nr{def_T}}}{=}
 \; 
 \frac{d-2}{d-1}
 \biggl[\,
  {p}^2_{ } - \frac{(\vec p\cdot \vec k)^2_{ }}{{k}^2_{ }}
 \,\biggr]^2_{ }
 \;. \la{angle}
\ee
All in all, denoting the sum of 
the diagrams (ii), (iii) and (iv) of \fig\ref{fig:diags} 
by ``$J$'', and substituting $\tau' \leftrightarrow \tau''$
in order to combine similar terms, we get
\ba
 \delta \P_\tensor^\rmii{$J$}(\tau,k)
 & = & 
 \meas{ (32\pi G)^2_{ } }
 \int \! \frac{{\rm d}^d_{ }\vec{p}}{(2\pi)^d_{ }}
 \frac{d-2}{d-1} 
 \,\biggl[\,
   p^2_{ } - \frac{(\vec p \cdot \vec k)^2_{ }}{k^2_{ }} 
 \,\biggr]^2_{ }
 \, J^{ }_{pqk}(\tau)
 \;, \la{pre_J} \\[3mm]
 J^{ }_{pqk}(\tau) 
 & \equiv & 
 \,\biggl\{ \, 
 \int_{-\infty}^{\tau} \! {\rm d}\tau'
 \int_{-\infty}^{\tau'} \! {\rm d}\tau''
 \,
   \bigl[\, 
   \widetilde \phi_k^{ }(\tau) \widetilde \phi_k^{*}(\tau'\,) 
 - \widetilde \phi_k^{*}(\tau) \widetilde \phi_k^{ }(\tau'\,)
 \,\bigr] \, 
 \widetilde \phi_k^*(\tau) \, \widetilde \phi_k^{ }(\tau''\,)
 \nn[3mm]
 & & \hspace*{0.0cm}
 -\,
 \int_{-\infty}^{\tau} \! {\rm d}\tau''
 \int_{-\infty}^{\tau''} \! {\rm d}\tau'
 \,
   \bigl[\, 
   \widetilde \phi_k^{ }(\tau) \widetilde \phi_k^{*}(\tau''\,) 
 - \widetilde \phi_k^{*}(\tau) \widetilde \phi_k^{ }(\tau''\,)
 \,\bigr] \, 
 \widetilde \phi_k^{ }(\tau) \, \widetilde \phi_k^{*}(\tau'\,)
 \, \biggr\}
 \nn[3mm]
 & & \hspace*{0.2cm}
 \times \;
 \Phi_p^*(\tau'\,) \,    
 \Phi_q^*(\tau'\,) \,
 \;
 \Phi^{ }_p(\tau''\,) \,
 \Phi^{ }_q(\tau''\,) \,  
 \;, \la{res_J}
\ea
where we have denoted $q \equiv |\vec{p+k}|$ or $q \equiv|\vec{k-p}|$, 
which are equivalent after $\vec{p}\leftrightarrow -\vec{p}$.

In the diagrams (v) and (vi) of \fig\ref{fig:diags}, evaluated with
the quartic vertex from the first line of \eq\nr{S_to_H}, 
we meet the polarization sum
\be
 \sum_{i,j,l}
 \sum_{\lambda,\lambda'}
 \epsilon^{\lambda}_{ij,\vec{k}}
 \epsilon^{\lambda'}_{ij,-\vec{k}}
 \epsilon^{\lambda'*}_{ml,-\vec{k}}
 \epsilon^{\lambda*}_{ln,\vec{k}}
 \;
 \overset{\rmii{\nr{eps_props}}}{=}
 \; 
 \sum_{i,j,l}
 \mathbbm{T}^{ }_{ij;ln}
 \mathbbm{T}^{ }_{ij;ml}
 \;
 \underset{\rmii{\nr{def_T}}}{
 \overset{\rmii{\nr{eps_props}} \vphantom{ |_q^b } }{=}}
 \;  
 \frac{(d-2)(d+1)}{2(d-1)}\,
 \mathbbm{K}^{ }_{mn}
 \;. 
\ee
This gets contracted with the scalar loop momenta, $p^{ }_m\, p^{ }_n$, 
yielding
\be
 \sum_{m,n} \mathbbm{K}^{ }_{mn}\, p^{ }_m\, p^{ }_n 
 \; = \; 
   p^2_{ } - \frac{(\vec p \cdot \vec k)^2_{ }}{k^2_{ }} 
 \;. 
\ee
Denoting the sum of 
the diagrams (v) and (vi) of \fig\ref{fig:diags} by ``$L$'', we find 
\ba
 \delta \P_\tensor^\rmii{$L$}(\tau,k)
 & = & 
 \meas{ (32\pi G)^2_{ } }
 \int \! \frac{{\rm d}^d_{ }\vec{p}}{(2\pi)^d_{ }}
 \frac{(d-2)(d+1)}{4(d-1)} 
 \,\biggl[\,
   p^2_{ } - \frac{(\vec p \cdot \vec k)^2_{ }}{k^2_{ }} 
 \,\biggr]
 \, L^{ }_{pk}(\tau)
 \;, \la{pre_L} \\[2mm]
 L^{ }_{pk}(\tau) 
 & \equiv & 
 {i}
 \int_{-\infty}^\tau \! {\rm d}\tau' 
 \, 
 \bigl[\, 
   \widetilde \phi_k^{*2}(\tau) \widetilde \phi\hspace*{0.3mm}_k^2(\tau'\,) 
 - \widetilde \phi\hspace*{0.3mm}_k^{2}(\tau)  \widetilde \phi_k^{*2}(\tau'\,)
 \,\bigr] 
 \, | \Phi^{ }_p(\tau'\,) |^2_{ }
 \;. \hspace*{9mm} \la{res_L}
\ea

Finally, we consider 
diagrams (v) and (vi) of \fig\ref{fig:diags}, but now with 
the $\xi$-dependent vertex from the second line of \eq\nr{S_to_H}. 
This time the polarization sum has the appearance
\be
 \sum_{i,j,m,n}
 \sum_{\lambda,\lambda'}
 \epsilon^{\lambda}_{ij,\vec{k}}
 \epsilon^{\lambda'}_{ij,-\vec{k}}
 \epsilon^{\lambda'*}_{mn,-\vec{k}}
 \epsilon^{\lambda*}_{mn,\vec{k}}
 \;
 \overset{\rmii{\nr{eps_props}}}{=}
 \; 
 \sum_{i,j,m,n}
 \mathbbm{T}^{ }_{ij;mn}
 \mathbbm{T}^{ }_{ij;mn}
 \;
 \underset{\rmii{\nr{def_T}}}{
 \overset{\rmii{\nr{eps_props}} \vphantom{ |_q^b } }{=}}
 \;  
 \frac{(d-2)(d+1)}{2}
 \;. 
\ee
Denoting this contribution  by ``$X$'', 
we obtain
\ba
 \delta \P_\tensor^\rmii{$X$}(\tau,k)
 & = & 
 \meas{ \,\xi\, (32\pi G)^2_{ } }
 \int \! \frac{{\rm d}^d_{ }\vec{p}}{(2\pi)^d_{ }}
 \frac{(d-2)(d+1)}{8} 
 \, X^{ }_{pk}(\tau)
 \;, \la{pre_X} \\[2mm]
 X^{ }_{pk}(\tau) 
 & \equiv & 
 {i}
 \int_{-\infty}^\tau \! {\rm d}\tau' 
 \, 
 \bigl[\, 
   \widetilde \phi_k^{*2}(\tau) \,
   (\partial^{ }_{\tau'} - k)
   \widetilde \phi\hspace*{0.3mm}_k^{ }(\tau'\,) \,
   (\partial^{ }_{\tau'} + k)
   \widetilde \phi\hspace*{0.3mm}_k^{ }(\tau'\,) 
 \nn[3mm]
 & & \hspace*{1.35cm}
 - \,\widetilde \phi\hspace*{0.3mm}_k^{2}(\tau) \,
   (\partial^{ }_{\tau'} - k)
   \widetilde \phi_k^{*}(\tau'\,)
   (\partial^{ }_{\tau'} + k)
   \widetilde \phi_k^{*}(\tau'\,)
 \,\bigr] 
 \, | \Phi^{ }_p(\tau'\,) |^2_{ }
 \;. \hspace*{9mm} \la{res_X}
\ea

\vspace*{3mm}

The expression for $J^{ }_{pqk}$ in \eq\nr{res_J} can be given
a more transparent appearance by pulling apart what we 
call the {\em classical 
contribution}. To this aim, we write 
\be
 J^{ }_{pqk}(\tau) \; = \; 
 J^\vac_{pqk}(\tau) \; + \; J^\cl_{pqk}(\tau)
 \;. \la{splitup}
\ee
The classical term is defined and determined in \app\ref{app:def_cl}, 
cf.\ \eq\nr{P_t_cl}. 
We also note that the dependence on the graviton mode functions, 
$\widetilde \phi^{ }_k$, can be expressed more compactly
in terms of Green's functions, 
$ \widetilde \Delta^{ }_k $ 
and 
$ \widetilde G^\rmii{R}_k $, defined in 
\eqs\nr{full_delta_k} and \nr{full_gr_k}. 
After renaming $\tau'\leftrightarrow \tau''$
in order to combine terms, 
we thus find
\ba
 J^\vac_{pqk}(\tau) 
 & 
 \underset{\rmii{\nr{splitup}}}{
 \overset{\rmii{\nr{res_J}} \vphantom{ |^b_q } }{=}} 
 & 
 2\, 
 \int_{-\infty}^{\tau} \! {\rm d}\tau'
 \int_{-\infty}^{\tau'} \! {\rm d}\tau''
 \, \widetilde G^\rmii{R}_k(\tau,\tau'\,)
 \, \widetilde \Delta^{ }_k(\tau,\tau''\,)
 \im\bigl[ 
 \Phi_p^*(\tau'\,) \, \Phi_q^*(\tau'\,) \,  
 \Phi^{ }_p(\tau''\,) \,\Phi^{ }_q(\tau''\,) \,  
 \bigr]
 \;, \nn
 \la{res_Jvac}
 \\[3mm]
 J^\cl_{pqk}(\tau) 
 & 
 \underset{\rmii{\nr{splitup}}}{
 \overset{\rmii{\nr{res_J}} \vphantom{ |^b_q } }{=}} 
 & 
 \int_{-\infty}^{\tau} \! {\rm d}\tau'
 \int_{-\infty}^{\tau'} \! {\rm d}\tau''
 \, \widetilde G^\rmii{R}_k(\tau,\tau'\,)
 \, \widetilde G^\rmii{R}_k(\tau,\tau''\,)
 \re\bigl[ 
 \Phi_p^*(\tau'\,) \, \Phi_q^*(\tau'\,) \,  
 \Phi^{ }_p(\tau''\,) \,\Phi^{ }_q(\tau''\,) \,  
 \bigr]
 \;, \nn
 \la{res_Jcl}
 \\[3mm]
 L^{ }_{pk}(\tau) 
 & 
 \underset{\rmii{ }}{
 \overset{\rmii{\nr{res_L}}}{=}} 
 & 
 -\, 2\,
 \int_{-\infty}^\tau \! {\rm d}\tau'
 \, \widetilde G^\rmii{R}_k(\tau,\tau'\,)
 \, \widetilde \Delta^{ }_k(\tau,\tau'\,)
 \, | \Phi^{ }_p(\tau'\,) |^2_{ }
 \;, \la{res_L_2}
 \\[3mm]
 X^{ }_{pk}(\tau) 
 & 
 \underset{\rmii{ }}{
 \overset{\rmii{\nr{res_X}}}{=}} 
 & 
 -\, 2\,
 \int_{-\infty}^\tau \! {\rm d}\tau'
 \, \bigl[ \, 
 \partial^{ }_{\tau'} \widetilde G^\rmii{R}_k(\tau,\tau'\,)
 \partial^{ }_{\tau'} \widetilde \Delta^{ }_k(\tau,\tau'\,)
 \; - \; 
 k^2_{ } \,
 \widetilde G^\rmii{R}_k(\tau,\tau'\,)
 \widetilde \Delta^{ }_k(\tau,\tau'\,)
 \, \bigr]
 \, | \Phi^{ }_p(\tau'\,) |^2_{ }
 \;. \nn \la{res_X_2}
\ea
We see how $J^\vac_{pqk}$ and $L^{ }_{pk}$ 
have similar graviton Green's functions, and this will play a role
in \se\ref{ss:cancel}. In contrast, 
$J^\cl_{pqk}$ has a different structure, and will be analyzed
in detail in \se\ref{se:match}. 

%
\subsection{Radial and angular integrals}
\la{ss:angular}

For further steps, the angular integrals 
in \eqs\nr{pre_J}, \nr{pre_L}, and \nr{pre_X} need to be carried out. 
We do this in $d$ dimensions, as is relevant for 
dimensional regularization. 

In $d$ dimensions, we write the integration measure as 
\be
 \int \! \frac{ {\rm d}^d_{ } \vec{p} }{(2\pi)^d_{ }} 
 \; = \; 
 \frac{4}{(4\pi)^{\frac{d+1}{2}}_{ } \Gamma(\frac{d-1}{2})}
 \int_0^{\infty} \! {\rm d}p \, p^{d-1}_{ } 
 \int_{-1}^{+1} \! {\rm d}z \, (1-z^2)^{\frac{d-3}{2}}_{ }
 \;, \la{d_dim_z}
\ee 
where $z = \vec{k}\cdot\vec{p}/(kp)$ 
parametrizes an angle with respect to an external vector. 
If there is no dependence on $z$, the integral yields
\be
 \int_{-1}^{+1} \! {\rm d}z \, (1-z^2)^{\alpha}_{ } 
 \;=\; 
 \frac{\Gamma(\frac{1}{2})\Gamma(1+\alpha)}{\Gamma(\frac{3}{2}+\alpha)} 
 \;, \la{z_int}
\ee
whence we find the measure factor in \eq\nr{P_t}, {\em viz.}
\be
 \int \! \frac{ {\rm d}^d_{ } \vec{p} }{(2\pi)^d_{ }} 
 \; 
 \underset{\scriptscriptstyle \alpha \;=\; \frac{d-3}{2 \vphantom{ |^a } } }{
 \overset{\rmii{\nr{d_dim_z},\nr{z_int}}\vphantom{\big | }}{=}} 
 \; 
 \frac{2}{(4\pi)^{\frac{d}{2}}_{ } \Gamma(\frac{d}{2})}
 \int_0^{\infty} \! {\rm d}p \, p^{d-1}_{ }  
 \;. \la{d_dim}
\ee

We can apply these expressions to \eqs\nr{pre_J} and \nr{pre_L}. 
If the $J^{ }_{pqk}$ were $z$-independent, \eq\nr{pre_J} would contain
\be
 \int \! \frac{{\rm d}^d_{ }\vec{p}}{(2\pi)^d_{ }}
 \frac{d-2}{d-1} 
 \, p^4_{ } \, (1 - z^2_{ })^2_{ }
 \;
 \underset{\rmii{\nr{z_int}}}{
 \overset{\rmii{\nr{d_dim_z}} \lift }{=}}
 \;
 \frac{2}{(4\pi)^{\frac{d}{2}}_{ } \Gamma(\frac{d}{2})}
 \frac{(d-2)(d+1)}{d(d+2)}
 \int_0^{\infty} \! {\rm d}p \, p^{d+3}_{ }  
 \;, \la{ang_J_1}
\ee
and for \eq\nr{pre_L} we would get
\be
 \int \! \frac{{\rm d}^d_{ }\vec{p}}{(2\pi)^d_{ }}
 \frac{(d-2)(d+1)}{4(d-1)} 
 \, p^2_{ } \,
 (1 - z^2_{ })
 \; 
 \underset{\rmii{\nr{z_int}}}{
 \overset{\rmii{\nr{d_dim_z}} \lift }{=}}
 \;  
 \frac{2}{(4\pi)^{\frac{d}{2}}_{ } \Gamma(\frac{d}{2})}
 \frac{(d-2)(d+1)}{4 d}
 \int_0^{\infty} \! {\rm d}p \, p^{d+1}_{ }  
 \;. \la{ang_L_1}
\ee
Equations \nr{ang_J_1} and \nr{ang_L_1}
are needed in \se\ref{ss:cancel}.

\vspace*{3mm}

More generally, if the integrand depends on $z$, 
we reparametrize the $z$-integral through 
$
 q \equiv |\vec{p+k}| = \sqrt{p^2_{ } + k^2_{ } + 2 p k z \vphantom{ |^t }}
$, 
as 
\be
 \int_{-1}^{+1} \! {\rm d}z \;=\;
 \int_{|p-k|}^{p+k} \! {\rm d}q \, \frac{{\rm d}z}{{\rm d}q}
 \; = \; 
 \int_{|p-k|}^{p+k} \! {\rm d}q \, \frac{q}{pk}
 \;.  \la{dz_dq}
\ee
Furthermore, we introduce new variables as 
\be
 r \; \equiv \; \frac{q+p}{k}
 \;, \quad
 a \; \equiv \; \frac{q-p}{k}
 \quad \Leftrightarrow \quad
 q \; = \; \frac{k}{2}(r+a)
 \;, \quad
 p \; = \; \frac{k}{2}(r-a) 
 \;, \la{qpm}
\ee
where $r$ refers to ``radial'' and $a$ to ``angular''. 
In terms of these, 
\be
 1 - z^2_{ }
 \; = \; 
  \frac{
 [\, (p+k)^2_{ } - q^2_{ } \,]^{ }_{ }
 [\, q^2_{ } - (p-k)^2_{ } \,]^{ }_{ }
 }{4 p^2_{ }k^2_{ }}
 \; \overset{\rmii{\nr{qpm}}}{=} \; 
 \frac{(r^2_{ }-1)(1-a^2_{ })}{(r-a)^2_{ }}
 \;. \la{ra}
\ee
A general 1-loop integral then has the form 
\ba
 && \hspace*{-1.5cm}
 \int \! \frac{ {\rm d}^d_{ } \vec{p} }{(2\pi)^d_{ }} \, f(p,|\vec{p+k}|)
 \la{full_meas} \\[2mm]
 & 
 \overset{\rmii{\nr{d_dim_z}}}{=} 
 &
 \frac{4}{(4\pi)^{\frac{d+1}{2}}_{ } \Gamma(\frac{d-1}{2})}
 \int_0^{\infty} \! {\rm d}p \, p^{d-1}_{ } 
 \int_{-1}^{+1} \! {\rm d}z \, (1-z^2_{ })^{\frac{d-3}{2}}_{ }
 \, f\Bigl(\, p, \sqrt{p^2_{ } + k^2_{ } + 2 p k z \vphantom{ |^t }} \,\Bigr)
 \nn[2mm]
 &
 \!\!
 \underset{\rmii{\nr{ra}}}{
 \overset{\rmii{\nr{dz_dq}} \vphantom{ |^b_q } }{=}} 
 \!\!
 & 
 \frac{2^{1-2d}_{ } k^d_{ }}{\pi^{\frac{d+1}{2}}_{ } \Gamma(\frac{d-1}{2}) }
 \int_{1}^{\infty} \!\!\!\! {\rm d}r 
 \int_{-1}^{+1} \!\!\!\! {\rm d}a
 \, (r^2_{ } - a^2_{ }) 
 \, (r^2_{ } - 1)^{\frac{d-3}{2}}_{ }
 \, (1 - a^2_{ })^{\frac{d-3}{2}}_{ }
 \, f\Bigl( {\textstyle \frac{k(r-a)}{2 \vphantom{ |^b }},
                        \frac{k(r+a)}{2 \vphantom{ |^b }} } \Bigr)
 \;. \nonumber
\ea 

%
\section{Connection to recent literature}
\la{se:checks}

%

Before proceeding to our main computations, 
in \se\ref{se:match}, we compare with some
recent literature~\cite{ema,madrid}. 
The goal is, on one hand, to find consistency
checks on our starting point, and on the other, to underline why 
we need to go beyond the current status.  

%
\subsection{Cancellation at the integrand level}
\la{ss:cancel}

According to ref.~\cite{ema}, in an 
expansion in $| k\tau | \ll 1$, there is no 
$k$-independent correction of $\rmO(H^4_{ }/(\pi^2\mpl^4))$ 
to the tensor power spectrum, $\P^{ }_\tensor(0,k)$. 
However, this statement relies on assumptions,
notably that neither IR nor UV 
divergences play a role. 
Concretely, the cancellation was observed between 
the $k \ll p$, $ | k\tau'' | \ll 1$ parts
of the {\em integrands} 
of $\delta \P^{\hspace*{0.3mm}\rmii{$J$},\vac}_\tensor(0,k)$ and 
$\delta \P^\rmii{$L$}_\tensor(0,k)$. Let us see under which 
approximation this cancellation can be found in our expressions, 
and why the argument is ultimately not valid for the 
full $\P^{ }_\tensor(0,k)$.  

We start by 
inspecting the equation for the scalar mode function, 
$\Phi^{ }_p(\tau)$, 
from \eq\nr{mode_eqn}. Following ref.~\cite{ema}, we 
perturb the equation with respect to the momentum, 
\be
 \bigl[\, 
 \partial_\tau^2 
 +
  p^2_{ }(1 + 2 \epsilon)
 + 
 \widehat m^2_{ }(\tau)
 \,\bigr]
 \Phi^{ }_{p\hspace*{0.3mm}(1+\epsilon)}(\tau)
 \; 
 \underset{\rmii{\nr{mode_eqn}}}{
 \overset{\rmii{\nr{e-l-2}} \lift }{=}} 
 \;
 0 
 \;. \la{pert_p}
\ee
Expressing the solution as 
$
 \Phi^{ }_{p(1+\epsilon)} = \Phi_p^{\zero} + \Phi_p^{\one} + ...
$, 
the first-order correction satisfies
\be
 \bigl[\, 
 \partial_\tau^2 
 +
  p^2_{ }
 + 
 \widehat m^2_{ }(\tau)
 \,\bigr]
  \Phi_p^{\one} 
 \; = \; - 2 p^2_{ } \epsilon \, \Phi^{\zero}_p
 \;, 
\ee
which can be integrated into 
\be
 \Phi^{\one}_p(\tau)
 \; = \; 
 -2 p^2_{ } \epsilon \int_{-\infty}^{\tau} \! 
 {\rm d}\tau' \, 
 [G^\rmii{R}_p(\tau,\tau'\,)]^{\zero}_{ } \, 
 \Phi^{\zero}_p(\tau ')
 \;. \la{Phi1}
\ee
Here the Green's function has the form
\be
 G^\rmii{R}_{p}(\tau,\tau'\,)
 \; = \; 
 i \hspace*{0.3mm} \theta(\tau - \tau'\,)\, 
 \bigl[\, 
 \Phi_p^{ }(\tau) 
 \Phi_p^{*}(\tau'\,) 
 - 
 \Phi_p^{*}(\tau) 
 \Phi_p^{ }(\tau'\,) 
 \,\bigr]
 \;, 
 \la{GR_vs_Phi}
\ee
evaluated with $\Phi^{\zero}_p$. 
It then follows that 
\be
 \partial^{ }_p \Phi^{ }_p(\tau)
 \; = \; 
 \lim_{\epsilon\to 0} 
 \frac{\Phi^{\one}_p(\tau)}{p\epsilon}
 \; = \; 
 - 2 i p 
 \int_{-\infty}^{\tau} \! 
 {\rm d}\tau' \, 
 \bigl[\, 
 \Phi_p^{ }(\tau) 
 \Phi_p^{*}(\tau'\,) 
 - 
 \Phi_p^{*}(\tau) 
 \Phi_p^{ }(\tau'\,) 
 \,\bigr]
 \Phi_p^{ }(\tau'\,)
 \;. 
\ee
Taking a complex conjugate of this relation, 
and renaming time
coordinates, shows that 
\be
 \partial^{ }_p \bigl[\, 
 \Phi_p^{*}(\tau'\,) 
 \Phi_p^{ }(\tau'\,) 
 \,\bigr]
 \; = \; 
 -4p
 \int_{-\infty}^{\tau'} \! 
 {\rm d}\tau'' \, 
 \im\bigl[ 
 \Phi_p^*(\tau'\,) \, \Phi_p^*(\tau'\,) \,  
 \Phi^{ }_p(\tau''\,) \,\Phi^{ }_p(\tau''\,) \,  
 \bigr]
 \;. \la{id_L}
\ee

We now use \eq\nr{id_L} in order to evaluate 
$
 \delta \P_\tensor^\rmii{$L$}
$
from \eq\nr{pre_L}.
Let us first carry out a partial integration in \eq\nr{ang_L_1},
\be
 \int_0^\infty \! {\rm d}p\, p^{d+1}_{ }
 \, L^{ }_{pk}(\tau)
 \; = \; 
 \int_0^\infty \! {\rm d}p\, \frac{{\rm d}}{{\rm d}p}
 \biggl( \frac{ p^{d+2}_{ } }{d+2} \biggr)
 \, L^{ }_{pk}(\tau)
 \; = \;
  - 
 \int_0^\infty \! {\rm d}p\,
 \frac{ p^{d+2}_{ } }{d+2}
 \partial^{ }_p  L^{ }_{pk}(\tau)
 \;, \la{ibp_L} 
\ee 
where the vanishing of the boundary terms is a property of
dimensional regularization. Combining \eqs\nr{pre_L} and \nr{res_L_2}
with \eqs\nr{id_L} and \nr{ibp_L}, yields
\ba
 \delta \P_\tensor^\rmii{$L$}(\tau,k)
 & 
 \underset{\rmii{\nr{id_L},\nr{ibp_L}}}{
 \overset{\rmii{\nr{pre_L},\nr{res_L_2}} \vphantom{ |^b_q } }{ = }} 
 &
 \meas{  (32\pi G)^2_{ } }
 \frac{2}{(4\pi)^{\frac{d}{2}}_{ } \Gamma(\frac{d}{2})}
 \frac{(d-2)(d+1)}{d(d+2)}
 \int_0^{\infty} \! {\rm d}p \, p^{d+3}_{ }  
 \nn[3mm]
 & & \hspace*{-2.5cm} \times \; (-2)
 \int_{-\infty}^\tau \! {\rm d}\tau'
 \, \widetilde G^\rmii{R}_k(\tau,\tau'\,)
 \, \widetilde \Delta^{ }_k(\tau,\tau'\,)
 \int_{-\infty}^{\tau'} \! 
 {\rm d}\tau'' \, 
 \im\bigl[ 
 \Phi_p^*(\tau'\,) \, \Phi_p^*(\tau'\,) \,  
 \Phi^{ }_p(\tau''\,) \,\Phi^{ }_p(\tau''\,) \,  
 \bigr]
 \;. \hspace*{6mm} \la{id_final}
\ea
Comparing with the contribution of $J^\vac_{pqk}$ from \eqs\nr{pre_J}, 
\nr{res_Jvac} and \nr{ang_J_1}, we observe
the cancellation that was reported in ref.~\cite{ema}. 

However, this cancellation is 
of approximate nature. First of all, it relies
on the assumption $k\ll p$, so that \eq\nr{ang_J_1} can be employed
in \eq\nr{pre_J}. Yet, the meaning of this approximation is 
unclear, because $k$ is fixed, but $p$ is
an integration variable, taking both IR ($p \sim k$) 
and UV ($p \gg k$) values. 
Second, comparing \eqs\nr{res_Jvac} and \nr{id_final}, 
the cancellation works if 
$\widetilde \Delta^{ }_k(\tau,\tau''\,)$ 
is independent of~$\tau''$, 
which is only true for $|k\tau|, |k\tau''| \ll 1$ (cf.\ \eq\nr{delta_k}).
This is not always the case, since $k\tau''$ is an 
integration variable (cf.\ \eq\nr{id_final}).
Third, the function $J^\cl_{pqk}$ from \eq\nr{res_Jcl} is not 
part of the cancellation, yet it gives a non-vanishing
contribution (cf.\ \eq\nr{sum_I}), which constitutes 
the focus of \se\ref{se:match}. 
For these reasons, we do not make use of this cancellation in our
actual computations, and the $k$-independent
part of the 1-loop
contribution to $\P^{ }_\tensor(0,k)$ does actually not vanish~\cite{madrid}.

%
\subsection{A sample loop integral in cutoff regularization}
\la{ss:sample_ct}

Full 1-loop computations of the tensor power spectrum 
from \eq\nr{P_t_0} have been 
reported for a conformally~\cite{conformal} and 
minimally~\cite{madrid,seoul} coupled scalar field, 
albeit only in the massless limit. In ref.~\cite{madrid}, 
the computation was carried out both 
in dimensional and in a specific version of
cutoff regularization. The statement there 
is markedly different from that in ref.~\cite{ema}: 
there {\em are} contributions of $\rmO(H^4_{ }/(\pi^2 \mpl^4))$
to the tensor power spectrum. They have the special property, however, 
that logarithms of $k\tau$ cancel, i.e.\ that there is no 
late-time divergence in the tensor power spectrum. The absence
of such an IR logarithm was related to diffeomorphism invariance,
forbidding the generation of a graviton mass, 
and it also implies that the result is in essence
dictated by the counterterms. 

Carrying out a full 1-loop computation 
requires the introduction of a regularization scheme. 
Envisaging an eventual numerical simulation of a late stage
of inflation, possibly incorporating 
reheating dynamics~\cite{dissip,fluctu}, we need to define
a scheme which renders the integrals finite in a strict sense. 
Therefore, we start by considering cutoff regularization. 

Regularization by a momentum cutoff 
breaks some of the symmetries of the
underlying theory, notably diffeomorphism invariance. However, 
a partial symmetry can still be maintained, namely 
dilatation invariance (cf.,\ e.g.,\ ref.~\cite{classic}). 
This symmetry is present if the cutoff, 
which we denote by $\mu$, 
is imposed on physical rather than comoving momenta, 
\be
 \frac{p}{a(\tau'\,)} \;\le\; \mu 
 \quad 
 \overset{\rmii{\nr{de_Sitter}}\vphantom{\big |}}{\Leftrightarrow} 
 \quad
 p \;\le\; - \frac{\mu}{\tau' H}
 \;. \la{p_cutoff}
\ee
This can be implemented 
concretely on the level of the mode functions, setting 
\be
 \Phi^{ }_p(\tau'\,) \;\longrightarrow\;
 \Phi^{ }_p(\tau'\,)\, \theta\biggl( \tau' + \frac{\mu}{p H} \biggr)
 \;. \la{Phi_cutoff}
\ee

\vspace*{3mm}

Let us now inspect the contribution
$
 \delta \P_\tensor^\rmii{$L$}(\tau,k)
$
from \eqs\nr{pre_L}, \nr{res_L_2} and \nr{ang_L_1}
for a massless minimally coupled scalar field, 
\ba
 \delta \P_\tensor^\rmii{$L$}(\tau,k)
 &
  \underset{\rmii{\nr{ang_L_1}}}{
  \overset{\rmii{\nr{pre_L},\nr{res_L_2}} \vphantom{\big | } }{=}}
 & -\,
 \meas{ (32\pi G)^2_{ } }
 \frac{(d-2)(d+1)}{2d} 
 \nn[2mm]
 & & \;\times\,
 \int_{-\infty}^\tau \! {\rm d}\tau'
 \, \widetilde G^\rmii{R}_k(\tau,\tau'\,)
 \, \widetilde \Delta^{ }_k(\tau,\tau'\,)
 \int \! \frac{{\rm d}^d_{ }\vec{p}}{(2\pi)^d_{ }}
 \, p^2_{ }
 \, | \Phi^{ }_p(\tau'\,) |^2_{ }
 \;. \la{P_L_combined}
\ea
In cutoff regularization, 
we can set $d\to 3$ in the prefactor and the integration
measure, and use the mode function from \eq\nr{phi_minimal}. Then the
integral over the mode function yields
\be
 \int \! \frac{{\rm d}^d_{ }\vec{p}}{(2\pi)^d_{ }}
 \, p^2_{ }
 \, | \Phi^{ }_p(\tau'\,) |^2_{ }
 \;
 \xrightarrow[m\,,\,\xi\,,\,\delta\;\to\;0]
             {\rmii{\nr{Phi_cutoff},\nr{phi_minimal}}\vphantom{\big | }}
 \; 
 \frac{1}{4\pi^2_{ }}
 \int_0^{-\frac{\mu}{\tau' H \vphantom{ |^b }}}
 \! {\rm d}p \, p^3_{ }\, \biggl[ 1 + \frac{1}{(p\tau'\,)^2_{ }} \biggr]
 \; = \; 
 \frac{\mu^4_{ } + 2 \mu^2_{ } H^2_{ }}
 {16\pi^2_{ }(\tau' H )^4_{ }}
 \;.
\ee
Subsequently, the time integral is doable, making use of 
the graviton Green's
functions from \eqs\nr{delta_k} and \nr{gr_k},
\be
 \int_{-\infty}^\tau \!\!\! {\rm d}\tau'
 \;  
    \frac{
    \widetilde G^{\rmii{R$(0)$}}_k(\tau,\tau'\,)
 \, \widetilde \Delta^{\rmii{$(0)$}}_k(\tau,\tau'\,) }{(\tau'\,)^{4}_{ }}
 \;
 \underset{\rmii{\nr{gr_k}}}{
 \overset{\rmii{\nr{delta_k}} \vphantom{ |^b_q } }{=}}
 \;
 \frac{H^4_{ }}{6 k^3_{ }}
 \bigl\{\, 
   2 - \re \bigl[\,
    e^{2ik\tau }_{ } (i + k \tau)^2_{ }
    E^{ }_1(2 i k \tau) 
     \,\bigr]
 \,\bigr\}
 \;, \la{gr_delta_int}
\ee
where $E^{ }_1$ is the exponential integral
(cf.\ \eq\nr{E1}).
All in all, we thereby obtain
\ba
 \P_\tensor^\rmii{$L$}(\tau,k)
 & 
 \underset{m\,,\,\xi\,,\,\delta \;\to\; 0}{
 \overset{\rmii{\nr{P_L_combined}--\nr{gr_delta_int}}\vphantom{ \big | }}{=}} 
 & 
 \;-\, \frac{(8\pi G)^2_{ }}{18 \pi^4_{ } }
 \bigl( \mu^4_{ } + 2 \mu^2_{ } H^2_{ } \bigr)
 \bigl\{\,
  2 - \re \bigl[\,
   e^{2ik\tau}_{ }(i + k \tau)^2_{ }
   E^{ }_1(2 i k \tau )  
          \,\bigr]
 \,\bigr\}
 \;. \hspace*{6mm} \la{P_L_cutoff}
\ea
Inserting 
$
  E^{ }_1(2 i k \tau) = 
  i\pi - \mbox{Ei}(-2 i k \tau ) 
$, 
\eq\nr{P_L_cutoff} agrees with \eq(90) of ref.~\cite{madrid}, 
after noting that their 
$\P^{ }_h$ needs to be multiplied
by two to get our $\P^{ }_\tensor$, 
to account for both helicities. 


Apart from $\P_\tensor^\rmii{$L$}$, 
ref.~\cite{madrid} also gave cutoff-regularized 
results for what we call $\P_\tensor^\rmii{$J$}$.
However, it abandoned \eq\nr{p_cutoff} for that case, 
omitting instead substitutions at $p\to +\infty$
via analytic continuation of the infinite-past time boundary, 
and regularizing time integrals 
via a ``most natural prescription'',  
avoiding coincident times in their \eq(92) 
(only partly in \eq(91)). 
Such a regularization breaks the underlying $\tau'\leftrightarrow\tau''$
symmetry, and introduces artificial imaginary parts to the power spectrum. 
For the considerations in our \se\ref{ss:cutoff}, we cannot rely
on a recipe, but need to stick to the well-defined
regularization from \eq\nr{Phi_cutoff}. 

%
\subsection{A sample loop integral in dimensional regularization}
\la{ss:sample_dr}

In dimensional regularization, 
the contribution $\P_\tensor^\rmii{$L$}$ can be determined
for any $m$ and $\xi$.
As a starting point, we return to \eq\nr{P_L_combined}. 
We can then insert the mode function, $\Phi^{ }_p(\tau'\,)$, 
from \eq\nr{phi_massive}. Rescaling $p \to z/(-\tau'\,)$, 
this leads to 
\ba
 \delta \P_\tensor^\rmii{$L$}(\tau,k)
 &
 \underset{\rmii{\nr{phi_massive}}}{
 \overset{\rmii{\nr{P_L_combined}} \vphantom{ |^b_q } }{=}} 
 & -\,
 \meas{ (32\pi G)^2_{ } }
 \frac{(d-2)(d+1)\pi}{4d (4\pi)^{\frac{d}{2}}_{ } \Gamma(\frac{d}{2})} 
 \int_{-\infty}^\tau \!\!\! {\rm d}\tau'
 \,  
    \frac{
    \widetilde G^\rmii{R}_k(\tau,\tau'\,)
 \, \widetilde \Delta^{ }_k(\tau,\tau'\,) }{(-\tau'\,)^{d+1}_{ }}
 \, c^{ }_{\,d+2,\nu}
 \;, \la{P_L_dr} \hspace*{9mm} \\[2mm]
 c^{ }_{\,d,\nu}
 & \equiv & 
 \int_0^\infty \! {\rm d}z\, z^{d-1 } 
 \bigl| H^{\one}_\nu(z) \bigr|^2_{ }
 \;, \la{c_def}
\ea
where $H^{\one}_\nu$ is a Hankel function of the first kind. 

The coefficient $c^{ }_{\,d,\nu}$ can be evaluated 
by making use of Nicholson's formula, 
\be
 \bigl| H^{\one}_\nu(z) \bigr|^2_{ }
 \; = \; 
 J^{\hspace*{0.3mm}2}_\nu(z) + Y^{\hspace*{0.3mm}2}_\nu(z)
 \; = \; 
 \frac{8}{\pi^2_{ }}
 \int_0^\infty \! {\rm d}t\, \cosh(2\nu\hspace*{0.3mm} t) K^{ }_0 (2 z \sinh t)
 \;, 
\ee
where in turn the modified Bessel function of the second kind, 
$K^{ }_0$, is given by
\be
 K^{ }_0(x) \; = \; 
 \int_0^\infty \! {\rm d}s 
 \, \frac{\cos(x s)}{\sqrt{s^2_{ } + 1 \vphantom{ t } }}
 \;. 
\ee
Changing orders of integration and 
substituting $z\to z/(s \sinh t)$, the three integrals get 
factorized, so that\footnote{%
 For a practical crosscheck, 
 we remark that the subtracted integral
 $
 [c^{ }_{\,d,\nu}]^{ }_\mathrm{subtr} \equiv 
 \int_0^\infty \! {\rm d}z \, z^{d-1}_{ }
 \bigl[\, 
 \bigl| H^{\one}_\nu(z) \bigr|^2_{ } - 2/(\pi z)
 \,\bigr]
 $, 
 which in dimensional regularization 
 has the same value as the original one, is absolutely 
 convergent in 
 a certain range of $d$ and $\nu$,
 and can thus even be evaluated numerically.  
 } 
\ba
 c^{ }_{\,d,\nu}
 & = &  
 \frac{8}{\pi^2_{ }}
 \int_0^\infty \!\!\! {\rm d}t 
    \, \frac{\cosh(2\nu\hspace*{0.3mm} t)}{\sinh^{d}_{ }(t)}
 \int_0^\infty \!\!\! {\rm d}s
    \, \frac{1}{ s^{d}_{ }\sqrt{s^2_{ } + 1 \vphantom{ t }} } 
 \int_0^\infty \!\!\! {\rm d}z
    \, z^{d-1}_{ }\cos(2 z)
 \nn[2mm]
 & = & 
 \frac{\Gamma(d)\Gamma(1-d)\Gamma(\frac{d}{2})\Gamma(\frac{1-d}{2})}
      {\pi^{5/2}_{ }}
 \,
 \biggl[
   \frac{\Gamma(\frac{d}{2}-\nu)}{\Gamma(1-\frac{d}{2} - \nu)}
 + 
   \frac{\Gamma(\frac{d}{2}+\nu)}{\Gamma(1-\frac{d}{2} + \nu)} 
 \biggr]
 \,
 \cos\biggl( \frac{d\pi}{2} \biggr) 
 \;. \hspace*{6mm} \la{c_d_nu}
\ea
For a massless minimally coupled scalar field, 
the parameters take the values
$d = 3 + \delta$ and $\nu = (3 + \delta)/2$, 
cf.\ \eq\nr{nu}. Then, in $c^{ }_{d+2,\nu}$, the
second term in the square 
brackets in \eq\nr{c_d_nu} contains $1/\Gamma(0)$,
and vanishes, whereas the first term yields a finite contribution,
\ba
 c^{ }_{\,5+\delta\,,\, \frac{3}{2} + \frac{\delta}{2} }
 & \overset{\rmii{\nr{c_d_nu}}}{=} & 
   -\frac{9}{4\pi} + \rmO(\delta) \;, 
   \quad 
   \mathrm{for} 
   \; 
   m = \xi = 0
   \;. 
   \la{res_c_min}
\ea

As for the time integral in \eq\nr{P_L_dr}, it is
convergent for $d\to 3$, and given by \eq\nr{gr_delta_int}. 
Putting together the results of the momentum and time
integrals, the final expression reads
\ba
  \delta \P_\tensor^\rmii{$L$}(\tau,k)
 & 
 \underset{\rmii{\nr{P_L_dr}}}{
 \overset{\rmii{\nr{gr_delta_int}} \vphantom{ |^b_q } }{=}} 
 &
  -\,
 \frac{(8\pi G)^2_{ } H^4_{ }}{\pi^2_{ }}
 \frac{c^{ }_{5,\nu}}{9\pi}
 \bigl\{\, 
   2 - \re \bigl[\,
    e^{2ik\tau }_{ } (i + k \tau)^2_{ }
    E^{ }_1(2 i k \tau) 
     \,\bigr]
 \,\bigr\}
 \la{res_P_L_dr}
 \\[2mm]
 & 
 \underset{\rmii{\nr{res_c_min}}}{
 \overset{\scriptscriptstyle |k\tau|\;\ll\; 1 \vphantom{ |_q }}{\approx}}
 &
 -\, \frac{(8\pi G)^2_{ } H^4_{ }}{\pi^4_{ }}
 \; 
 \frac{ 
 \ln|2 k \tau| + \gammaE^{ }
 -2  
 }{4}
 \;. \la{res_L_dr}
\ea
Recalling 
$
  E^{ }_1(2 i k \tau) = 
  i\pi - \mbox{Ei}(-2 i k \tau ) 
$, 
the $\tau$-dependence of \eq\nr{res_P_L_dr} 
agrees with \eq(78) of ref.~\cite{madrid}, 
once the latter is multiplied by 2, 
in order to account for both helicity states.

\vspace*{3mm}

We end this section by remarking that the contribution 
$\P^\rmii{$X$}_\tensor$, from \eq\nr{pre_X}, can be evaluated
in a similar way as $\P^\rmii{$L$}_\tensor$. Given that it is
proportional to $\xi$, it appears to be absent in the
minimally coupled case, $\xi = 0$.
However,  if we put $m^2_{ }/H^2_{ } = 0$, and only then take
$\xi \to 0$, the first term in the square brackets
in \eq\nr{c_d_nu} has
a pole $\sim 1/\xi$, combining with the prefactor
to yield a finite contribution. 
The term only vanishes if we keep $m^2_{ }/H^2_{ }\neq 0$
while we take $\xi\to 0$, demonstrating the subtle nature
of these integrals. 
Nevertheless, the $\tau$-dependence from
\eq\nr{res_X_2} has the same form as at leading order, 
$1 + k^2_{ }\tau^2_{ }$, so this does not produce 
any qualitatively new structures. 

%
\section{Divergence structure of the classical contribution}
\la{se:match}

1-loop computations of the tensor power spectrum
induced by massless spectator scalar 
fields suggest that after the cancellation of 
power divergences, 
the result is suppressed by $H^2_{ }/\mpl^2$ compared with
the tree-level result, 
and therefore small~\cite{conformal,madrid,seoul}. 
However, the real world
is not massless, and could experience more 
complicated dynamics than these toy models.  
For example, as the universe thermalizes, 
its scalar fluctuations couple 
to hydrodynamic temperature and velocity perturbations. 
Such phenomena need normally  
to be studied numerically, but we should also maintain
control over the UV modes. 
If a framework is renormalizable, 
this can be achieved by {\em matching} the theoretically
well-defined dimensionally regularized UV computation, 
to a concrete cutoff-regularized IR framework. 
However, we show in the following that, perhaps surprisingly, 
the intuitively suggestive classical framework
(cf.\ \app\ref{app:def_cl}) 
contains {\em additional} divergences that are {\em not} 
present in the full theory. Nevertheless, with a concrete
cutoff regularization, these divergences are rendered finite
(even for a massless field), with the price that the outcome depends strongly 
on the cutoff, and needs therefore to be ``renormalized''. 
In the present section, we stay in de Sitter spacetime
with a minimally coupled massless scalar field, returning
to an outlook on necessary extensions in \se\ref{se:concl}. 

%
\subsection{Cutoff regularization}
\la{ss:cutoff}

A would-be numerical simulation requires a framework in which 
all integrals are manifestly finite. 
In non-linear studies this is often achieved with the help of a lattice
regularization, but here we choose the simpler path
of a cutoff regularization. A momentum cutoff was already introduced
in \se\ref{ss:sample_ct}, cf.\ \eq\nr{p_cutoff}. We now compute 
${J}^\cl_{pqk}$, from \eq\nr{res_Jcl}, and 
the associated 
$
  \delta \P_\tensor^{\scriptscriptstyle J,\mathrm{cl}}
$,
from \eq\nr{pre_J}, by slightly generalizing this procedure. 

Before proceeding, 
let us recall the change in physics that
is caused by a momentum cutoff. 
The cutoff is imposed on physical rather than comoving momenta.
At any time, it eliminates the contribution of very large comoving momenta, 
which we refer to as the UV domain. However, due to cosmological redshift,
small comoving momenta were large physical momenta at early times. The
smaller the comoving momenta that we consider at the observation time, 
the longer we have to 
integrate over them.
These long time integrations,
which we refer to as the IR domain,
yield another potential divergence (see also ref.~\cite{angelo}).  

To be concrete, 
according to \eq\nr{p_cutoff}, at a given time $\tau''$, 
we should restrict to comoving momenta 
$p,q < -\mu / (\tau'' H)$. 
Thanks to the time ordering in \eq\nr{res_Jcl}, 
which implies $|\tau''| \ge |\tau'|$, 
this guarantees that $ p,q < -\mu / (\tau' H) $. 
It is convenient to enlarge the domain a bit, restricting to 
\be
 p + q \; \le \; -\frac{2 \mu}{\tau'' H}
 \quad 
 \underset{\rmii{\nr{full_meas}}}{
 \overset{\rmii{\nr{qpm}}\vphantom{ \big | }}{\Leftrightarrow}} 
 \quad
 1 \; \le \; 
 r 
 \; \le \; -\frac{2\mu}{\tau'' k H} 
 \; \le \; -\frac{2\mu}{\tau' k H} 
 \; \le \; -\frac{2\mu}{\tau k H} 
 \;. \la{bounds}
\ee
The absolute lower bound on the times is 
$
 \tau,\tau', \tau'' \; \ge \; - { 2 \mu }/{ (k H) }
$, 
with the smallest momentum transfers ($r\approx 1$)
contributing the longest.
Streamlining the notation with the shorthands
\be
 r^{ }_\rmi{max} \; \equiv \; -\frac{2\mu}{\tau k H}
 \;, \quad
 \tau^{ }_\rmi{min} \; \equiv \; -\,\frac{2\mu}{r k H}
 \;, \la{shorthands}
\ee
the full measure reads
\ba
 \int_{1}^{r^{ }_\rmii{max}} \! {\rm d}r
 \int_{-1}^{+1} \! {\rm d}a
 \int_{\tau^{ }_\rmii{min}}^{\tau} \! {\rm d}\tau' 
 \int_{\tau^{ }_\rmii{min}}^{\tau'} \! {\rm d}\tau''
 \;. \la{t_limits}
\ea
We note that at the upper boundary of the $r$-integration, 
$r\to r^{ }_\rmi{max}$, mode functions enter the integration domain
only just before the evaluation time, i.e.\ 
$\lim_{r\to r^{ }_\rmii{max}}\tau^{ }_\rmi{min}(r) = \tau^{ }_{ }$.

In order to evaluate ${J}^\cl_{pqk}$
with the measure from \eq\nr{t_limits}, 
we define the indefinite integral 
\ba
 Z(\tau,\tau'\,) 
 & \equiv & 
 \int_{ 
      }^{ } \! {\rm d}\tau'
 \, \widetilde G^\rmii{R$(0)$}_k(\tau,\tau'\,)
 \, \Phi^{\zero}_p(\tau'\,)
 \, \Phi^{\zero}_q(\tau'\,)
 \; = \; 
 \frac{W^{ }_+(\tau,\tau'\,) - W^{ }_-(\tau,\tau'\,)}{i}
 \;, 
 \la{def_Z} \\[2mm]
 W^{ }_{\pm}(\tau,\tau'\,)
 & 
 \overset{\scriptscriptstyle \tau \; > \; \tau' \lift }{\equiv} 
 & 
 \frac{H^2_{ } e^{-i(p+q)\tau}_{ }}{4 k^3_{ }\sqrt{p q}}
 e^{i(p+q\pm k)(\tau - \tau'\,)}_{ }
 (1 \mp i k \tau) \beta^{ }_\pm (\tau'\,) 
 \; + \; c^{ }_{\pm}(\tau) 
 \;, \la{def_W}
\ea
where $\widetilde G^\rmii{R$(0)$}_k$ is from \eq\nr{gr_k}, 
$\Phi^{\zero}_p$ is from \eq\nr{def_Phi0}, and 
$c^{ }_\pm$ are integration constants. For a massless 
minimally coupled scalar field, $\Phi^{\zero}_p$ takes 
the form in \eq\nr{phi_minimal}, and 
\ba
 \beta^{ }_{\pm}(\tau'\,)
 & 
 \overset{m\,,\,\xi\;\to\;0 \vphantom{\big | }}{\equiv} 
 & 
 \frac{1}{p q \tau'}
 + \frac{i(p q \pm k p \pm k q) \mp p q k \tau'}{p q (p+q\pm k)}
 \pm \frac{i k}{(p+q\pm k)^2_{ }}
 \;. \la{res_beta}
\ea
Making use of $Z(\tau,\tau'\,)$, we obtain
\ba
  J^{\cl\zero}_{pqk}(\tau) 
 & 
 \underset{\rmii{\nr{shorthands},\nr{def_Z}}}{
 \overset{\rmii{\nr{res_Jcl}} \vphantom{ |^b_q} }{ \equiv }} 
 & 
 \int_{\tau^{ }_\rmii{min}}^{\tau} \!\!\! {\rm d}\tau'
 \, 
 \re\bigl\{
 \, 
 \overbrace{
 \widetilde G^\rmii{R$(0)$}_k(\tau,\tau'\,)
 \Phi_p^{*\zero}(\tau'\,) 
 \Phi_q^{*\zero}(\tau'\,) 
 }^{ \partial Z^*_{ }(\tau,\tau'\,)/\partial\tau' }
 \, \bigl[\,
       Z(\tau,\tau'\,) - Z(\tau,\tau^{ }_\rmi{min})
    \,\bigr]
 \, \bigr\}
 \nn[2mm]
 & = & 
 \int_{\tau^{ }_\rmii{min}}^{\tau} \!\!\! {\rm d}\tau'
 \, 
 \frac{\partial}{\partial\tau'}
 \biggl\{\,
   \frac{|Z(\tau,\tau'\,)|^2_{ }}{2}
   - \re \bigl[\, Z^*_{ }(\tau,\tau'\,) Z(\tau,\tau^{ }_\rmi{min}) \,\bigr]
   + c^{ }_0(\tau) 
 \,\biggr\}
 \nn[2mm]
 & = & 
 \frac{|Z(\tau,\tau) - Z(\tau,\tau^{ }_\rmii{min})|^2_{ }}{2} 
 \;, \la{ex_Z}
\ea
where $c^{ }_0 \in \mathbbm{R}$ 
is another integration constant, which has dropped out from the result. 
Note that $c^{ }_0$ would remain present if we were to compute the 
integral naively, by ignoring the lower boundary.
The power spectrum corresponding
to \eq\nr{ex_Z} is obtained from \eq\nr{pre_J}, as 
\ba
 \delta \P_\tensor^{\scriptscriptstyle J,\mathrm{cl}}(\tau,k)
 & 
 \underset{\rmii{\rmii{\nr{ra},\nr{full_meas}}}}{
 \overset{\rmii{\nr{pre_J}} \vphantom{ |^b_q} }{=}} 
 & 
 \frac{k^{10}_{ }}{2\pi^2_{ } \mpl^4}
 \la{P_J_cl_3} \\[2mm]
 &
 &
 \hspace*{-3mm}
 \times\,
 \int_{1}^{r^{ }_\rmii{max} 
           } \!\!\!\! {\rm d}r 
 \int_{-1}^{+1} \!\!\!\! {\rm d}a
 \, (r^2_{ } - a^2_{ }) 
 \, (r^2_{ } - 1)^{2}_{ }
 \, (1 - a^2_{ })^{2}_{ }
 \, \times \, J^{\cl\zero}_{pqk}(\tau)
 \Big|^{p \to \frac{k(r-a)}{2}}_{q\to \frac{k(r+a)}{2} } 
 \;. \nonumber
\ea

Introducing the shorthand notations
\be
 \beta^{ }_{\pm} \; \equiv \; \beta^{ }_{\pm}(\tau)
 \;, \quad
 \beta^{ }_{\pm\rmi{min}} \; \equiv \; \beta^{ }_{\pm}(\tau^{ }_\rmi{min})
 \;, \quad
 \theta^{ }_{\pm} \; \equiv \; (r\pm 1) k (\tau - \tau^{ }_\rmi{min} )
 \;, \la{beta_pm}
\ee
the results from \eqs\nr{ex_Z} and \nr{P_J_cl_3} can be expressed as 
\ba
 \frac{
    \delta \P_\tensor^{\scriptscriptstyle J,\mathrm{cl}}(\tau,k)
      }{
     H^4_{ } / (8 \pi^2_{ }\mpl^4) 
      }
 &
 \underset{\rmii{\nr{def_W},\nr{beta_pm}}}{
 \overset{\rmii{\nr{ex_Z},\nr{P_J_cl_3}}\vphantom{\big | }}{=}} 
 &
 \frac{k^2_{ }}{2}
 \int_1^{r^{ }_\rmii{max}} \!\! {\rm d}r\, (r^2_{ } - 1)^2_{ }
 \int_{-1}^{+1} \! {\rm d}a \, (1 - a^2_{ })^2_{ } 
 \, \F(\cdot)
 \;, 
 \la{J_cl_cutoff} \\[2mm]
 \F(\cdot) 
 & 
 \equiv 
 & 
 \bigl|\,
 (1 - i k \tau)(  \beta^{ }_+ - \beta^{ }_{+\rmi{min}}   )  
 - 
 (1 + i k \tau)(  \beta^{ }_- - \beta^{ }_{-\rmi{min}}   )
 \,\bigr|^2_{ }
 \la{calF} \\[2mm]
 & & 
 + \,2\, 
  \re\bigl\{ 
  \bigl[\, 1 - e^{i ( \theta^{ }_+ - \theta^{ }_- )}_{ } \,\bigr] 
  \, 
                (1 - i k \tau)^2_{ }
                 \beta^{ }_{+\rmi{min}}
                 \beta^*_{-\rmi{min}} 
         \bigr\}
 \nn[2mm]
 & & 
 + \,2\, 
 \re\bigl\{ 
 \bigl(\, 1- e^{-i \theta^{ }_+}_{ } \,\bigr)
  \, \bigl[
    (1 + k^2_{ }\tau^2_{ }) \beta^{ }_{+} 
  - 
    (1 + i k \tau)^2_{ } \beta^{ }_{-} 
         \bigr]
  \, \beta^{*}_{+\rmi{min}} \bigr\}
 \nn[2mm]
 & & 
 + \,2\,
  \re\bigl\{
  \bigl(\, 1 - e^{-i \theta^{ }_-}_{ } \,\bigr)
  \, \bigl[
    (1 + k^2_{ }\tau^2_{ }) \beta^{ }_{-} 
  - 
    (1 - i k \tau)^2_{ } \beta^{ }_{+} 
  \bigr]
  \, \beta^{*}_{-\rmi{min}}
  \bigr\}
 \;, \nonumber 
\ea
where the argument
``$\cdot$'' stands for the integration variables, $r$ and $a$, 
and the two dimensionless parameters appearing
in the integrand, $k\tau$ and $\mu/H$
(the latter via $k \tau^{ }_\rmii{min}$).
We note that the angular integral, over $a$, can be carried out 
in all the terms, however the resulting expressions 
are in general intransparent, 
so we refrain from showing them here. 

\begin{figure}[t]

\hspace*{-0.1cm}
\centerline{%
 \epsfysize=7.3cm\epsfbox{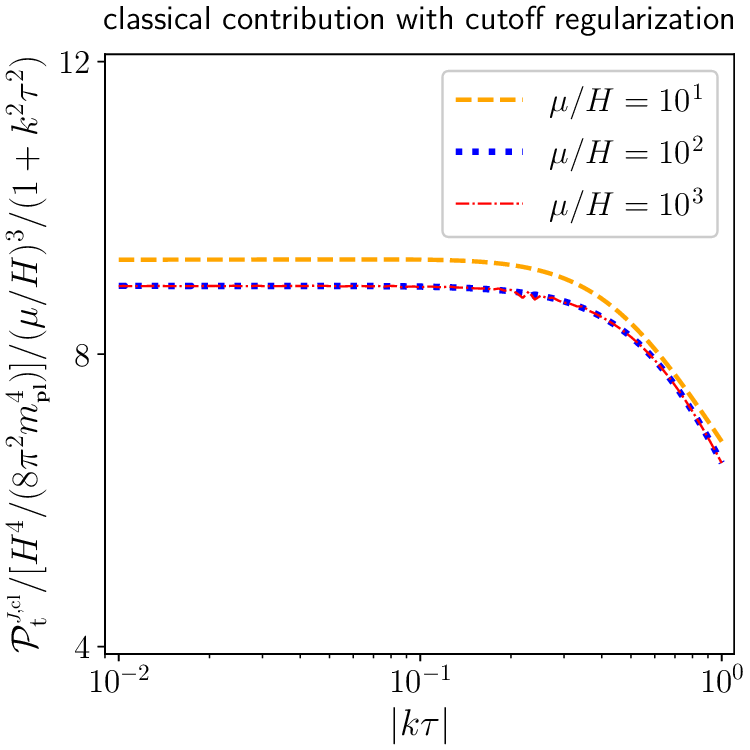}%
 \hspace{0.4cm}%
 \epsfysize=7.3cm\epsfbox{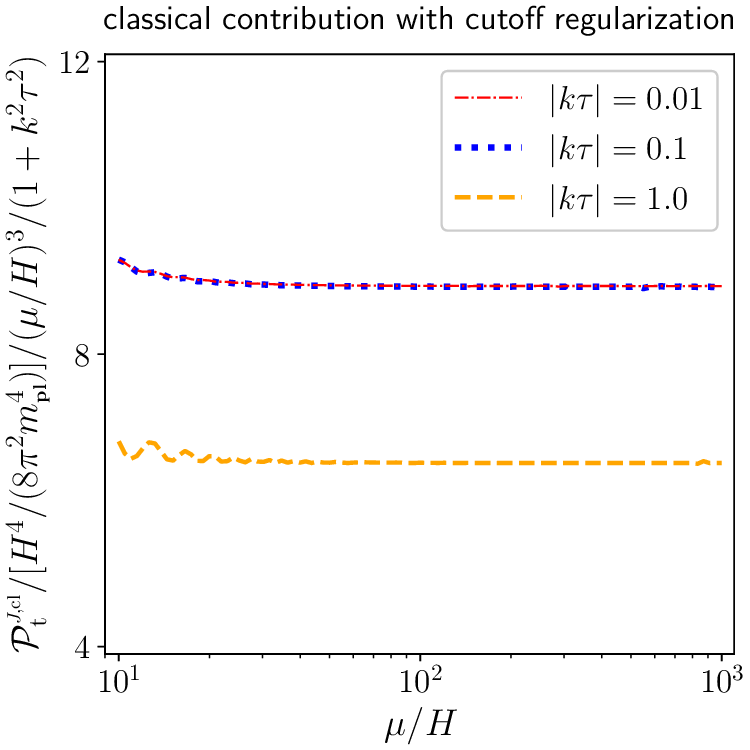}%
}

\caption[a]{\small
  We plot the result from \eq\nr{J_cl_cutoff} for 
  $ 
  {
    \delta \P_\tensor^{\scriptscriptstyle J,\mathrm{cl}}(\tau,k)
      } / {
    [H^4_{ } / (8 \pi^2_{ }\mpl^4)]
      }
  $
  as a function of  
  $|k \tau| \in (0.01 ... 1.0 )$ for 
  $\mu/H \in \{ 10,100, 1000\}$, and 
  as a function of $\mu/H \in (10 ... 1000)$ for 
  $|k\tau| \in \{ 0.01,0.1,1.0 \}$, 
  normalized to $(\mu/H)^3_{ }(1 + k^2_{ }\tau^2_{ })$.
  The origin of the cubic dependence on $\mu/H$
  is explained in \se\ref{ss:ct}.
  }

\la{fig:Jcl_bare}
\end{figure}

In \fig\ref{fig:Jcl_bare},
we plot the result from \eq\nr{J_cl_cutoff} 
as a function of 
$|k \tau| \in (0.01 ... 1.0 )$ for 
$\mu/H \in \{ 10,100, 1000\}$, 
and as a function of  
$\mu/H \in (10 ... 1000 )$ for 
$|k\tau| \in \{ 0.01,0.1,1.0 \}$.
The results have been normalized to 
$(\mu/H)^3_{ } (1 + k^2_{ }\tau^2_{ })$,
showing that they 
become constant for $|k \tau| \ll 1$, 
and 
scale nearly perfectly as $(\mu/H)^3_{ }$.  
We return to how the power divergence 
can be understood in \se\ref{ss:ct}, 
after having first shed light on the $r$-integrand, 
by re-considering it in dimensional regularization, 
where the scale $\mu/H$ is absent.

%
\subsection{Dimensional regularization}
\la{ss:dr}

Next, we determine the classical contribution to the tensor
power spectrum, 
$
  \delta \P_\tensor^{\scriptscriptstyle J,\mathrm{cl}}
$,
in dimensional regularization. From the outset, 
it is important to stress that
dimensional regularization only concerns
the momentum integration measure, and the procedure also requires
{\em another} prescription, namely how to deal with
the lower boundaries of the time integrals
($\tau',\tau''\to-\infty$). 
A straightforward 
possibility is to add an exponential convergence factor into
the time dependences.
In analogy with \eq\nr{Phi_cutoff}, 
we thus set
\be
 \Phi^{ }_p(\tau'\,) \;\longrightarrow\;
 \Phi^{ }_p(\tau'\,)\, e^{ \frac{\omega \tau'}{2} }_{ }
 \;, \quad \omega = 0^+_{ } 
 \;. \la{Phi_expo}
\ee
Since the mode functions appear as a product, 
$
 \Phi^{ }_p(\tau'\,) \Phi^{ }_q(\tau'\,)
$, the convergence factors turns into $e^{\omega \tau'}_{ }$, 
and $\omega$ cuts off times 
$\tau' \ll - 1/\omega$. 
Comparing with the bounds mentioned below \eq\nr{bounds} we see that, 
parametrically, $\omega/k \leftrightarrow H/(2 \mu)$.
Conceptually, 
$\omega$ serves as an IR regulator.
In practical
terms, the presence of $\omega > 0$ implies that the lower-boundary
substitution can be omitted in \eq\nr{ex_Z}, provided 
that the indefinite integral has been properly identified
(we explain this in some detail in \se\ref{ss:vac}, where 
the issue is particularly delicate). 

As for the side of the spatial momentum integrals, we
adopt the starting point from \eq\nr{pre_J}. 
Inserting $1-z^2_{ }$ from \eq\nr{ra}, 
and re-expressing the integration measure according to 
\eq\nr{full_meas}, we consider 
\ba
 \delta \P_\tensor^{\scriptscriptstyle J,\mathrm{cl}}(\tau,k)
 \!\!
 & 
 \underset{\rmii{\nr{ra},\nr{full_meas}}}{
 \overset{\rmii{\nr{pre_J}} \vphantom{ |^b_q } }{=}} 
 & 
 \!\!
 \A 
 \int_{1}^{\infty} \!\!\!\! {\rm d}r 
 \int_{-1}^{+1} \!\!\!\! {\rm d}a
 \, (r^2_{ } - a^2_{ }) 
 \, (r^2_{ } - 1)^{\frac{d+1}{2}}_{ }
 \, (1 - a^2_{ })^{\frac{d+1}{2}}_{ }
 \; J^\rmi{cl}_{pqk}(\tau)
 \Big|^{p \to \frac{k(r-a)}{2}}_{q\to \frac{k(r+a)}{2} } 
 \;, \hspace*{8mm} 
 \la{P_J_cl_1} 
 \\[3mm]
 \A 
 \hspace*{-3mm}
 & \equiv & 
 \hspace*{-3mm}
 \meas{ }
 \frac{ (32\pi G)^2_{ } }
      {  (4\pi)^{\frac{d+1}{2}}_{ } \Gamma(\frac{d-1}{2}) }
 \frac{(d-2) k^{d+4}_{ } }
      {(d-1) 2^{d+2}_{ } }
 \;. \la{def_A}
\ea 

Our goal is to determine the $1/\delta$ divergences of 
\eq\nr{P_J_cl_1}. For this purpose, it is sufficient to adopt the 
$\rmO(\delta^0_{ })$ part of $ J^{\cl}_{pqk} $, 
as given by \eq\nr{ex_Z}, i.e.\ 
$ J^{\cl\zero}_{pqk} = |Z(\tau,\tau)|^2_{ }/2$ 
(this is explained in more detail
in the paragraph below \eq\nr{sum_I}).
From \eqs\nr{def_W} and \nr{res_beta}, we then see that 
\ba
 J^{\cl\zero}_{pqk}(\tau) 
 &
    \overset{\rmii{\nr{ex_Z}}  \vphantom{ |^b_q } }
   {\underset{\omega \; \to \; 0}{=}} 
 & 
 \frac{H^4_{ }}{32 k^6_{ } p q}
 \bigl|\,
 (1 - i k \tau)\,  \beta^{ }_+ 
 - 
 (1 + i k \tau)\,  \beta^{ }_- 
 \,\bigr|^2_{ }
 \la{res_ZZ} \\[2mm]
 & 
 \underset{m\,,\,\xi\;\to\;0 \vphantom{ | }}{
 \overset{\rmii{\nr{def_W},\nr{res_beta}}\vphantom{ \big | }}{=}} 
 &  
 \frac{ H^4_{ } }
 {4 k^{10}_{ } (r^2_{ } - a^2_{ }) (r^2_{ } - 1 )^2_{ } }
 \biggl[\,
   \frac{8(1 + k^2_{ }\tau^2_{ })}{(r^2_{ } - 1)^2_{ }}
 + \, \frac{32(1 + k^2_{ }\tau^2_{ })}{(r^2_{ } - 1)(r^2_{ } - a^2_{ })}
 \nn[2mm]
 & & \hspace*{4mm} 
 + \, \frac{32(1 + a^2_{ }k^2_{ }\tau^2_{ })}{(r^2_{ } - a^2_{ })^2_{ }}
 + \, \frac{48 k^2_{ }\tau^2_{ }}{ r^2_{ } - a^2_{ } }
 + \, 2 k^4_{ }\tau^4_{ }
 \,\biggr]
 \;, \hspace*{6mm} 
 \la{Jcl_res} 
\ea
where the prefactor has been chosen to cancel the $r$-dependent
measure terms from \eq\nr{P_J_cl_1} for $d=3$. We remark that,  
apart from the last contribution in \eq\nr{Jcl_res}, 
which vanishes for $|k\tau|\to 0$, 
all terms are UV-finite at large $r$. 

Even if the $|k\tau|\to 0$ limit of \eq\nr{Jcl_res}
is UV-finite, the first two terms are IR-divergent. 
Expanding the integrand around $r = 1$, we find that 
\be
 J^{\cl\zero}_{pqk}(\tau) 
 \;
 \underset{r\;\approx\;1}{
 \overset{\rmii{\nr{Jcl_res}} \vphantom{ |^b_q } }{\approx}} 
 \;
 \frac{ H^4_{ } (1 + k^2_{ }\tau^2_{ }) }
 {2 k^{10}_{ } (r^2_{ } - a^2_{ }) (r^2_{ } - 1 )^2_{ } }
 \biggl[\,
 + \, \frac{1}{(r-1)^2}  + \frac{7 + a^2_{ }}{(r-1)(1-a^2_{ })}
 + ... 
 \,\biggr]
 \;. \la{Jcl_ir}
\ee
Later on, 
in \eq\nr{Jvac_ir},  
we show how this IR divergence is cancelled
by the corresponding terms of the vacuum contribution. 

All the terms in \eq\nr{Jcl_res} can be integrated in dimensional 
regularization. 
The logarithmic IR divergences, 
manifest in 
\eq\nr{Jcl_ir}, yield poles in $1/\delta^{ }_\rmii{IR}$, 
where the subscript $(...)^{ }_\rmii{IR}$ 
acts as a reminder of their origin.  
We list the individual integrals
originating from \eq\nr{Jcl_res} in \app\ref{app:I_J}.
Their sum yields 
\ba
 \delta \P_\tensor^{\scriptscriptstyle J,\mathrm{cl}}(\tau,k)
 & 
 \underset{\rmii{\nr{def_A}}}{
 \overset{\rmii{\nr{P_J_cl_1}} \vphantom{ |^b_q } }{ = }} 
 & 
 \overbrace{
 \biggl[\,
 \frac{ H^4_{ }}{8 \pi^2_{ }\mpl^4 } \, + \, \rmO(\delta)
 \,\biggr]
 }^{ 
 { \A H^4_{ }} / ({ 4\hspace*{0.3mm} k^{10}_{ } })  
 }
 \; 
 \biggl[\,
 \sum_i \mathcal{I}^{ }_i(\tau,k)
 \; + \; \rmO(1)
 \,\biggr]
 \;, \la{P_J_cl_2}
 \\[3mm]
 \sum_i \mathcal{I}^{ }_i(\tau,k)
 & 
 \overset{\rmii{\nr{i1}--\nr{i5}}  \vphantom{ |^b_q }  }{=} 
 & 
 \frac{192\,(1 + k^2_{ }\tau^2_{ })}{5}
 \biggl(\, \frac{1}{\delta^{ }_\rmii{IR}} + 2 \ln 2 \,\biggr)
 \nn[2mm]
 & & \quad - \,
 \frac{16\,(291 + 91 k^2_{ }\tau^2_{ } + 10 k^4_{ }\tau^4_{ })}{75}
 + \rmO(\delta)
 \;. \la{sum_I}
\ea
The term of $\rmO(1)$ is incomplete
in \eq\nr{P_J_cl_2}, since we have
not accounted for the $\rmO(\delta)$-corrections to the mode functions. 
However, the $1/\delta$ divergence is unambiguous, since there are no
$1/\delta^{\hspace*{0.3mm}2}_{ }$ divergences 
in 1-loop integrals, which could combine
with the $\rmO(\delta)$-corrections, 
to yield further $1/\delta$ terms. 
It is worth stressing
that due to its IR sensitivity, 
$
 \delta \P_\tensor^{\scriptscriptstyle J,\mathrm{cl}}
$ 
in \eq\nr{P_J_cl_2} 
is {\em not} suppressed by $k^3_{ }$, as was assumed
in ref.~\cite{ema}, where it was omitted for that reason.

%
\subsection{Cancellation of IR divergences between
the classical and vacuum parts}
\la{ss:vac}

We now turn to the vacuum contribution, 
$ J^{\rmi{vac}\zero}_{pqk} $,
and the corresponding power spectrum, 
$
 \delta \P_\tensor^{\scriptscriptstyle J,\mathrm{vac}}
$, 
showing that it cancels the IR divergence found in 
\eq\nr{Jcl_ir}.   
We also take this as an opportunity to illustrate the subtle
nature of the time integrals, and the role that the additional
regulator introduced in \eq\nr{Phi_expo} plays
in dimensional regularization. 

To understand the issue, we consider an integral similar to that
in \eq\nr{ex_Z}, 
\be
 \I^{ }_\omega (\tau) \; \equiv \; 
 \int_{-\infty}^{\tau} \! {\rm d}\tau' 
 \, f^{ }_\omega (\tau'\,) 
 \, \partial^{ }_{\tau'} f^{ }_\omega(\tau'\,) 
 \;, 
 \quad
 f^{ }_\omega(\tau'\,) \; = \; e^{\hspace*{0.3mm}\omega\tau'}_{ } \, 
 \Bigl(  
   a \tau' + b + \frac{c}{ \tau' }
 \Bigr)
 \;, \quad 
 \omega \; = \; 0^+_{ }
 \;. \la{subtle}
\ee
The integral can obviously be carried out, with the result
$ 
 \I^{ }_\omega(\tau) = f^2_\omega(\tau) / 2
$.
However, suppose that we put $\omega\to 0$ {\em before}
carrying out the integral, 
\ba
 \lim_{\omega\to 0}
 \, f^{ }_\omega (\tau'\,) 
 \, \partial^{ }_{\tau'} f^{ }_\omega(\tau'\,) 
 & = & 
 a^2_{ } \tau' + a b - \frac{b c}{\tau^{\prime\hspace*{0.3mm}2}_{ }}
 - \frac{c^2_{ }}{ \tau^{\prime\hspace*{0.3mm}3}_{ } }
 \nn[2mm]
 & = & 
 \partial^{ }_{\tau'}
 \biggl(\, 
 \frac{a^2_{ } \tau^{\prime\hspace*{0.3mm}2}_{ }}{2}
 + a b \tau' 
 + \frac{bc}{\tau'}
 + \frac{c^2_{ }}{2 \tau^{\prime\hspace*{0.3mm}2}_{ }}
 + \mbox{constant}
 \,\biggr)
 \;. \la{constant}
\ea
If we substitute at the upper boundary in \eq\nr{subtle}, 
omitting the constant from \eq\nr{constant}, we miss 
the $\tau$-independent terms,
$a c + b^2_{ }/2$, from the correct result, $f^2_{0}(\tau)/2$. 
This would not be a problem with a {\em finite} domain, 
because then the constant cancels between the substitutions. 

To avoid the problem, there are two possibilities. One is that we 
indeed keep $\omega > 0$ while carrying out the integral
in \eq\nr{subtle}, and 
take $\omega\to 0^+_{ }$ only afterwards. Unfortunately, 
this yields lengthy expressions. The other is that 
we identify the dangerous terms, with no explicit $\tau'$-dependence, 
as factorized structures, 
$\sim f^{ }_\omega \partial^{ }_{\tau'} f^{ }_\omega $.
Then that integral is easy to carry out with $\omega > 0$, 
yielding  
$f^2_\omega(\tau)/2$. In the following we illustrate the latter 
approach. 

To proceed, 
we insert the mode functions at zeroth order in $\delta$, 
and define 
\be
 Y(\tau,\tau'\,) 
 \; \equiv \; 
 \int_{ 
      }^{ } \! {\rm d}\tau'
 \, \widetilde \Delta^\rmii{$(0)$}_k(\tau,\tau'\,)
 \, \Phi^{\zero}_p(\tau'\,)
 \, \Phi^{\zero }_q(\tau'\,)
 \; = \; 
 \frac{W^{ }_+(\tau,\tau'\,) + W^{ }_-(\tau,\tau'\,)}{2}
 \la{def_Y}
 \;, 
\ee
where $\widetilde \Delta^\rmii{$(0)$}_k$ is from \eq\nr{delta_k}, 
and $W^{ }_\pm$ are from \eq\nr{def_W}
(but now with additional contributions from $\omega$). 
Making use of $Y(\tau,\tau'\,)$, we obtain
\be
  J^{\vac\zero}_{pqk}(\tau) 
 \;
 \underset{\rmii{\nr{shorthands},\nr{def_Y}}}{
 \overset{\rmii{\nr{res_Jvac}} \vphantom{ |^b_q} }{ \equiv }} 
 \;
 2\, 
 \int_{-\infty}^{\tau} \!\!\! {\rm d}\tau'
 \, 
 \im\bigl\{
 \, 
 \overbrace{
 \widetilde G^\rmii{R$(0)$}_k(\tau,\tau'\,)\,
 \Phi_p^{*\zero}(\tau'\,) 
 \Phi_q^{*\zero}(\tau'\,) 
 }^{ \partial Z^*_{ }(\tau,\tau'\,)/\partial\tau' }
 \, \bigl[\,
       Y(\tau,\tau'\,) - Y(\tau,-\infty)
    \,\bigr]
 \, \bigr\}
 \;. \la{ex_Y}
\ee
The convergence factor $\omega > 0$ guarantees that 
$Y(\tau,-\infty) = 0$. We write 
$
 \partial Z(\tau,\tau'\,)/\partial\tau'
$
in a form similar to \eqs\nr{def_Z} and \nr{def_Y}, 
\ba
 \frac{\partial Z(\tau,\tau'\,) }{\partial \tau'}
 & \overset{\rmii{\nr{def_Z}}}{=} & 
 \, \widetilde G^\rmii{R$(0)$}_k(\tau,\tau'\,)
 \, \Phi^{\zero}_p(\tau'\,)
 \, \Phi^{\zero }_q(\tau'\,)
 \; \overset{\rmii{\nr{def_Z}}}{=} \;
 \frac{
     \partial^{ }_{\tau'} W^{ }_+(\tau,\tau'\,)
     -
     \partial^{ }_{\tau'} W^{ }_-(\tau,\tau'\,)
     }{i}  
 \;, 
 \la{d_Z} \hspace*{9mm} \\[2mm]
 \frac{\partial W^{ }_{\pm}(\tau,\tau'\,) }{\partial \tau'}
 & 
 \overset{ }{ = } 
 & 
 \frac{H^2_{ } e^{-i(p+q)\tau + \omega \tau'}_{ }}{4 k^3_{ }\sqrt{p q}}
 e^{i(p+q\pm k)(\tau - \tau'\,)}_{ }
 (1 \mp i k \tau)\, \alpha^{ }_\pm (\tau'\,) 
 \;, \la{def_alpha}
\ea
where the coefficient functions read
\ba
 \alpha^{ }_{\pm}(\tau'\,)
 & 
 \overset{\rmii{\nr{phi_minimal},\nr{gr_k}}\vphantom{\big | }}{
 \underset{m\,,\,\xi\;\to\;0 }{ \equiv }} 
 & 
 \pm\, i k \tau' 
 + 1 \pm \frac{k(p+q)}{pq}
 - \frac{i(p+q\pm k)}{p q \tau'}
 - \frac{1}{p q (\tau'\,)^2_{ }}
 \;. \la{res_alpha}
\ea
Inserting these, we get
\ba
  J^{\vac\zero}_{pqk}(\tau) 
 &
 \underset{\rmii{\nr{def_Y},\nr{def_alpha}}}{
 \overset{\rmii{\nr{ex_Y}} \vphantom{ |^b_q} }{ = }} 
 &
 \frac{H^4_{ }}{16 k^6_{ } p q}
 \int_{-\infty}^{\tau} \!\!\! {\rm d}\tau' \, e^{2 \omega  \tau'}_{ }
 \,  \Bigl\{ 
   \, (1 + k^2_{ }\tau^2_{ })
   \re\bigl[\,
      ( \alpha^*_+ \, \beta^{ }_+ 
   \;-\;  
      \alpha^*_- \, \beta^{ }_- \,) (\tau'\,) 
      \,\bigr]
 \nn[2mm]
 & & 
 \, + \, 
     \cos[2 k (\tau - \tau'\,)]
  \, \re \bigl[\,
     (1 + i k \tau)^2 
     \bigl(\,
       \alpha^*_{+} \, \beta^{ }_{-} 
   \, - \, 
      \alpha^{ }_{-} \, \beta^{*}_{+} 
     \,\bigr)(\tau'\,)
     \,\bigr]
 \nn[2mm]
 & & 
 \, + \, 
     \sin[2 k (\tau - \tau'\,)]
  \, \im \bigl[\,
     (1 + i k \tau)^2 
     \bigl(\,
       \alpha^*_{+} \, \beta^{ }_{-} 
   \, - \, 
      \alpha^{ }_{-} \, \beta^{*}_{+} 
     \,\bigr) (\tau'\,) 
     \,\bigr]
  \;\Bigr\}
  \;. \hspace*{6mm} \la{res_Jvac_1}  
\ea

Now, the problems reside in the first term
of \eq\nr{res_Jvac_1}, which would have no explicitly
$\tau'$-dependent prefactor, if we were to set $\omega\to 0$
in the integrand. Luckily, these particular terms have 
a similar
structure as in \eq\nr{ex_Z}. 
The reason is that while 
$Z$ (cf.\ \eq\nr{def_Z}) 
and $Y$ (cf.\ \eq\nr{def_Y})
are different, they are made of 
the same parts, 
\be
 Z \; \overset{\rmii{\nr{def_Z}}}{=} \; \frac{W^{ }_+ - W^{ }_-}{i}
 \;, \quad
 Y \; \overset{\rmii{\nr{def_Y}}}{=} \; \frac{W^{ }_+ + W^{ }_-}{2}
 \;. \la{schem}
\ee
Therefore, the classical and vacuum parts contain
\ba
 J^{\cl\zero}_{pqk}(\tau) 
 &
 \overset{\rmii{\nr{ex_Z}}}{=}
 & 
 \int_{-\infty}^{\tau} \!\!\! {\rm d}\tau'
 \, \re[(\partial^{ }_{\tau'} Z^*_{ }) Z]
 \
 \overset{\rmii{\nr{schem}}}{\supset}
 \ 
 \int_{-\infty}^{\tau} \!\!\! {\rm d}\tau' \, 
 \re \bigl[\,
   ( \partial^{ }_{\tau'} W_+^* ) W^{ }_+
 \; + \; 
   ( \partial^{ }_{\tau'} W_-^* ) W^{ }_-
 \,\bigr]
 \;,
 \nn
  \la{differ}
 \\
 J^{\vac\zero}_{pqk}(\tau) 
 &
 \overset{\rmii{\nr{ex_Y}}}{=}
 & 
 2 
 \int_{-\infty}^{\tau} \!\!\! {\rm d}\tau'
 \, \im[(\partial^{ }_{\tau'} Z^*_{ }) Y]
 \
 \overset{\rmii{\nr{schem}}}{\supset}
 \ 
 \int_{-\infty}^{\tau} \!\!\! {\rm d}\tau' \, 
 \re \bigl[\,
   ( \partial^{ }_{\tau'} W_+^* ) W^{ }_+
 \; - \; 
   ( \partial^{ }_{\tau'} W_-^* ) W^{ }_-
 \,\bigr]
 \;. \nonumber
\ea
These particular terms can be integrated in both cases, and yield
structures like in \eq\nr{res_ZZ}. Therefore \eq\nr{res_Jvac_1}
turns into
\ba
  J^{\vac\zero}_{pqk}(\tau) 
 &
 \underset{\rmii{\nr{res_ZZ},\nr{differ}}}{
 \overset{\rmii{\nr{res_Jvac_1}} \vphantom{ |^b_q} }{ = }} 
 &
 \frac{H^4_{ }}{16 k^6_{ } p q}
 \biggl\{\, 
   (1 + k^2_{ }\tau^2_{ })
   \frac{ |\beta^{ }_+|^2_{ } - |\beta^{ }_-|^2_{ } }{2}
 \; + \; 
 \int_{-\infty}^{\tau} \!\!\! {\rm d}\tau' \, e^{2 \omega  \tau'}_{ }
 \nn[2mm]
 & &
   \times \,  \Bigl[ 
   \, \cos[2 k (\tau - \tau'\,)]
  \, \re \bigl[\,
     (1 + i k \tau)^2 
     \bigl(\,
       \alpha^*_{+} \, \beta^{ }_{-} 
   \, - \, 
      \alpha^{ }_{-} \, \beta^{*}_{+} 
     \,\bigr) (\tau'\,)
     \,\bigr]
 \la{res_Jvac_2}  \\[2mm]
 & &
 \hspace*{1mm} 
 \, + \, 
     \sin[2 k (\tau - \tau'\,)]
  \, \im \bigl[\,
     (1 + i k \tau)^2 
     \bigl(\,
       \alpha^*_{+} \, \beta^{ }_{-} 
   \, - \, 
      \alpha^{ }_{-} \, \beta^{*}_{+} 
     \,\bigr) (\tau'\,)
     \,\bigr]
  \;\Bigr] \, \biggr\} 
  \;. \hspace*{6mm} \nonumber
\ea
Note that to compute the first term explicitly, 
we would need the correct $\omega$-dependent $\beta^{ }_\pm$'s, 
following from \eq\nr{def_alpha}.
The remaining integrals in \eq\nr{res_Jvac_2}
can be carried out without problems, 
even with $\omega\to 0$, 
since the trigonometric functions give them time
dependence. 

Subsequently, inserting $\alpha^{ }_\pm$, $\beta^{ }_\pm$ 
from \eqs\nr{res_beta} and \nr{res_alpha}, we obtain
\ba
  J^{\vac\zero}_{pqk}(\tau) 
 & 
 \overset{ \rmii{\nr{res_Jvac_2}}  \vphantom{ |^b_q} }{
 \underset{\rmii{\nr{res_beta},\nr{res_alpha}}}{=}}
 & 
 \frac{ H^4_{ } r }
 {8 k^{10}_{ } (r^2_{ } - a^2_{ }) (r^2_{ } - 1 )^2_{ } }
 \biggl\{\,
  5 r^2_{ }+ 61 + k^2_{ }\tau^2_{ }(5 r^2_{ }+ 13)
  + 2 k^4_{ }\tau^4_{ }(r^2_{ } - 3)
  \nn[2mm]
 & & 
 - \,  
   \frac{16\, (1 + k^2_{ }\tau^2_{ })}{(r^2_{ } - 1)^2_{ }}
 + \frac{8\, (1 + k^2_{ }\tau^2_{ })}{ r^2_{ } - 1 }
 - \, \frac{64\, (1 + k^2_{ }\tau^2_{ })}{(r^2_{ } - 1)(r^2_{ } - a^2_{ })}
 \la{J_vac_1}  \\[2mm]
 & & 
  - \,
   \frac{32 a^2_{ }\,
   [\, 5 - 3 a^2_{ } + k^2_{ }\tau^2_{ }(3  - a^2_{ }) \,]}
    {(r^2_{ } - a^2_{ })^2_{ }}
 \; - \,
   \frac{8\, [\, 13 -21 a^2_{ } + k^2_{ }\tau^2_{ } (11 - 7a^2_{ }) \,]}
    { r^2_{ } - a^2_{ } }
 \,\biggr\}
 \;. \nonumber 
\ea

The integrals of the terms in \eq\nr{J_vac_1} 
can be carried out in dimensional regularization. 
We list the individual contributions in \app\ref{app:I_J}.
Their sum yields 
\ba
 \delta \P_\tensor^{\scriptscriptstyle J,\mathrm{vac}}(\tau,k)
 & 
 \underset{\rmii{\nr{def_A}}}{
 \overset{\rmii{\nr{P_J_cl_1}}  \vphantom{ |^b_q} }{=}} 
 & 
 \overbrace{
 \biggl[\,
 \frac{ H^4_{ }}{16 \pi^2_{ }\mpl^4 } \, + \, \rmO(\delta)
 \,\biggr]
 }^{ 
 { \A H^4_{ }} / ({ 8\hspace*{0.3mm} k^{10}_{ } })  
 }
 \; 
 \biggl[\,
 \sum_i \mathcal{J}^{ }_i(\tau,k)
 \; + \; \rmO(1)
 \,\biggr]
 \;, \la{P_J_vac_2}
 \\[3mm]
 \sum_i \mathcal{J}^{ }_i(\tau,k)
 & 
 \overset{\rmii{\nr{j1}--\nr{j3}}\vphantom{\big | }}{=} 
 & 
 \rmO(\delta)
 \;. \la{sum_J}
\ea
The term of $\rmO(1)$ in \eq\nr{P_J_vac_2} 
is incomplete, since we have
not accounted for the $\rmO(\delta)$-corrections to the mode functions, 
however the absence of a divergence is unambiguous. 


Let us now return to \eq\nr{J_vac_1}, 
and analyze its IR structure, similarly
to \eq\nr{Jcl_ir}. We obtain
\be
  J^{\vac\zero}_{pqk}(\tau) 
 \;
 \underset{r\;\approx\;1}{
 \overset{\rmii{\nr{J_vac_1}} \vphantom{ |^b_q} }{\approx}} 
 \;
 \frac{ H^4_{ } (1 + k^2_{ }\tau^2_{ }) }
 {2 k^{10}_{ } (r^2_{ } - a^2_{ }) (r^2_{ } - 1 )^2_{ } }
 \biggl[\,
 - \frac{1}{(r-1)^2} - \frac{7 + a^2_{ }}{(r-1)(1-a^2_{ })}
 + ... 
 \,\biggr]
 \;. \la{Jvac_ir}
\ee
We see that the IR divergences cancel 
between \eqs\nr{Jcl_ir} and \nr{Jvac_ir}. 
This means that in reality, 
$ 
  \delta \P_\tensor^{\scriptscriptstyle J,\mathrm{vac}}
$
does contain logarithmic divergences, but they originate both
from IR (``$1/\delta^{ }_\rmii{IR}$'') 
and UV domains (``$1/\delta^{ }_\rmii{UV}$''),
and cancel against each other within the vacuum contribution,  
\be
 \delta \P_\tensor^{\scriptscriptstyle J,\mathrm{vac}}(\tau,k)
 \;
 \underset{\rmii{\nr{sum_J},\nr{Jvac_ir}}}{
 \overset{\rmii{\nr{sum_I}} \vphantom{ |^b_q} }{=}} 
 \;
 \frac{(8\pi G)^2_{ }H^4_{ }}{\pi^4_{ }} 
 \biggl[
   \frac{3\, (1 + k^2_{ }\tau^2_{ })}{40}
   \biggl( \frac{1}{\delta^{ }_\rmii{UV}}
    - \frac{1}{\delta^{ }_\rmii{IR}} \biggr)
  + \rmO(1)
 \biggr]
 \;. \la{P_J_vac_final}
\ee
The sum of the classical and vacuum contributions yields
\be
 \delta \P_\tensor^{\scriptscriptstyle J,\mathrm{cl}}(\tau,k)
 \; + \; 
 \delta \P_\tensor^{\scriptscriptstyle J,\mathrm{vac}}(\tau,k)
 \;
 \underset{\rmii{\nr{P_J_vac_final}}}{
 \overset{\rmii{\nr{sum_I}}}{=}} 
 \;
 \frac{(8\pi G)^2_{ }H^4_{ }}{\pi^4_{ }} 
 \biggl[
   \frac{3\, (1 + k^2_{ }\tau^2_{ })}{40}
   \frac{1}{\delta^{ }_\rmii{UV}}
  + \rmO(1)
 \biggr]
 \;. \la{P_J_final}
\ee
We have factored out the same coefficient
as in ref.~\cite{madrid}, helping to 
see that the $1/\delta$ pole agrees with their \eq(79), 
once the latter is multiplied by two, to account for both
helicities.  

To summarize, the IR divergence in 
$ \delta \P_\tensor^{\scriptscriptstyle J,\mathrm{cl}} $
(cf.\ \eq\nr{sum_I}) gets cancelled, if we rather compute
the full 1-loop tensor power spectrum (cf.\ \eq\nr{P_J_final}). 
It is therefore not physical. 
 As a deeper reason for this cancellation,
 we can envisage a connection to the KLN theorem~\cite{kln1,kln2}, with
 the role of real corrections played by the classical part, 
 and that of virtual corrections by
 the vacuum part.

%
\subsection{Analysis of the power divergence in cutoff regularization}
\la{ss:ct}

In \fig\ref{fig:Jcl_bare}, 
we have shown the cutoff-regularized result for the classical
contribution to the tensor power spectrum, demonstrating that it diverges
as $(\mu/H)^3_{ }$ for $\mu/H \gg 1$. 
In \ses\ref{ss:dr} and \ref{ss:vac}, we have analyzed
the classical and vacuum contributions
in dimensional regularization, 
showing that the classical contribution is sensitive to small momentum
transfer, $r\approx 1$. In this section, 
we explain how the cutoff-regularized result differs from the dimensionally
regularized one; how the cubic divergence in 
\fig\ref{fig:Jcl_bare} originates; 
and what can perhaps be done about it. 

To guide the eye, we first plot the integrand 
relevant for \eq\nr{J_cl_cutoff}, 
{\it viz.}
\be
 \mathcal{G}(\cdot)
 \;
 \underset{\rmii{\nr{calF}}  }{
 \overset{\rmii{\nr{J_cl_cutoff}} \vphantom{ | } }{\equiv }}
 \;
 \frac{k^2_{ }}{2}
 \, (r^2_{ } - 1)^2_{ }
 \int_{-1}^{+1} \! {\rm d}a \, (1 - a^2_{ })^2_{ } 
 \, \F(\cdot)
 \;, 
 \la{calG} 
\ee
as a function of $r$. The result is shown in \fig\ref{fig:Jcl_igd},
for $\mu/H = 100$ and $|k\tau| = 0.1$. 
Three features
are important: unlike \eq\nr{Jcl_ir}, the integrand is finite
as $r\to 1^+_{ }$; for intermediate~$r$, it is rapidly
oscillating, with an average magnitude $\sim (\mu/H)^2_{ }$; 
at large~$r$, there is a cancellation, and the integrand vanishes. 
In the following, we elaborate on these features, and how they combine
to produce the observed divergence in \fig\ref{fig:Jcl_bare}. 

\begin{figure}[t]

\hspace*{-0.1cm}
\centerline{%
 \epsfysize=7.3cm\epsfbox{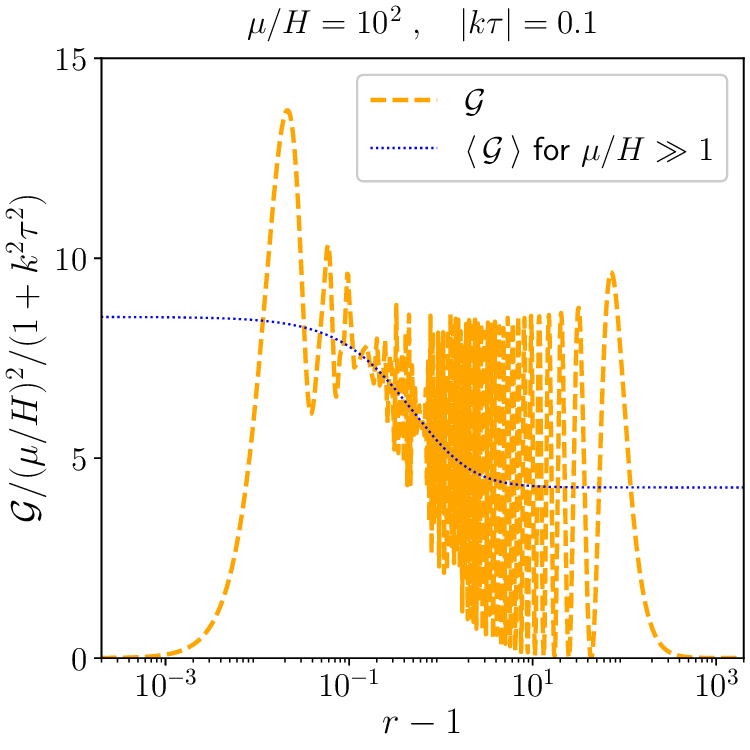}%
}

\caption[a]{\small
  The $r$-integrand from \eq\nr{calG}, denoted by~$\mathcal{G}$,
  for $\mu/H = 100$ and $|k\tau| = 0.1$, as well as its
  oscillation average, $\langle\mathcal{G}\rangle$, 
  from \eq\nr{aveG}.
  The key features of this plot are discussed in the text. 
  }

\la{fig:Jcl_igd}
\end{figure}

\bi

\item {\em Small $r$}. 
When $r\approx 1$, the coefficient $\beta^{ }_-$ 
diverges as $\sim 1/(r-1)^2_{ }$ (cf.\ \eq\nr{res_beta}).
Given that $\beta^{ }_-$ appears quadratically in 
$\F$ (cf.\ \eq\nr{calF}), and the measure in \eq\nr{calG}
is quadratic in $(r-1)$, we could expect $\mathcal{G}$ to diverge
as $\sim 1/(r-1)^2_{ }$. This is indeed observed in the dimensionally
regularized expression, in \eq\nr{Jcl_ir}. However, in the 
cutoff-regularized expression, the divergences $1/(r-1)^2_{ }$
and $1/(r-1)$ cancel, and even the finite term cancels, so that
$\mathcal{G}$ goes to zero as $r\to 1^+_{ }$. This cancellation
relies on the angle $\theta^{ }_-$ being small in \eq\nr{calF}.
According to \eq\nr{beta_pm}, this happens when
\be
 \theta^{ }_- \; \ll \; 1 
 \; 
 \overset{\rmii{\nr{beta_pm}}\vphantom{\big | }}{\Longrightarrow}
 \; 
 r - 1 \; \ll \; \frac{1}{|k\tau^{ }_\rmii{min}|}
 \; 
 \overset{\rmii{\nr{shorthands}}}{\approx} 
 \; 
 \frac{H}{2\mu}
 \;. \la{small_r}
\ee

\item {\em Intermediate $r$}.
The small-$r$ cancellation is lifted when $\theta^{ }_- \sim 1$. 
Then the dominant terms
come from the part proportional to $k\tau'$ in \eq\nr{res_beta}, 
with $\tau'\to \tau^{ }_\rmii{min}$ and 
$|k\tau^{ }_\rmii{min}| \gg 1$. Averaging furthermore
the exponential phase factors to zero in \eq\nr{calF}, we find
\be
 \langle\, \mathcal{G}(\cdot) \,\rangle
 \; 
 \underset{\rmii{\nr{calG}} }{
 \overset{\rmii{\nr{calF}}\vphantom{ | }}{=}} 
 \; 
 \frac{16 (r^2_{ } + 1)\,
          (1 + k^2_{ }\tau^2_{ })\, 
               k^2_{ }\tau^2_\rmii{min}}{15}
 \; + \; \rmO(1)
 \;, \la{aveG}
\ee
where $\langle ... \rangle$ indicates 
the oscillation average, and $k \tau^{ }_\rmii{min}$
is from \eq\nr{shorthands}. 
Eq.~\nr{aveG} gives the overall
scale of $\mathcal{G}$, 
as shown with the dotted blue curve in \fig\ref{fig:Jcl_igd}.

\item {\em Large $r$}.
At large $r$, when $\tau^{ }_\rmii{min}(r)\to\tau$ (cf.\ \eq\nr{shorthands}), 
another cancellation between the oscillating and 
non-oscillating terms in \eq\nr{calF} takes place. Concretely, 
expanding the integrand $\F(\cdot)$
in $k (\tau - \tau^{ }_\rmii{min})$, 
the zeroth and first-order terms vanish. 
Thus the cancellation becomes effective when 
\be
 \theta^{ }_+ - \theta^{ }_- 
 \; 
 \overset{\rmii{\nr{beta_pm}}}{=} 
 \; 
  2 k (\tau - \tau^{ }_\rmi{min})
 \; 
 \ll
 \;
  1
 \; 
 \underset{\scriptscriptstyle |k\tau|\;\ll\; 1}{
 \overset{\rmii{\nr{shorthands}}\vphantom{\big | }}{\Longrightarrow}} 
 \; 
 r \; \gg \; \frac{4\mu}{H}
 \;. \la{large_r}
\ee

\ei
The overall divergence of 
$
 \delta \P_\tensor^{\scriptscriptstyle J,\mathrm{cl}}
$ 
is obtained by multiplying the magnitude of the integrand, 
from \eq\nr{aveG}, with the extent of the integration range, 
from \eq\nr{large_r}, yielding $\sim (\mu/H)^3_{ }$. 

It may be wondered if the cubically divergent contribution to 
$
 \delta \P_\tensor^{\scriptscriptstyle J,\mathrm{cl}}
$ 
could be treated as a counterterm, 
to be subtracted from the results of simulations. We note, however, that
such power divergences are ``non-universal'', depending on how precisely
the momentum cutoff is implemented.
Therefore, in practical terms, it may be useful to implement
the cancellation on a non-perturbative level. 
This means that rather than addressing the absolute value of  
$
 \delta \P_\tensor^{\scriptscriptstyle J,\mathrm{cl}}
$, 
we pose questions about how 
$
 \delta \P_\tensor^{\scriptscriptstyle J,\mathrm{cl}}
$
depends on relevant model parameters. Most simply, this could
be achieved by using the very same framework (in particular with
identical momentum cutoffs) for two different computations, one
of them with a simple enough self-interaction potential that the 
result can be estimated analytically, 
and the other for the actual case of interest. Then, only 
the difference of the two results bears physical significance. 

%
\section{Conclusions and outlook}
\la{se:concl}

The purpose of this paper has been to
clarify the relation between quantum and classical 
computations of the inflationary tensor power spectrum. 
We have worked in de Sitter spacetime throughout, 
considering thus times well before Hubble horizon re-entry.
With the example of a spectator scalar field, we have shown
that the second-order classical contribution, which is 
closely related to scalar-induced gravitational waves (SIGW), 
can be viewed as a part of a full 1-loop computation.
However, though intuitively attractive and a target of 
possible numerical simulations, the classical contribution
does {\em not} isolate the low-energy aspects of the problem, 
as an effective theory would do. Rather, 
in dimensional regularization, 
we have shown that it suffers from  
unphysical IR divergences from small 
momentum transfer (cf.\ \eq\nr{Jcl_ir}). These small momenta
contribute to the signal for a very long time. 
If we instead sum together the 
classical and vacuum contributions, to  
obtain the full 1-loop quantum result, then the IR divergences cancel 
(cf.\ \eq\nr{Jvac_ir}). Left over 
are finite terms and the usual UV divergences, 
which can be renormalized
with well-established counterterms. 

Despite the issues associated with the classical contribution, 
it should not be given up lightly, because it offers a possible 
avenue towards non-perturbative (i.e.\ non-Gaussian) numerical simulations. 
With this in mind, we have re-investigated 
its divergence structure in the presence of 
a well-defined momentum cutoff. In order to maintain dilatation invariance, 
the cutoff is imposed on physical rather than comoving momenta. 
Due to inflationary redshift, small comoving 
momenta were large physical momenta
in the distant past, and are therefore only evolved for
a finite amount of time. 
We have shown that this renders the classical contribution 
IR finite on its own, with the price that it is strongly
dependent on the cutoff (cf.\ \se\ref{ss:cutoff}). To extract physical
information from such a result, requires a subtraction procedure,
for which we have suggested
a non-perturbative implementation (cf.\ \se\ref{ss:ct}). 
It remains to be seen, however, whether such a procedure
can be turned into a practical tool. 
  
Though we have reported several concrete results in this paper, 
our main points have been on the conceptual side. 
On technical aspects, much further work could be envisaged. 
Apart from test simulations, 
let us list a few of the
most imminent possible extensions:
\bi

\item
We have referred to classicality only on the gravitational side
(it would require 
physical phenomena, 
such as thermal scatterings, to decohere the scalar side).
However,  
given that the divergences of the classical contribution originate
from the IR domain of small momentum transfer, 
it could be asked if effective
descriptions could be used 
for the scalar fields (cf., e.g., ref.~\cite{usr}). 
The usual suspects
are an effective theory of inflation~\cite{ieft1,ieft2}; 
soft de Sitter effective theory~\cite{sdseft1,sdseft2};  
and their open effective theory extensions~\cite{oeft}.

\item
Independent of the formalism, it would be important to 
determine corrections to 
$
 \delta \P_\tensor^{\scriptscriptstyle J,\mathrm{cl}}
$
from the parameters $\xi$ and $m^2_{ }/H^2_{ }$, which introduce
additional ``scales'' to the problem, and could lead to
a more natural distinction between 
domains than a massless field. 
While easily 
done for integrals in which the time and momentum dependences
get factorized (cf.\ \eq\nr{c_d_nu}), this is cumbersome
for non-factorized cases, such as the classical contribution.
It might be worth exploring whether 
Wick rotation to imaginary time
(cf.,\ e.g.,\ ref.~\cite{explicit}), 
or Mellin-Barnes representation of Hankel functions
(cf.,\ e.g.,\ refs.~\cite{mb0,mb}), 
could offer tools for this generalization.

\item
The computation should be extended to 
an inflaton rather than a spectator field, whereby 
scalar metric perturbations need to be included. Then
general relativistic gauge invariance can be used as a nice
consistency check of the framework.

\item
Going beyond de Sitter spacetime, it would be valuable
to determine corrections from $\dot{H}/H^2_{ } \neq 0$, which 
should be manageable at least in an effective theory setup.

\item
Turning to the reheating epoch, 
the computations should ultimately be formulated with a 
general time dependence of the Hubble rate, $H$, and by
including plasma perturbations 
as additional degrees of freedom~\cite{sigw_theory}. 

\item
Finally, on the side of dimensional regularization, 
we recall that contributions
of $\rmO(\delta)$ should be included in the mode functions, to 
obtain the finite parts of the results
(cf.,\ e.g.,\ refs.~\cite{classic,madrid}). Though these terms are 
of limited phenomenological interest, 
since they are numerically very small, for mathematical
completeness it would be nice to account for them. However, 
as shown by \eqs\nr{c_d_nu} and \nr{nu}, if a way is found
to incorporate $\xi$ and $m^2_{ }/H^2_{ }$, which we consider
to be a more pressing challenge (see above), 
we should also be able to include all orders of $\delta$ at once. 

\ei
We hope to return to some of these topics in future work. 

%
\section*{Acknowledgements}

M.L.\ thanks Anna Tokareva for inspiring discussions in early 2025, 
and the Swiss National Science Foundation (SNSF) for supporting 
her visit to Bern, under grant IZSEZ0-232400.

\newpage

%
\appendix
\renewcommand{\thesection}{\Alph{section}}
\renewcommand{\thesubsection}{\Alph{section}.\arabic{subsection}}
\renewcommand{\theequation}{\Alph{section}.\arabic{equation}}

%
\section{Non-linear parametrization of gravitational perturbations}
\la{app:A}

\normalsize

In the main body of the text, we have parametrized the metric, 
$g^{ }_{\mu\nu}$, 
according to \eq\nr{g_mn_nl}. 
Such a ``non-linear'' parametrization is often used
in 1-loop computations, as it simplifies some 
of the expressions. 
On the other hand, 
in the classical study of scalar-induced gravitational waves, 
a ``linear'' parametrization is frequently adopted, 
\be
 g^\rmii{(linear)}_{\mu\nu}
 \; \equiv \; 
 a^2_{ } \, \bigl(\, \eta^{ }_{\mu\nu} + h^{ }_{\mu\nu}\,\bigr)
 \;, \la{g_mn}
\ee
where 
$
 \eta^{ }_{\mu\nu} \equiv 
 \eta^{\mu\nu}_{ } \equiv 
 \mathop{\mbox{diag}}(\mbox{$-$$+$$+$$+$})
$. 
An advantage of the linear parametrization is that 
it makes it straightforward to analyze gauge dependence.
For the benefit of readers accustomed to the linear 
parametrization, we briefly summarize here the main aspects
of the non-linear one, staying in $d=3$ for brevity
(relevant $d$-dimensional generalizations are given in \app\ref{app:mode}). 

In the non-linear parametrization, we employ the 
metric in \eq\nr{g_mn_nl} and the inverse metric in \eq\nr{g^mn_nl}.
To compute the Ricci scalar, it is convenient to denote
\be
 h^\tensor_{ij} \; \equiv \; 2 \vartheta^\tensor_{ij}
 \;, \la{vartheta}
\ee
whereby factors of $1/2$ can be avoided. Showing all terms before
inserting special properties of the tensor channel, 
we get
\ba
 R 
 \;
 & 
 \overset{\rmii{\nr{g_mn_nl}} \vphantom{ |^b_q } }{
 \underset{\rmii{\nr{g^mn_nl}}}{=}} 
 & 
 \;
 \frac{1}{a^2_{ }} \biggl\{ \, 
 6(\H' + \H^2_{ })
 + 
 2 \bigl[ 
  (\partial^{ }_\tau + 3 \H) 
  \cancel{ \vartheta^{\tensor\prime}_{kk} }
 + \cancel{ \vartheta^\tensor_{kl,kl} }  
 - \cancel{ \vartheta^\tensor_{kk,ll} }
 \bigr]
 \nn[3mm]
 & & \qquad 
 \; + \, 
     \vartheta^{\tensor\prime}_{ij} \vartheta^{\tensor\prime}_{ij} 
 -  \vartheta^\tensor_{ij,k} \vartheta^\tensor_{ij,k}
 \nn[3mm]
 & & \qquad
 \; + \, 
 \cancel{ \vartheta^{\tensor\prime}_{kk}
 \vartheta^{\tensor\prime}_{ll} }
 \; - \; 
 \cancel{
 \vartheta^{\tensor}_{kk,i}
 \vartheta^{\tensor}_{ll,i} 
 } 
 \; + \;
 4 \vartheta^{\tensor}_{ij} 
 \cancel{
  \vartheta^{\tensor}_{kk,ij}
 }
 \; - \;
 4 \vartheta^{\tensor}_{ij} 
 \cancel{
  \vartheta^{\tensor}_{ik,jk}
 }
 \nn[3mm]
 & & \qquad
 \; + \, 
 4
 \cancel{ 
  \vartheta^{\tensor}_{ij,i} 
 }
 \cancel{
  \vartheta^{\tensor}_{kk,j}
 }
 \; - \, 
 2
 \cancel{
  \vartheta^{\tensor}_{ij,i} 
 }
 \cancel{
  \vartheta^{\tensor}_{jk,k}
 }
 \biggr\}
 \; + \; \rmO(\vartheta^3_{ })
 \;. \la{R_nl}
\ea
We also need the determinant of the metric, which is trivial
thanks to the tracelessness of the perturbation, 
as well as 
$
 \det\exp = \exp\tr
$, 
\be
 g 
 \; \equiv \; 
 \det g^{ }_{\mu\nu}
 \;
  \overset{\rmii{\nr{g_mn_nl}}}{=} 
 \;
 - \, a^8_{ }  
 \;. \la{det_g}
\ee
Consequently, the non-$\varphi$ part of \eq\nr{S} gives 
the effective Lagrangian 
$\mathcal{L} = \bar{\mathcal{L}} + \delta\mathcal{L}$, with  
\ba
 \delta \L 
 &
 \underset{\rmii{\nr{R_nl},\nr{det_g}}}{
 \overset{\rmii{\nr{S}} \vphantom{ |^b_q } }{=}}
 & 
 \frac{a^2_{ }}{16\pi G}
 \bigl(\, 
 \vartheta^{\tensor\prime}_{ij} \hspace*{0.3mm} \vartheta^{\tensor\prime}_{ij} 
 \; - \; 
 \vartheta^\tensor_{ij,k} \hspace*{0.3mm} \vartheta^\tensor_{ij,k}  
 \,\bigr)
 \; + \; \rmO(\vartheta^3_{ })
 \;. \la{del_L_nl}
\ea
The Euler-Lagrange equation then directly yields the gravitational
wave equation, 
\be
 \vartheta^{\tensor\prime\prime}_{ij} 
 + 2 \H \vartheta^{\tensor\prime}_{ij}
 - \nabla^2_{ } \vartheta^\tensor_{ij} 
 \;
 \overset{\rmii{\nr{del_L_nl}}}{=}
 \; 
 0
 \; + \; 
 \mbox{(non-linear terms)}
 \;. \la{eom_ht_nl}
\ee

\newpage

%
\section{Mode functions in $d$ spatial dimensions}
\la{app:mode}

\normalsize

In order to evaluate the 2-point
gravitational correlator in \eq\nr{G_t}, 
field operators of the interaction picture are needed. 
As the field operators
evolve with the non-interacting Hamiltonian, 
$\hat H^{ }_{\scriptscriptstyle 0}$, they satisfy classical 
evolution equations. 
The Euler-Lagrange equation for $\varphi$, originating from 
\eq\nr{S}, reads 
\be
 \varphi\hspace*{0.3mm}{}''
 + (d-1) \H \varphi\hspace*{0.3mm}{}' - \nabla^2_{ }\varphi
 + a^2_{ } \bigl(\, m^2_{ } + \xi \bar R \,\bigr) \varphi \; = \; 0
 \;, \la{eom_varphi}
\ee
where 
$
 \H 
 \equiv  
 {a'} / {a}
$, 
and $\bar R$ denotes the Ricci scalar evaluated
at the background level. 
We refer to the case $m^2_{ } = \xi = 0$
as a {\em minimally coupled massless scalar field}.
The tensor perturbation from \eq\nr{rescale}, $\varh^\tensor_{ij}$, 
also obeys \eq\nr{eom_varphi}, just with $m^2_{ } = \xi = 0$
(cf.\ \eq\nr{eom_ht_nl}). 

For transparent quantization, it is helpful to 
rescale the fields into dimensionless (``conformal'') ones
according to \eq\nr{hat_phi}, 
i.e.\ 
$ 
 \widehat\varphi 
 \; \equiv \; 
 a^{\frac{d-1}{2}}_{ } \hspace*{0.3mm} \varphi
$. 
Then the evolution equation becomes
\be
 \widehat{\varphi}\hspace*{0.4mm}'' - \nabla^2_{ }\widehat\varphi
 + 
 \overbrace{
 \biggl[
 \,
 a^2_{ } \bigl(\, m^2_{ } + \xi \bar R \,\bigr)
 - \frac{d-1}{2}\frac{a''}{a} 
 - \frac{(d-1)(d-3)}{4}\biggl(\frac{a'}{a}\biggr)^2_{ }
 \, 
 \biggr]
 }^{ \; \equiv \; \widehat m^2_{ }(\tau) }
 \,\widehat\varphi
 \; 
 \underset{\rmii{\nr{hat_phi}}}{
 \overset{\rmii{\nr{eom_varphi}} \vphantom{ |^b_q } }{=}}
 \; 
 0
 \;. \la{e-l-2}
\ee
The canonical commutator that is expressed in physical coordinates as 
$
 [\varphi(t,a\hspace*{0.3mm}\vec{x}),
  \dot{\varphi}(t,a\hspace*{0.3mm}\vec{y})]
 = 
 i \delta^{(d)}_{ }(a\hspace*{0.3mm}\vec{x} - a\hspace*{0.3mm}\vec{y})
$,
becomes
\be
 \bigl[\,
  \widehat\varphi(\tau,\vec{x})\,,
  \,\widehat\varphi\hspace*{0.4mm}'(\tau,\vec{y}) 
 \,\bigr]
 \; = \; 
 i \delta^{(d)}_{ }(\vec x - \vec y)
 \;, \la{commutator}
\ee
and similarly for $ {\widehat\varh}^\tensor_{ij} $.

We now define a scalar mode expansion as 
\ba
 \widehat\varphi\hspace*{0.3mm}(\tau,\vec{x})
 & \equiv & 
 \int \! \frac{{\rm d}^d_{ }\vec{p}}{\sqrt{(2\pi)^d_{ }}}
 \,\biggl[\, 
   w_\vec{p}^{ }\,
  \Phi^{ }_p(\tau)\,
   e^{ i \vec p\cdot \vec x }_{ }
 \; + \; 
   w_\vec{p}^\dagger\,
   \Phi\hspace*{0.3mm}{}^*_p(\tau)\,
   e^{- i \vec p\cdot \vec x }_{ }
 \,\biggr]
 \;. \la{mode_exp_phi}
\ea
The creation and annihilation operators are normalized as 
\be
 \bigl[\, 
  w_\vec{p}^{ } \,,\, w^\dagger_\vec{q}
 \,\bigr]
 \; = \; 
 \delta^{(d)}_{ }(\vec p - \vec q)
 \;. \la{w_x}
\ee
Then we see that \eq\nr{commutator} is satisfied, if
the Wronskian of the mode function obeys
\be
 2 \im \bigl[\,
 \Phi^{ }_p(\tau)
 \, \partial^{ }_\tau
 \Phi\hspace*{0.3mm}{}^*_p(\tau)
 \,\bigr]
 \; = \; 1
 \;. \la{wronskian}
\ee
The annihilation operator, $w^{ }_\vec{p}$, is defined
by demanding that $\Phi_p^{ }(\tau)$ display a forward time evolution
in the infinite past, 
$
 \Phi^{ }_p(\tau\to-\infty) \sim e^{-ip\tau}/\sqrt{2p \vphantom{|}}
$.

For the gravitational perturbations, the structure of 
the vertices (cf.\ \eq\nr{S_to_H})
is such that it is convenient to go back to canonical
normalization after quantization, 
\ba
 \varh^\tensor_{ij}(\tau,\vec{x})
 & \equiv & 
 \int \! \frac{{\rm d}^d_{ }\vec{k}}{\sqrt{(2\pi)^d_{ }}}
 \sum_{\lambda }
 \,\biggl[\, 
   x_\vec{k}^{\lambda}\,
  \epsilon^\lambda_{ij,\vec k}
  \,\widetilde\phi^{ }_k(\tau)\,
   e^{ i \vec k\cdot \vec x }_{ }
 \; + \; 
   x_\vec{k}^{\lambda\dagger}\,
   \epsilon^{\lambda\hspace*{0.3mm}*}_{ij,\vec k}
   \,\widetilde\phi\hspace*{0.3mm}{}^*_k(\tau)\,
   e^{- i \vec k\cdot \vec x }_{ }
 \,\biggr]
 \;, \la{mode_exp_hij}
\ea
where 
\be
 \widetilde\phi^{ }_k(\tau)
 \; 
 \underset{\rmii{\nr{mode_exp_phi}}}{
 \overset{\rmii{\nr{hat_phi}} \vphantom{ |^b_q } }{\equiv}} 
 \; 
 a^{\frac{1-d}{2} }_{ } \, \Phi^{ }_k(\tau) |^{ }_{m \; = \; \xi \; = \; 0}
 \;, \la{widetilde_phi}
\ee
and the creation and annihilation operators are normalized as 
\be
 \bigl[\, 
  x_\vec{k}^{\lambda} \,,\, x^{\lambda'\dagger}_\vec{q}
 \,\bigr]
 \; = \; 
 \delta^{\lambda\lambda'}_{ }
 \delta^{(d)}_{ }(\vec k - \vec q)
 \;. \la{x_k}
\ee
The polarization tensors, 
$ \epsilon^\lambda_{ij,\vec k} $, 
are symmetric in $i \leftrightarrow j$, 
traceless and transverse, and 
satisfy the orthonormality and completeness relations 
\be
 \sum_{i,j}
 \epsilon^{\lambda}_{ij,\vec{k}}
 \epsilon^{\lambda'\hspace*{0.3mm}*}_{ij,\vec{k}}\,
 \; = \;
 \delta^{\lambda\lambda'}_{ } 
 \;, \quad
 \sum_{\lambda}
 \epsilon^{\lambda}_{ij,\vec{k}}\,
 \epsilon^{\lambda\hspace*{0.3mm}*}_{mn,\vec{k}}
 \;
 = 
 \;
 \mathbbm{T}^{ }_{ij;mn}
 \;. \la{eps_props}
\ee
Here the traceless 
($
 \delta^{ }_{ij} \mathbbm{T}^{ }_{ij;mn} = 
 0
$)
and 
transverse 
($
 k^{ }_i \mathbbm{T}^{ }_{ij;mn} = 
 0 
$)
projector reads 
\be 
 \mathbbm{T}^{ }_{ij;mn}
 \; 
 \equiv
 \; 
 \frac{
   \mathbbm{K}^{ }_{im} 
   \mathbbm{K}^{ }_{jn}
 +  
   \mathbbm{K}^{ }_{in} 
   \mathbbm{K}^{ }_{jm}
 }{2}
 - \frac{
   \mathbbm{K}^{ }_{ij} 
   \mathbbm{K}^{ }_{mn} 
   }{d-1}
 \;, \quad
 \mathbbm{K}^{ }_{ij} 
 \; 
 \equiv 
 \; 
 \delta^{ }_{ij}
  - \frac{k^{ }_i k^{ }_j}{ {k}^2_{ } }
 \;. \la{def_T}
\ee

For the further steps, we specialize to de Sitter spacetime, where 
\be
 a
 \; \equiv \;
 -\, \frac{1}{H\tau}
 \;, \quad
 \frac{a'}{a} 
 \; = \;
 - \, 
 \frac{1}{\tau}
 \;, \quad
 \frac{a''}{a} 
 \; = \;
 \frac{2}{\tau^2_{ }}
 \;, \quad
 \tau \; \in \; (-\infty,0)
 \;. \la{de_Sitter}
\ee
Inserting the $d$-dimensional expression for the Ricci scalar, 
$ 
 \bar R 
  =  
      { 2 d\hspace*{0.3mm} a'' } / { a^3_{ }} 
 + 
     { d(d-3) a^{\prime\hspace*{0.3mm}2}_{ } } / { a^4_{ } }
$, 
the equation for $\Phi^{ }_p$ takes the form 
\be
 \Phi\hspace*{0.3mm}{}''_p 
 +
  p^2_{ }\Phi^{ }_p
 + 
 \overbrace{
 \biggl[\,
 \frac{m^2_{ }}{H^2_{ }}
  + (d + 1) \biggl(\,  d \hspace*{0.3mm} \xi - \frac{d-1}{4} \,\biggr) 
 \,\biggr]
 }^{\rm const}
 \,
 \frac{\Phi^{ }_p}{\tau^2_{ }}
 \; 
 \underset{\rmii{\nr{de_Sitter}}}{
 \overset{\rmii{\nr{e-l-2}} \vphantom{ |^b_q } }{=}}
 \; 
 0
 \;. \la{mode_eqn}
\ee

A simple solution of \eq\nr{mode_eqn}, normalized according to 
\eq\nr{wronskian}, can be found if 
we consider 
three spatial dimensions, $d=3$, 
the massless limit, $m=0$,
and choose a minimal coupling, $\xi = 0$. Then 
\ba
 \Phi_p^{\zero}(\tau)
 & \equiv &
  \lim_{d\,\to\,3}
  \Phi^{ }_p(\tau)
 \;, \la{def_Phi0} \\[3mm]
 \lim_{m\,\to\, 0\,,\,\xi\,\to\,0}
  \Phi^{\zero}_p(\tau)
 & = & 
 \frac{e^{-i p \tau}_{ }}{\sqrt{2p \vphantom{ | }}}
 \biggl(\, 
  1 - \frac{i}{p\tau}
 \,\biggr)
 \;. \la{phi_minimal} 
\ea
For the gravitational mode function, the rescaling by $a^{(1-d)/2}_{ }$
from \eq\nr{widetilde_phi} implies that 
\be
 \widetilde \phi^{\hspace*{0.4mm}\zero}_k (\tau) 
 \; = \; 
 i H \, 
 \frac{e^{-i k \tau}_{ }}{\sqrt{2k^3_{ }}}
 \bigl(\, 
  1 + i k \tau
 \,\bigr)
 \;. \la{tilde_phik}
\ee

For a general $d$, $m$, or $\xi$, 
the mode function can be represented
in terms of a Hankel function, 
$H^{\one}_\nu = J^{ }_\nu + i Y^{ }_\nu$,
where $J^{ }_\nu$ and $Y^{ }_\nu$ are Bessel functions, and 
\be 
 \nu \; \equiv \; 
 \sqrt{\frac{d^{\hspace*{0.3mm}2}_{ }}{4}
 - \frac{m^2_{ }}{H^2_{ }} - d(d+1)\xi }
 \;. \la{nu}
\ee 
The asymptotics of the Hankel functions are
usually displayed with positive argument, 
\be
  H^{\one}_\nu(s) 
  \; 
  \overset{s\to +\infty}{\approx}
  \; 
  \sqrt{\frac{2}{\pi s}}
  \, 
  e^{i\bigl( s - \frac{\nu\pi}{2} - \frac{\pi}{4} \bigr) }_{ }
 \;. \la{H1}
\ee
With this expression we see that 
the solution satisfying the correct initial condition is 
\be
 \Phi^{ }_p(\tau) 
 \; = \; 
 \frac{\sqrt{-\pi\tau}}{2}
 \, e^{i \frac{\pi}{2} ( \nu + \frac{1}{2} ) }_{ }
 H^{\one}_{\nu}(-p\tau)
 \;, \quad 
 \tau \;\le\; 0
 \;. \la{phi_massive}
\ee

\vspace*{3mm}

In dimensionally regularized computations, 
we write $d=3 + \delta$ and expand in $\delta\ll 1$. 
In principle, the series can be obtained by
expanding in the index $\nu$ from \eq\nr{nu}, however 
$
 \partial^{ }_\nu H^{\one}_{\nu\vphantom{l}}
$
is not readily evaluated analytically. Instead, we can work out
the expansion directly from \eq\nr{mode_eqn}, 
where $d$ appears in the last term. If we consider a small perturbation
thereof, we can write the equation as 
\be
 \biggl(\, \mathcal{D} + \frac{\epsilon}{\tau^2_{ }} \,\biggr)
 \Phi^{ }_p \; = \; 0
 \;, \la{eps_exp}
\ee
where $\mathcal{D}$ is the unperturbed differential operator. 
Eq.~\nr{eps_exp} can be solved in analogy with \eq\nr{pert_p}.
Writing the solution as 
$
 \Phi^{ }_p = \Phi_p^{\zero} + \Phi_p^{\one} + ...
$, 
the first-order correction satisfies
\be
 \mathcal{D} \Phi_p^{\one} 
 \; = \; - \frac{\epsilon}{\tau^2_{ }} \, \Phi^{\zero}_p
 \;, \la{Phi1_def}
\ee
which can be integrated into 
\be
 \Phi^{\one}_p(\tau)
 \; = \; 
 -\epsilon \int_{-\infty}^{\tau} \! 
 \frac{ {\rm d}\tau' }{(\tau^{\prime}_{ })^2_{ }} \, 
 [G^\rmii{R}_p(\tau,\tau'\,)]^{\zero}_{ } \, 
 \Phi^{\zero}_p(\tau ')
 \;. \la{Phi1_res}
\ee
Here the zeroth-order Green's function has the form in \eq\nr{GR_vs_Phi}, 
evaluated with $\Phi^{\zero}_p$. 

In some cases, the $\rmO(\delta)$-correction 
can be determined analytically. For 
a massless minimally coupled scalar field, 
with 
$\nu = 3/2 + \delta/2$ and 
$\epsilon=-3\hspace*{0.3mm}\delta/2$, 
\eqs\nr{phi_massive} and \nr{Phi1_res} yield
\ba
 \lim_{m\,\to\, 0\,,\,\xi\,\to\,0}
 \Phi^{\one}_p(\tau) 
 & 
 \overset{\rmii{\nr{phi_massive}}}{
 \underset{}{=}} 
 & 
 \frac{\delta}{2}
 \, \partial^{ }_{\nu}
 \biggl[\, 
 \frac{\sqrt{-\pi\tau}}{2}
 e^{i \frac{\pi}{2} ( \nu + \frac{1}{2} ) }_{ }
 H^{\one}_{\nu}(-p\tau)
 \,\biggr]^{ }_{\nu \; = \; 3/2}
 \nn[3mm]
 & \overset{\rmii{\nr{Phi1_res}}}{=} & 
 \frac{\delta}{2}
 \frac{e^{-i p \tau}_{ }}{\sqrt{2 p \vphantom{|}}}
 \, \frac{ 
 2 + (1 - i p \tau ) e^{2 i p \tau}_{ } E^{ }_1(2 i p \tau)
 }{i p \tau}
 \;, \la{Phi1_min}
\ea
where $E^{ }_1$ is an exponential integral (cf.\ \eq\nr{E1}).
Similarly, the correction to 
$
 \widetilde \phi^{\hspace*{0.4mm}\zero}_k
$, 
defined according to \eq\nr{widetilde_phi}, reads
\be
 \widetilde \phi^{\hspace*{0.4mm}\one}_k (\tau) 
 \;
 \underset{\rmii{\nr{Phi1_min}}}{
 \overset{\rmii{\nr{widetilde_phi}} \vphantom{ |^b_q } }{=}}
 \; 
 \frac{\delta}{2}\,
 i H \, 
 \frac{e^{-i k \tau}_{ }}{\sqrt{2k^3_{ }}}
 \Bigl[\, 
 \bigl(\, 
  1 + i k \tau
 \,\bigr) \ln(-H\tau)
 + 2 + 
 \bigl(\, 
  1 - i k \tau
 \,\bigr)
 e^{2 i k \tau }_{ } E^{ }_1(2 i k \tau) 
 \,\Bigr]
 \;. \la{tilde_phik_1}
\ee
Recalling the small-$|z|$ behaviour
$
 E^{ }_1(z) = - \gamma^{ }_\rmii{E} - \ln z + \rmO(z)
$, 
we see that the logarithmic small-$|\tau|$ divergence cancels
from $ \widetilde \phi^{\hspace*{0.4mm}\one}_k (\tau) $. 

Equation~\nr{tilde_phik_1} gives as an opportunity to 
return to the discussion in the paragraph below \eq\nr{rescale}, 
concerning the appearance of the renormalization scale, $\mu$.
As mentioned there, the field $\varh^\tensor_{ij}$, and subsequently
also the mode function $\widetilde \phi^{\hspace*{0.4mm}\one}_k$,
are always accompanied by a factor   
$
 \mu^{-\delta/2}_{ }  
$.
Moreover, each momentum integral brings in a factor 
$k^{\hspace*{0.2mm}d}_{ } = 
k^{\hspace*{0.2mm}3+\delta}_{ }$, from which the part 
$k^{\hspace*{0.2mm}\delta/2}_{ }$ 
can be associated with the mode function
(since a propagator involves two mode functions). 
Once we expand $(k/\mu)^{\hspace*{0.2mm}\delta/2}_{ }$ to $\rmO(\delta)$, 
the cancellation of $\ln|\tau|$ in \eq\nr{tilde_phik_1}
is accompanied by a cancellation of $\ln k$, and 
the remaining $\ln H$ turns into $\ln(H/\mu)$. 
Further logarithms can come from scalar mode functions
(cf.\ \eq\nr{Phi1_min} and the discussion
below \eq\nr{E1}), and they combine
into the type $\ln|k\tau|$. 
The latter have been observed to cancel
against the corresponding logarithm 
in \eq\nr{res_L_dr}~\cite{madrid}.


Finally, we remark that 
the corrections arising through the function $E^{ }_1$ in 
\eqs\nr{Phi1_min}--\nr{tilde_phik_1} can be important in loop
computations, because they change the large-$p$ 
behaviour of the mode functions. Employing the asymptotic representation
\ba
 E^{ }_1(z) 
 \; 
 \overset{\scriptscriptstyle \mathrm{Re} z\; \ge \; 0 \vphantom{\big |}}{=} 
 \; 
 \int_1^\infty \! {\rm d}t \, \frac{e^{ -t z}_{ }}{t}
 & 
 \overset{\scriptstyle t\; \to \; 1+t \vphantom{\big | }}{\simeq} 
 &
 e^{-z}_{  }
 \int_0^\infty \! {\rm d}t
 \, \bigl( 1 - t + t^2_{ } - t^3_{ } + \ldots \bigr) \, e^{ -t z}_{ }
 \nn[2mm]
 & = & 
 e^{ -z}_{ } \, \biggl(
  \frac{1}{z} - \frac{1}{z^2_{ }} + \frac{2}{z^3_{ }} - \frac{6}{z^4_{ }} 
  + \ldots \biggr) 
 \;, \la{E1}
\ea
we see that $E^{ }_1$ has a tail of power-suppressed momentum dependences
(with factorially growing coefficients). 
The loops contributing to the tensor power spectrum are 
power-divergent. One of the power-suppressed terms in \eq\nr{E1} 
therefore leads to a logarithmic integral, which gives
a $1/\delta$ pole in dimensional regularization. 

\newpage

%
\section{Relation of mode functions and Green's functions}
\la{app:greens}

\normalsize

If a computation is carried out with 
path integrals rather than the canonical formalism, 
the basic objects appearing are various types of Green's functions.
Denoting $\X \equiv (\tau,\vec{x})$ and $\Y \equiv (\tau',\vec{y})$, 
2-point functions with different time orderings are defined as 
\ba
 \Delta(\X,\Y)
 &  
 \equiv 
 & 
 \big\langle \, 0 \, \big|\, 
 \tfr{1}{2}
 \bigl[\, 
       \widehat\varphi(\X) \,\widehat\varphi(\Y)
     + \widehat\varphi(\Y) \,\widehat\varphi(\X) 
 \,\bigr]
 \,\big| \, 0 \, \big\rangle
 \; 
 \equiv
 \; 
  G^{ }_{rr}(\X,\Y)
 \;, \la{def_Delta}
 \\[3mm]
 \varrho(\X,\Y)
 &  
 \equiv 
 & 
 \big\langle \, 0 \, \big|\, 
 \tfr{1}{2}
 \bigl[\,
       \widehat\varphi(\X) \,\widehat\varphi(\Y)
     - \widehat\varphi(\Y) \,\widehat\varphi(\X)
 \,\bigr] 
 \,\big| \, 0 \, \big\rangle
 \;, \la{def_rho}
 \\[3mm]
 G^\rmii{R}_{ }(\X,\Y)
 &  
 \equiv 
 & 
 + \hspace*{0.3mm} 2 i \hspace*{0.3mm} \theta(\tau - \tau'\,)\, \varrho(\X,\Y) 
 \; 
 \equiv
 \;
 i \, G^{ }_{ra}(\X,\Y)
 \;, \la{def_GR}
 \\[3mm]
 G^\rmii{A}_{ }(\X,\Y)
 &  
 \equiv 
 & 
 - \hspace*{0.3mm} 2 i \hspace*{0.3mm} \theta(\tau' - \tau)\, \varrho(\X,\Y) 
 \; 
 \equiv
 \;
 i \, G^{ }_{ar}(\X,\Y)
 \;. \la{def_GA}
\ea
Here $\Delta$ is the so-called statistical function, 
$\varrho$ is the spectral function, 
$G^\rmii{R,A}_{ }$ are the retarded and advanced correlators, 
respectively, and 
$G^{ }_{xy}$, with $x,y \in \{r,a\}$, refer to the 
Schwinger-Keldysh formalism in the $r/a$ basis. 
In that case, $G^{ }_{aa}(\X,\Y) \equiv 0$.\footnote{%
 In the cosmological literature, the index $r$ is sometimes called
 classical (``$\mathrm{cl}$''), the index $a$ quantum 
 (``$\mathrm{q}$''); for a review, see appendix~A
 of ref.~\cite{clas15}. Thus the statistical function, $\Delta = G^{ }_{rr}$, 
 describes classical fluctuations, as underlined by its
 anticommutator nature. It can play an important role 
 also in thermal field theory~\cite{sch}. \la{fn:cl}
 }

The mode functions and Green's functions have relations to each 
other, obtained by inserting \eq\nr{mode_exp_phi} into
\eqs\nr{def_Delta}--\nr{def_GA}. To express the relations in 
a compact fashion, we go to a
mixed time-momentum presentation, and
indicate the momentum, $p \equiv |\vec{p}|$, as a subscript. Then products
of mode functions can be converted into Green's functions, via 
\ba
 \Phi_p^{ }(\tau) 
 \Phi_p^{*}(\tau'\,) 
 & = &  
 \Delta^{ }_p(\tau,\tau'\,) + 
 \varrho^{ }_{\hspace*{0.3mm}p}(\tau,\tau'\,)
 \;. \la{conversion}
\ea
The statistical and spectral functions have specific
symmetry properties, 
$
 \Delta^{ }_p(\tau',\tau) = \Delta^{ }_p(\tau,\tau'\,)
$
and 
$
 \varrho^{ }_p(\tau',\tau) = -\hspace*{0.3mm} \varrho^{ }_p(\tau,\tau'\,)
$.
If the time integrals involve time ordering, so that the integrand
contains $ \theta(\tau - \tau'\,) $ 
or $ \theta(\tau' - \tau) $, 
the spectral function
can be replaced by a retarded or advanced correlator, in accordance
with \eqs\nr{def_GR} and \nr{def_GA}. We adopt 
an implicit notation in which the momenta $p$ and $q$ imply that
we consider the propagator of $\widehat\varphi$, whereas $k$ refers to 
$\varh^\tensor_{ij}$.


For the graviton mode functions, we make use of the normalization
from \eq\nr{tilde_phik}, and indicate the momentum as $k$. Then, 
at zeroth order in dimensional regularization, 
\ba
 \widetilde\Delta^\rmii{$(0)$}_k(\tau,\tau'\,)
 & 
 \equiv
 &  
 \fr{1}{2} 
 \Bigl[\,
  \widetilde \phi^{\hspace*{0.4mm}\zero}_k (\tau) 
  \, 
  \widetilde \phi^{*\hspace*{0.4mm}\zero}_k (\tau'\,) 
  \; + \; 
  \widetilde \phi^{*\hspace*{0.4mm}\zero}_k (\tau) 
  \, 
  \widetilde \phi^{\hspace*{0.4mm}\zero}_k (\tau'\,) 
 \,\Bigr]
 \la{full_delta_k} \\[3mm]
 & 
 \overset{\rmii{\nr{tilde_phik}}}{
 \underset{ }{=}} 
 &  
 \frac{ H^2_{ } }{2 k^3_{ }}  \, 
 \re\bigl[\, 
  {e^{ik(\tau - \tau'\,)}_{ } (1 - i k \tau)(1+i k\tau'\,)}
 \,\bigr]
 \;, \la{delta_k} \\[3mm]
 \widetilde G^{\rmii{R$(0)$}}_k(\tau,\tau'\,)
 & 
 \equiv
 &  
 i \,\theta(\tau - \tau'\,)\,
 \Bigl[\,
  \widetilde \phi^{\hspace*{0.4mm}\zero}_k (\tau) 
  \, 
  \widetilde \phi^{*\hspace*{0.4mm}\zero}_k (\tau'\,) 
  \; - \; 
  \widetilde \phi^{*\hspace*{0.4mm}\zero}_k (\tau) 
  \, 
  \widetilde \phi^{\hspace*{0.4mm}\zero}_k (\tau'\,) 
 \,\Bigr]
 \la{full_gr_k} \\[3mm]
 & 
 \overset{\rmii{\nr{tilde_phik}}}{
 \underset{ }{=}} 
 & 
 \frac{ H^2_{ } }{ k^3_{ } } \,\theta(\tau - \tau'\,)\,
 \im\bigl[\, 
  {e^{ik(\tau - \tau'\,)}_{ } (1 - i k \tau)(1+i k\tau'\,)} 
 \,\bigr]
 \;. \hspace*{6mm} \la{gr_k}
\ea

\newpage

%
\section{Definition of a classical contribution to the tensor power spectrum}
\la{app:def_cl}

\normalsize

In this appendix, we determine what we call the classical
contribution to the tensor power spectrum. It is appropriate to
stress, however, that the notion 
``classical'' has been employed in various 
meanings in the literature 
(cf.,\ e.g., refs.~\cite{clas1,clas14,clas15,clas2} 
for enlarged notions of classicality, 
     to some of which we return 
     below \eq\nr{P_t_cl}). 
In our hybrid implementation, classicality refers solely 
to the gravitational-wave
side, while the source term is treated quantum-mechanically,  
{\em without} approximations such as decoherence.  
That said, on the linear level, 
the quantum-mechanical mode functions can still be solved from 
classical time-evolution equations, albeit in complexified field
space (a practical implementation of this approach, 
including gradual decoherence by a 
thermal environment, has been worked
out in refs.~\cite{dissip,fluctu}).   

Suppose that we compute the second-order contribution to the tensor
power spectrum from general relativity. 
To avoid clutter, we work in $d=3$ dimensions here. 
Adopting the 
normalization from \eq\nr{hat_phi}, and deriving the Euler-Lagrange
equation with the vertex from \eq\nr{S_I_1}, the evolution equation
reads
\be
 \biggl( \partial_\tau^2 - \nabla^2_{ } - \frac{a''}{a} \biggr)
 {\widehat\varh}^\tensor_{ij}(\tau,\vec{x})
 \; = \; 
 \frac{\sqrt{8\pi G}}{a(\tau)}\,
 \mathbbm{T}^{ }_{ij;mn} \, 
 \widehat\varphi^{ }_{,m}(\tau,\vec{x})
 \widehat\varphi^{ }_{,n}(\tau,\vec{x})
 \;. \la{evo_cl}
\ee
To streamline the notation, we suppress the indication
of whether the tensor projector, 
$\mathbbm{T}^{ }_{ij;mn}$, is in coordinate
or momentum space. 
Going to Fourier space in the spatial directions, and determining
a special solution with the Green's functions method, we find
\be
 {\widehat\varh}^\tensor_{ij}(\tau,\vec{k})
 \; = \; 
 \int_{-\infty}^\tau \! {\rm d}\tau' \, 
 G^\rmii{R}_k(\tau,\tau'\,) \, 
 \frac{\sqrt{8\pi G}}{a(\tau'\,)}\,
 \mathbbm{T}^{ }_{ij;mn} \, 
 \int_\vec{x} 
 \widehat\varphi^{ }_{,m}(\tau',\vec{x})
 \widehat\varphi^{ }_{,n}(\tau',\vec{x})
 \, e^{-i \vec{k}\cdot\vec{x}}_{ }
 \;, \la{sol_cl}
\ee
where we have abbreviated the spatial integral as 
$\int_\vec{x} \equiv \int \! {\rm d}^3_{ }\vec{x}$. 

In order to specify 
the scalar fields, we need to choose their
initial conditions. At early times, the scalar fields
are deep inside the Hubble horizon. We impose
quantum-mechanical initial conditions at this time. 
To this aim, it is helpful to represent the scalar fields
in the form of a mode expansion, 
and view their classical evolution equation as that for the
mode functions. In principle 
the time evolution can however take a different
form than in the quantum theory, notably it could include 
a stochastic Langevin evolution, coupling the scalar mode function
to a heat bath and its hydrodynamic modes~\cite{fluctu}. 
In any case, for the quantum side, we insert the mode expansion from 
\eq\nr{mode_exp_phi} for $\widehat\varphi$, yielding
\ba
 {\widehat\varh}^\tensor_{ij}(\tau,\vec{k})
 &
  \underset{\rmii{\nr{mode_exp_phi}}}
  {\overset{\rmii{\nr{sol_cl}} \vphantom{ |^b_q } }{=}}
 & 
 \int_{-\infty}^\tau \! {\rm d}\tau' \, 
 G^\rmii{R}_k(\tau,\tau'\,) \, 
 \frac{\sqrt{8\pi G}}{a(\tau'\,)} \, 
 \mathbbm{T}^{ }_{ij;mn} \, 
 \int \! {\rm d}^3_{ }\vec{p}  
 \int \! {\rm d}^3_{ }\vec{q}  
 \nn[3mm]
 & & \; \times \, \Bigl\{ \; 
 (-p^{ }_m\, q^{ }_n)
 \, \delta^{(3)}_{ }(\vec{p+q-k})
 \, w^{ }_\vec{p}
 \, w^{ }_\vec{q}
 \, \Phi^{ }_p(\tau'\,) 
 \, \Phi^{ }_q(\tau'\,) 
 \nn[3mm]
 & & \quad \; + \, 
 (p^{ }_m\, q^{ }_n)
 \, \delta^{(3)}_{ }(\vec{p-q+k})
 \, w^{\dagger}_\vec{p}
 \, w^{ }_\vec{q}
 \, \Phi^{*}_p(\tau'\,) 
 \, \Phi^{ }_q(\tau'\,) 
 \nn[3mm]
 & & \quad \; + \, 
 (p^{ }_m\, q^{ }_n)
 \, \delta^{(3)}_{ }(\vec{p-q-k})
 \, w^{ }_\vec{p}
 \, w^{\dagger}_\vec{q}
 \, \Phi^{ }_p(\tau'\,) 
 \, \Phi^{*}_q(\tau'\,) 
 \nn[3mm]
 & & \quad \; + \, 
 (-p^{ }_m\, q^{ }_n)
 \, \delta^{(3)}_{ }(\vec{p+q+k})
 \, w^{\dagger}_\vec{p}
 \, w^{\dagger}_\vec{q}
 \, \Phi^{*}_p(\tau'\,) 
 \, \Phi^{*}_q(\tau'\,)
 \;\Bigr\}
 \;.  \la{ht_Phi_1}
\ea

We can then extract the power spectrum. The original definition, 
from \eqs\nr{G_t} and \nr{P_t}, can in momentum space be expressed as
\be 
 \frac{
 \langle\, 0(t) \,| \,
                 h^\tensor_{ij}(t,\vec{k})
               \,
                 h^\tensor_{ij}(t,\vec{r}) 
          \,  |\, 0(t) \,\rangle
 }
 {
 \langle\, 0(t) \,| 
        \, 0(t) \,\rangle 
 }
 \; \overset{\rmii{\nr{G_t}}}{=} \; 
 (2\pi)^3_{ }\delta^{(3)}_{ }(\vec{k+r})
 \, G^{ }_{\tensor}(t,k)
 \;. \la{G_t_k}
\ee
We need to recall the changes of normalization from 
\eqs\nr{rescale} and \nr{hat_phi}. 
Non-trivial quantum expectation values, for $\vec{k}\neq\vec{0}$,
originate from 
\ba
 &  & 
      \delta^{(3)}_{ }(\vec{p+q-k})
    \,\delta^{(3)}_{ }(\vec{p'+q'+r})
    \,\langle 0 | \,
      w^{ }_\vec{p}
   \, w^{ }_\vec{q} 
   \, w^{\dagger}_{\vec{p}'} 
   \, w^{\dagger}_{\vec{q}'} 
    \,| 0 \rangle
 \nn[3mm]
 & = & 
      \delta^{(3)}_{ }(\vec{p+q-k})
    \,\delta^{(3)}_{ }(\vec{p'+q'+r})
    \,\bigl[\,
       \delta^{(3)}_{ }(\vec{q - p'}) 
     \,\delta^{(3)}_{ }(\vec{p - q'}) 
     \;+\; 
       \delta^{(3)}_{ }(\vec{p - p'}) 
     \,\delta^{(3)}_{ }(\vec{q - q'}) 
    \,\bigr]
 \nn[3mm]
 &  & 
 \xrightarrow[]
             {\scriptscriptstyle \int\!{\rm d}^3_{ }\vec{p}'
                                 \int\!{\rm d}^3_{ }\vec{q}'}
 \;
 2\, \delta^{(3)}_{ }(\vec{p+q-k}) \delta^{(3)}_{ }(\vec{k + r})
 \;. \la{ht_Phi_3}
\ea
All in all this yields
\ba
 \P^\cl_\tensor(\tau,{k})
 &
 \underset{\rmii{\nr{rescale},\nr{hat_phi}}}{
 \overset{\rmii{\nr{ht_Phi_1}--\nr{ht_Phi_3}} \vphantom{ |^b_q} }{=}}
 & 
 \frac{k^3_{ }}{2\pi^2_{ }}
 \frac{32\pi G}{a^2_{ }(\tau)}
 \int_{-\infty}^\tau \! {\rm d}\tau' \, 
 G^\rmii{R}_k(\tau,\tau'\,) \, 
 \int_{-\infty}^\tau \! {\rm d}\tau'' \, 
 G^\rmii{R}_k(\tau,\tau''\,) 
 \,\frac{16\pi G}{a(\tau'\,)a(\tau''\,)}
 \nn[3mm]
 & & \; \times \,  
 \int\! \frac{{\rm d}^3_{ }\vec{p}}{(2\pi)^3_{ }}
 \; 
 \underbrace{
 \mathbbm{T}^{ }_{ij;mn}
 \, p^{ }_i\, q^{ }_j \, p^{ }_m\, q^{ }_n 
 }_{ 
 \frac{1}{2}
 \bigl[\,
  {p}^2_{ } - \frac{(\vec p\cdot \vec k)^2_{ }}{{k}^2_{ }}
 \,\bigr]^2_{ }
 }
 \;
 \Phi_p^*(\tau''\,) \, \Phi_q^*(\tau''\,) \,  
 \Phi^{ }_p(\tau'\,) \, \Phi^{ }_q(\tau'\,) \,  
 \;. \la{P_t_cl}
\ea 
Subsequently, 
splitting the time domain into two halves, and renaming
$\tau' \leftrightarrow \tau''$ in one of them, 
yields \eq\nr{res_Jcl}. 

\vspace*{3mm}

We note that the graviton Green's functions in \eq\nr{P_t_cl} are
real and causal, as is appropriate for a classical solution. In addition, 
the scalar source is 
rendered real by the $\tau' \leftrightarrow \tau''$ symmetrization, 
as visible in \eq\nr{res_Jcl}. This is a clear manifestation of the 
``classical'' nature of this contribution. 
Yet another aspect is that
taking the real part in \eq\nr{res_Jcl} turns the scalar source into 
an anticommutator, which is the time ordering 
that has a classical limit
 (cf.\ the discussion in \app\ref{app:greens}, with the operators
 in \eq\nr{def_Delta} interpreted as composite objects rather 
 than elementary fields). 

 However, it is possible to adopt further-going notions of classicality than
 what we have done above. Physically speaking, the scalar fields can be
 approximated as {\em fully} classical only if they are decohered 
 by additional phenomena, 
 such as thermal damping and noise, which is a gradual dynamical 
 process, rather than an on/off notion; for such a comprehensive
 description, the framework
 itself must leave room for initial quantum coherence~\cite{fluctu}. 
 On the other hand, formally, the classical limit can be defined
 by taking $\hbar\to 0$ in a given correlation function and density
 matrix~\cite{clgt3}. However, in a quantum field theory
 involving loop corrections, this limit is typically
 singular, reflecting the Rayleigh-Jeans divergence. 
 Yet another possibility, of a more practical nature, is to 
 associate classical fluctuations with the propagator 
 $\Delta = G^{ }_{rr}$ of the 
 Schwinger-Keldysh formalism 
 in the $r/a$ basis~\cite{clas15} (cf.\ \eq\nr{def_Delta}).
 In the following, we display some steps of the last approach
 in dimensional regularization, 
 noting how it unfortunately complicates rather than
 simplifies our loop-level expressions. 

 As a starting point, we take the source term from 
 \eq\nr{P_t_cl}, as written in \eq\nr{res_Jcl}.
 Inserting the decomposition from \eq\nr{conversion}, and paying
 attention to time ordering in order to recognize retarded 
 propagators, we find
 \be
  \re\bigl[ 
  \Phi_p^*(\tau'\,) \, \Phi_q^*(\tau'\,) \,  
  \Phi^{ }_p(\tau''\,) \,\Phi^{ }_q(\tau''\,) \,  
  \bigr]
   \; \overset{\scriptscriptstyle \tau' \; > \; \tau''}{=} \;
   \Delta^{ }_p(\tau',\tau''\,)\, \Delta^{ }_q(\tau',\tau''\,) 
   - \frac{
   G^\rmii{R}_p(\tau',\tau''\,)\, G^\rmii{R}_q(\tau',\tau''\,)
     }{4}
  \;. \la{decompo}  
 \ee
 If we now focus on the first term of \eq\nr{decompo}, it can be written as 
 \ba
   \Delta^{ }_p(\tau',\tau''\,)\, \Delta^{ }_q(\tau',\tau''\,) 
 & = & 
 \frac{1}{2} \bit \Bigl\{\, 
  \re\bigl[ 
  \Phi_p^*(\tau'\,) \, \Phi_q^*(\tau'\,) \,  
  \Phi^{ }_p(\tau''\,) \,\Phi^{ }_q(\tau''\,) \,  
  \bigr]
 \nn[2mm]
 & & \quad + \, 
  \re\bigl[ 
  \Phi_p^*(\tau'\,) \, \Phi_q^{ }(\tau'\,) \,  
  \Phi^{ }_p(\tau''\,) \,\Phi^*_q(\tau''\,) \,  
  \bigr]
  \,\Bigr\}
 \;. \la{clcl}
 \ea
 The first line of \eq\nr{clcl}
 yields one half of the original $ J^\cl_{pqk} $ 
 from \eq\nr{res_Jcl}, whereas the second line
 amounts to a term in which the function 
 $Z(\tau,\tau')$ from \eq\nr{def_Z} gets replaced with 
 \be
   \bar{Z}(\tau,\tau'\,) 
 \; \equiv \; 
 \int_{ 
      }^{ } \! {\rm d}\tau'
 \, \widetilde G^\rmii{R$(0)$}_k(\tau,\tau'\,)
 \, \Phi^{\zero}_p(\tau'\,)
 \, \Phi^{\zero *}_q(\tau'\,)
  \;. \la{def_barZ}
 \ee
 In view of \eq\nr{phi_minimal}, we see that, apart from the square root
 prefactor, we just need to replace $q\to -q$ in our previous computation. 
 Going through the same steps as before, and adopting in the end the 
 variables from \eq\nr{qpm}, this implies that we need to 
 exchange $r\leftrightarrow -a$ in most of the terms in \eq\nr{Jcl_res}. 
 Specifically, defining a ``classical-classical''
 (for both the source and the gravitational-wave propagation) 
 contribution as 
 \be
  J^{\cl,\cl}_{pqk}(\tau) 
  \; 
     \underset{\rmii{\nr{decompo}}}{
     \overset{\rmii{\nr{res_Jcl}} }{\equiv}} 
  \;
 \int_{-\infty}^{\tau} \! {\rm d}\tau'
 \int_{-\infty}^{\tau'} \! {\rm d}\tau''
 \, \widetilde G^\rmii{R}_k(\tau,\tau'\,)
 \, \widetilde G^\rmii{R}_k(\tau,\tau''\,)
 \Delta^{ }_p(\tau',\tau''\,)\, \Delta^{ }_q(\tau',\tau''\,) 
 \;, 
 \la{res_Jclcl}
 \ee
 the result becomes 
 \ba
  J^{\cl,\cl\zero}_{pqk}(\tau) 
   &
    \overset{\rmii{\nr{Jcl_res}}  \lift }
   {\underset{\rmii{\nr{clcl}}}{=}} 
   & 
  \frac{ H^4_{ } }
  {8 k^{10}_{ } (r^2_{ } - a^2_{ })}
  \,\biggl\{\, 
  \frac{1}{ (r^2_{ } - 1 )^2_{ } }
  \biggl[\,
    \frac{8(1 + k^2_{ }\tau^2_{ })}{(r^2_{ } - 1)^2_{ }}
  + \, \frac{32(1 + k^2_{ }\tau^2_{ })}{(r^2_{ } - 1)(r^2_{ } - a^2_{ })}
 \nn[2mm]
  & & \hspace*{34mm} 
  + \, \frac{32(1 + a^2_{ }k^2_{ }\tau^2_{ })}{(r^2_{ } - a^2_{ })^2_{ }}
  + \, \frac{48 k^2_{ }\tau^2_{ }}{ r^2_{ } - a^2_{ } }
  + \, 2 k^4_{ }\tau^4_{ }
  \,\biggr]
  \nn[2mm]
  & & \hspace*{24mm} + \, 
  \frac{1}{ (1 - a^2_{ })^2_{ } }
  \biggl[\,
    \frac{8(1 + k^2_{ }\tau^2_{ })}{(1 - a^2_{ })^2_{ }}
  + \, \frac{32(1 + k^2_{ }\tau^2_{ })}{(1 - a^2_{ })(r^2_{ } - a^2_{ })}
 \nn[2mm]
  & & \hspace*{34mm} 
  + \, \frac{32(1 + r^2_{ }k^2_{ }\tau^2_{ })}{(r^2_{ } - a^2_{ })^2_{ }}
  - \, \frac{48 k^2_{ }\tau^2_{ }}{ r^2_{ } - a^2_{ } }
  + \, 2 k^4_{ }\tau^4_{ }
  \,\biggr]
  \,\biggr\}
  \;. \hspace*{6mm} 
  \la{Jclcl_res} 
 \ea
 This is integrated with the measure from \eq\nr{P_J_cl_1},
 with the weight 
 $ 
  (r^2_{ } - a^2_{ }) (r^2_{ } - 1)^2_{ } (1 - a^2_{ })^2_{ }
 $
 in the vicinity of $d=3$ dimensions. 

%
%
%
%
%

 We now see that the second structure in \eq\nr{Jclcl_res}, absent 
 from \eq\nr{Jcl_res}, introduces {\em additional divergences} 
 compared with those in \eq\nr{Jcl_ir}. First of all, the new
 terms introduce UV divergences at large $r$. Second, the new
 terms contain divergences at the boundaries of the angular $a$
 integration domain,  
 i.e.\ from collinear or anti-collinear momenta (cf.\ \eq\nr{ra}).
 Incidentally,  
 it is a deficit of the classical approximation that it does not distinguish
 between collinear and anti-collinear limits,
 though they are physically distinct, since
 it does not distinguish between particles and antiparticles. 
 And finally, the old IR divergences remain present,
 from the first structure in 
 \eq\nr{Jclcl_res}, though their magnitude has been halved.

 To conclude, using the purely classical source term
 (cf.\ \eq\nr{clcl}), defined in the 
 diagrammatic sense of the Schwinger-Keldysh formalism in the $r/a$ basis
 (cf.\ footnote~\ref{fn:cl} on p.~\pageref{fn:cl}), 
 complicates rather than simplifies the 1-loop integrals, 
 compared with the definition that we adopt in \eq\nr{P_t_cl}.
 As possible reasons, we suspect that
 the additional large-$r$ singularities might
 reflect the Rayleigh-Jeans divergences of classical field theory, 
 and the additional $|a|=1$ divergences the absence of 
 a particle-antiparticle notion.
 However, these diagnoses are only heuristic. That said, 
 the fact that summing together all the 
 terms implements cancellations, is easy to 
 appreciate on general grounds, as explained below \eq\nr{P_J_final}.

\newpage

%
\section{Momentum integrals in $d$ spatial dimensions}
\la{app:I_J}

\normalsize

In this appendix, we work out the momentum integrals leading
to \eqs\nr{sum_I} and \nr{sum_J}.
The integrals $\I^{ }_i$, 
originating from \eq\nr{Jcl_res}, with the measure
from \eq\nr{P_J_cl_1}, read
\ba
 \frac{ 
   \I^{ }_{\hspace*{0.3mm}1}(\tau,k)
  }{ 8 (1 + k^2_{ }\tau^2_{ }) } 
 & \equiv & 
 \int_{1}^{\infty} \!\!\!\! {\rm d}r 
 \int_{-1}^{+1} \!\!\!\! {\rm d}a
 \, (r^2_{ } - 1)^{\frac{d-3}{2}}_{ }
 \, (1 - a^2_{ })^{\frac{d+1}{2}}_{ } 
 \, \frac{1}{ (r^2_{ } - 1)^2_{ } } 
 \nn[2mm]
 & =  &  
  \frac{
                         \Gamma(\frac{d+3}{2})
                         \Gamma(\frac{d-5}{2})
                         \Gamma(3-\frac{d}{2})
                         }{2 \Gamma(2 + \frac{d}{2})}
 \;
 \underset{\delta \; \ll \; 1 \vphantom{|} }{
 \overset{d\;=\;3+\delta \vphantom{|} }{\approx}}
 \;
 - \frac{8}{15} \biggl(\frac{1}{\delta} + 2 \ln 2 \biggr) 
 + \frac{154}{225}
 \;, \la{i1} \\[3mm]
 \frac{ 
 \I^{ }_{\hspace*{0.3mm}2}(\tau,k)
 }{ 32 (1 + k^2_{ }\tau^2_{ }) } 
 & \equiv &
 \int_{1}^{\infty} \!\!\!\! {\rm d}r 
 \int_{-1}^{+1} \!\!\!\! {\rm d}a
 \, (r^2_{ } - 1)^{\frac{d-3}{2}}_{ }
 \, (1 - a^2_{ })^{\frac{d+1}{2}}_{ } 
 \, 
    \overbrace{
    \frac{1}{(r^2_{ } - 1)(r^2_{ }- a^2_{ })}
    }^{
  \textstyle
  \bigl(\,  \frac{1}{r^2_{ } - 1 \vphantom{|^b_q} }
          - \frac{1}{r^2_{ }- a^2_{ }  \vphantom{|^b_q} } \,\bigr)
  \frac{1}{1 - a^2_{ }  \vphantom{|^b_q} }
   }
 \nn[2mm]
 &
 \underset{\delta \; \ll \; 1 \vphantom{|} }{
 \overset{d\;=\;3+\delta \vphantom{|} }{\approx}}
 & 
 \int_{1}^{\infty} \!\!\!\! {\rm d}r 
 \int_{-1}^{+1} \!\!\!\! {\rm d}a
 \, (r^2_{ } - 1)^{\frac{d-5}{2}}_{ }
 \, (1 - a^2_{ })^{\frac{d-1}{2}}_{ }
 \;-\; 
 \int_{1}^{\infty} \!\!\!\! {\rm d}r 
 \int_{-1}^{+1} \!\!\!\! {\rm d}a
 \, \frac{1 - a^2_{ }}{r^2_{ } - a^2_{ }}
 \hspace*{-6mm} 
 \nn[2mm] 
 & = & 
                  \frac{
                         \Gamma(\frac{d+1}{2})
                         \Gamma(\frac{d-3}{2})
                         \Gamma(2-\frac{d}{2})
                         }{2 \Gamma(1 + \frac{d}{2})}
 \;+\; 1 - \frac{\pi^2_{ }}{4} 
 \nn[2mm]
 &
 \underset{\delta \; \ll \; 1 \vphantom{|} }{
 \overset{d\;=\;3+\delta \vphantom{|} }{\approx}}
 & 
   \frac{4}{3} \biggl(\frac{1}{\delta} + 2 \ln 2 \biggr) 
  - \frac{1}{9}  - \frac{\pi^2_{ }}{4} 
 \;, \\[3mm]
 \frac{
   \I^{ }_{\hspace*{0.3mm}3}(\tau,k)
  }{32} 
 & \equiv &
 \int_{1}^{\infty} \!\!\!\! {\rm d}r 
 \int_{-1}^{+1} \!\!\!\! {\rm d}a \,
    (r^2_{ } - 1)^{\frac{d-3}{2}}_{ }
 \, (1 - a^2_{ })^{\frac{d+1}{2}}_{ }
 \, \frac{ 1 + a^2_{ }k^2_{ }\tau^2_{ } 
 }{  (r^2_{ } - a^2_{ })^2_{ }  } 
 \nn[2mm]
 & 
 \underset{\delta \; \ll \; 1 \vphantom{|} }{
 \overset{d\;=\;3+\delta \vphantom{|} }{\approx}}
 & 
 - 2 + \frac{\pi^2_{ }}{4 }
 + k^2_{ }\tau^2_{ } \biggl( \frac{4}{3} - \frac{\pi^2_{ }}{8} \biggr)
 \;, \\[3mm]
 \frac{
   \I^{ }_{\hspace*{0.3mm}4}(\tau,k)
 }{ 48 k^2_{ }\tau^2_{ } } 
 & \equiv &
  \int_{1}^{\infty} \!\!\!\! {\rm d}r 
 \int_{-1}^{+1} \!\!\!\! {\rm d}a \,
    (r^2_{ } - 1)^{\frac{d-3}{2}}_{ }
 \, (1 - a^2_{ })^{\frac{d+1}{2}}_{ }
 \, \frac{ 1 
 }{  r^2_{ } - a^2_{ }  } 
 \;
 \underset{\delta \; \ll \; 1 \vphantom{|} }{
 \overset{d\;=\;3+\delta \vphantom{|} }{\approx}}
 \; 
  - \frac{4}{3} + \frac{\pi^2_{ }}{4} 
 \;, \la{i4} \hspace*{6mm} \\[3mm]
 \frac{
   \I^{ }_{\hspace*{0.3mm}5}(\tau,k)
 }{ 2 k^4_{ }\tau^4_{ } } 
 & \equiv &
 \int_{1}^{\infty} \!\!\!\! {\rm d}r 
 \int_{-1}^{+1} \!\!\!\! {\rm d}a
 \, (r^2_{ } - 1)^{\frac{d-3}{2}}_{ }
 \, (1 - a^2_{ })^{\frac{d+1}{2}}_{ } 
 \nn[2mm]
 & =  & 
   \frac{
                         \Gamma(\frac{d+3}{2})
                         \Gamma(\frac{d-1}{2})
                         \Gamma(1-\frac{d}{2})
                         }{2 \Gamma(2 + \frac{d}{2})}
 \; 
 \underset{\delta \; \ll \; 1 \vphantom{|} }{
 \overset{d\;=\;3+\delta \vphantom{|} }{\approx}}
 \; 
 - \frac{16}{15} 
 \;. \la{i5} 
\ea
The terms proportional to $\pi^2_{ }$ cancel in the sum, 
and we obtain \eq\nr{sum_I}. 


In the case of the vacuum contribution, the integrals originate
from \eq\nr{J_vac_1}. The measure is again given by \eq\nr{P_J_cl_1}, 
but the appearance of an additional factor $r$ in \eq\nr{J_vac_1}, 
compared with \eq\nr{Jcl_res}, leads to different integrals. 
In fact, the first several terms, where the integrand can be expressed
as a positive or negative power of $r^2_{ } - 1$, vanish. 
This can be seen by writing it as 
\be
 \int_1^\infty \! {\rm d}r\, r \, 
 \, (r^2_{ } - 1)^{\frac{d-3}{2}}_{ } (r^2_{ } - 1)^n_{ }
 \;
 \underset{d\; = \; 3 + \delta}{
 \overset{z\; = \; r^2_{ }}{=}}  
 \;
 \frac{1}{2}
 \int_1^\infty \! {\rm d}z\, (z-1)^{n + \frac{\delta}{2}}_{ }
 \; 
 \overset{\scriptscriptstyle z \; = \;  1 \;+\; y }{=} 
 \; 
 \frac{1}{2}
 \int_0^\infty \! {\rm d}y\, y^{n + \frac{\delta}{2}}_{ }
 \;.
\ee
As the integral has no scale, it vanishes in dimensional regularization. 

Proceeding to the non-vanishing terms in \eq\nr{J_vac_1}, 
we start by considering
\ba
 \frac{
  \J^{ }_1(\tau,k)
 }
 {-64 (1 + k^2_{ }\tau^2_{ })}
 & \equiv & 
    \int_1^\infty \!{\rm d}r \, r\,
    (r^2_{ } - 1)^{\frac{d-3}{2}}_{ }
    \int_{-1}^{+1} \! {\rm d}a \, 
    (1 - a^2_{ })^{\frac{d + 1}{2} }_{ }
    \; \frac{1}{(r^2_{ } - 1)(r^2_{ } - a^2_{ })}
 \nn[2mm]
 & \overset{\scriptscriptstyle r^2_{ } \; = \;  1 \;+\; y }{=} & 
 \frac{1}{2} 
    \int_0^\infty \! {\rm d}y \, y^{\frac{d-5}{2}}_{ }
    \int_{-1}^{+1} \! {\rm d}a \, 
    (1 - a^2_{ })^{\frac{d+1}{2} }_{ }
    \; \frac{1}{y + 1 - a^2_{ }}
 \nn[2mm]
 & \overset{\scriptscriptstyle y \; = \; x\,(1 \;-\; a^2_{ })}{=} & 
 \underbrace{
 \frac{1}{2} 
    \int_{-1}^{+1} \! {\rm d}a \, 
    (1 - a^2_{ })^{d - 2}_{ }
  }_{
 \textstyle
 \frac{\Gamma(1/2)\Gamma(d-1)}{2 \Gamma(d-{1}/{2}) \vphantom{ |^b_q }}
  }
 \; 
 \underbrace{
    \int_0^\infty \! {\rm d}x \, 
    \frac{ x^{\frac{d - 5}{2}}_{ } }{  x + 1 }
 }_{
 \textstyle
 \frac{\pi}{\cos(d \pi/2) \vphantom{ |^b_q } } 
 }
 \nn[2mm]
 & 
 \; 
 \underset{\delta \; \ll \; 1 \vphantom{|} }{
 \overset{d\;=\;3+\delta \vphantom{|} }{\approx}}
 & 
   \frac{4}{3} \biggl(\frac{1}{\delta} + 2 \ln 2 \biggr) 
  - \frac{20}{9}
 \;. \la{j1}
\ea
The other two integrals can be solved with the same substitutions, 
yielding 
\ba
 \frac{
  \J^{ }_2(\tau,k)
 }
 {-32 }
 & \equiv &
    \int_1^\infty \!{\rm d}r \, r\,
    (r^2_{ } - 1)^{\frac{d-3}{2}}_{ }
    \int_{-1}^{+1} \! {\rm d}a \, 
    (1 - a^2_{ })^{\frac{d + 1}{2} }_{ }
    \; \frac{
         a^2_{ }\,
         [\,
       5 - 3 a^2_{ } + k^2_{ }\tau^2_{ } (3 - a^2_{ } ) 
         \,]
            }{(r^2_{ } - a^2_{ })^2_{ }}
 \nn[2mm]
 & 
    \overset{ \scriptscriptstyle r^2_{ } \; = \;  1 \;+\; y }{
    \underset{\scriptscriptstyle y \; = \; x\,(1 \;-\; a^2_{ })}{=}} 
 & 
 \underbrace{
 \frac{1}{2} 
    \int_{-1}^{+1} \! {\rm d}a \, 
    (1 - a^2_{ })^{d - 2}_{ }
   \, 
     a^2_{ }\,
         [\,
       5 - 3 a^2_{ } + k^2_{ }\tau^2_{ } (3 - a^2_{ } )
         \,]
  }_{
 \textstyle
 \frac{[d(5 + 3 k^2_{ }\tau^2_{ }) - 2]
  \Gamma(1/2)\Gamma(d-1)}{4 \Gamma(d+{3}/{2}) \vphantom{ |^b_q }}
  }
 \; 
 \underbrace{
    \int_0^\infty \! {\rm d}x \, 
    \frac{ x^{\frac{d - 3}{2}}_{ } }{  (x + 1)^2_{ } }
 }_{
 \textstyle
 \frac{(d-3)\pi}{2 \cos(d \pi/2) \vphantom{ |^b_q } } 
 }
 \nn[2mm]
 & 
 \underset{\delta \; \ll \; 1 \vphantom{|} }{
 \overset{d\;=\;3+\delta \vphantom{|} }{\approx}}
 & 
   \frac{4 (13 + 9 k^2_{ }\tau^2_{ })}{105}
 \;, \la{j2} \\[3mm]
 \frac{
  \J^{ }_3(\tau,k)
 }
 {-8 }
 & \equiv &
    \int_1^\infty \!{\rm d}r \, r\,
    (r^2_{ } - 1)^{\frac{d-3}{2}}_{ }
    \int_{-1}^{+1} \! {\rm d}a \, 
    (1 - a^2_{ })^{\frac{d + 1}{2} }_{ }
    \; \frac{
       13 - 21 a^2_{ } + k^2_{ }\tau^2_{ } ( 11 - 7 a^2_{ } )
            }{ r^2_{ } - a^2_{ } }
 \nn[2mm]
 & 
    \overset{ \scriptscriptstyle r^2_{ } \; = \;  1 \;+\; y }{
    \underset{\scriptscriptstyle y \; = \; x\,(1 \;-\; a^2_{ })}{=}} 
 & 
 \underbrace{
 \frac{1}{2} 
    \int_{-1}^{+1} \! {\rm d}a \, 
    (1 - a^2_{ })^{d - 1}_{ }
   \, 
         [\,
       13 - 21 a^2_{ } + k^2_{ }\tau^2_{ } ( 11 - 7 a^2_{ } )
         \,]
  }_{
 \textstyle
 \frac{[13 d - 4 + (11 d + 2) k^2_{ }\tau^2_{ } ]
  \Gamma(1/2)\Gamma(d)}{2 \Gamma(d+{3}/{2}) \vphantom{ |^b_q }}
  }
 \; 
 \underbrace{
    \int_0^\infty \! {\rm d}x \, 
    \frac{ x^{\frac{d - 3}{2}}_{ } }{  x + 1 }
 }_{
 \textstyle
 \frac{-\pi}{\cos(d \pi/2) \vphantom{ |^b_q } } 
 }
 \nn[2mm]
 & 
 \underset{\delta \; \ll \; 1 \vphantom{|} }{
 \overset{d\;=\;3+\delta \vphantom{|} }{\approx}}
 & 
  - \frac{32 (1 + k^2_{ }\tau^2_{ })}{3}
    \biggl(\frac{1}{\delta} + 2 \ln 2 \biggr) 
  + \frac{16\,(311 + 323 k^2_{ }\tau^2_{ })}{315}
 \;. \la{j3}
\ea
Summing together, all the terms shown cancel,
and we obtain \eq\nr{sum_J}. 

\newpage


\small{

}

\end{document}